\documentclass[%
 superscriptaddress,
 amsmath,amssymb,
 aps,
prstab,
floatfix,
]{revtex4-2}

\usepackage{graphicx}
\usepackage{dcolumn}
\usepackage{bm}
\usepackage{siunitx}
\usepackage{xcolor}
\usepackage{hyperref}
\usepackage{csquotes}

\begin{document}

\preprint{APS/PRAB}

\title{Ultrahigh brightness beams from plasma photoguns}

\author{A.~F.~Habib}
\email[Corresponding author: ]{ahmad.habib@strath.ac.uk}
\affiliation{Scottish Universities Physics Alliance, Department of Physics, University of Strathclyde, Glasgow, UK} 
\affiliation{Cockcroft Institute, Sci-Tech Daresbury, Daresbury, Cheshire, UK}

\author{T.~Heinemann}\thanks{This author contributed equally to this work}
\affiliation{Scottish Universities Physics Alliance, Department of Physics, University of Strathclyde, Glasgow, UK} 
\affiliation{Cockcroft Institute, Sci-Tech Daresbury, Daresbury, Cheshire, UK}
\affiliation{Deutsches Elektronen-Synchrotron DESY, Hamburg, Germany}

\author{G.~G.~Manahan}
\affiliation{Scottish Universities Physics Alliance, Department of Physics, University of Strathclyde, Glasgow, UK} 
\affiliation{Cockcroft Institute, Sci-Tech Daresbury, Daresbury, Cheshire, UK}
\author{L.~Rutherford}
\affiliation{Scottish Universities Physics Alliance, Department of Physics, University of Strathclyde, Glasgow, UK} 
\affiliation{Cockcroft Institute, Sci-Tech Daresbury, Daresbury, Cheshire, UK}

\author{D.~Ullmann}
\affiliation{Scottish Universities Physics Alliance, Department of Physics, University of Strathclyde, Glasgow, UK} 
\affiliation{Cockcroft Institute, Sci-Tech Daresbury, Daresbury, Cheshire, UK}
\affiliation{Institute for Optics and Quantum Electronics, University of Jena, 07743 Jena, Germany}

\author{P.~Scherkl}
\affiliation{Scottish Universities Physics Alliance, Department of Physics, University of Strathclyde, Glasgow, UK} 
\affiliation{Cockcroft Institute, Sci-Tech Daresbury, Daresbury, Cheshire, UK}

\author{A.~Knetsch}
\affiliation{Deutsches Elektronen-Synchrotron DESY, Hamburg, Germany}

\author{A.~Sutherland}
\affiliation{Scottish Universities Physics Alliance, Department of Physics, University of Strathclyde, Glasgow, UK} 
\affiliation{Cockcroft Institute, Sci-Tech Daresbury, Daresbury, Cheshire, UK}
\affiliation{SLAC National Accelerator Laboratory, Menlo Park, California, USA}

\author{A.~Beaton}
\affiliation{Scottish Universities Physics Alliance, Department of Physics, University of Strathclyde, Glasgow, UK} 
\affiliation{Cockcroft Institute, Sci-Tech Daresbury, Daresbury, Cheshire, UK}

\author{D.~Campbell}
\affiliation{Scottish Universities Physics Alliance, Department of Physics, University of Strathclyde, Glasgow, UK} 
\affiliation{Cockcroft Institute, Sci-Tech Daresbury, Daresbury, Cheshire, UK}
\affiliation{Ludwig-Maximilians-Universität München, Am Coulombwall 1, 85748 Garching, Germany}

\author{L.~Boulton}
\affiliation{Scottish Universities Physics Alliance, Department of Physics, University of Strathclyde, Glasgow, UK} 
\affiliation{Cockcroft Institute, Sci-Tech Daresbury, Daresbury, Cheshire, UK}
\affiliation{Deutsches Elektronen-Synchrotron DESY, Hamburg, Germany}

\author{A.~Nutter}
\affiliation{Scottish Universities Physics Alliance, Department of Physics, University of Strathclyde, Glasgow, UK} 
\affiliation{Cockcroft Institute, Sci-Tech Daresbury, Daresbury, Cheshire, UK}
\affiliation{Helmholtz-Zentrum Dresden–Rossendorf, Dresden, Germany}

\author{O.~S.~Karger}
\affiliation{Department of Experimental Physics, University of Hamburg, Hamburg, Germany}

\author{M.~D.~Litos}
\affiliation{Center for Integrated Plasma Studies, Department of Physics, University of Colorado, Boulder, Colorado, USA}

\author{B.~D.~O'Shea}
\affiliation{SLAC National Accelerator Laboratory, Menlo Park, California, USA}

\author{G.~Andonian}
\affiliation{Department of Physics and Astronomy, University of California Los Angeles, USA}
\affiliation{Radiabeam Technologies, Santa Monica, CA 90404, USA}

\author{D.~L.~Bruhwiler}
\affiliation{RadiaSoft LLC, Boulder, CO 80301, USA}

\author{J.~R.~Cary}
\affiliation{Center for Integrated Plasma Studies, Department of Physics, University of Colorado, Boulder, Colorado, USA}
\affiliation{Tech-X Corporation, Boulder, USA}

\author{M.~J.~Hogan}
\affiliation{SLAC National Accelerator Laboratory, Menlo Park, California, USA}

\author{V.~Yakimenko}
\affiliation{SLAC National Accelerator Laboratory, Menlo Park, California, USA}

\author{J.~B.~Rosenzweig}
\affiliation{Department of Physics and Astronomy, University of California Los Angeles, USA}

\author{B.~Hidding}
\affiliation{Scottish Universities Physics Alliance, Department of Physics, University of Strathclyde, Glasgow, UK} 
\affiliation{Cockcroft Institute, Sci-Tech Daresbury, Daresbury, Cheshire, UK}

\date{\today}

\begin{abstract}
\noindent
Plasma photocathodes open a path towards tunable production of well-defined, compact electron beams with normalized emittance and brightness many orders of magnitude better than state-of-the-art. Such beams could have far-reaching impact for applications such as light sources, but also open up new vistas on high energy physics and high field physics.
We report on challenges and details of the proof-of-concept demonstration of a plasma photocathode in 90$^\circ$ geometry at SLAC FACET within the \enquote{E-210: Trojan Horse} program. 
Using this experience, alongside theoretical and simulation-supported advances, we discuss the upcoming \enquote{E-310: Trojan Horse-II} program at FACET-II. 

\end{abstract}

\maketitle

\section{\label{sec:intro}Introduction}

\noindent
Today's particle accelerators power high-energy particle colliders, synchrotron radiation facilities and x-ray free-electron lasers (X-FELs), and thus enable groundbreaking discoveries in fundamental physics, material and life science. These breakthroughs are driven by the generation of high-quality electron beams that are compact in both real and in phase space.
A compound figure of merit that expresses such quality features is the beam brightness, either in the form of the 5D brightness $B_{\mathrm{5D}}=2I/\epsilon_{\mathrm{n}}^2$, where $I$ and $\epsilon_{\mathrm{n}}$ are the beam current and the transverse normalised emittance, respectively, or the 6D brightness $B_{\mathrm{6D}}=B_{\mathrm{5D}}/0.1 \% \Delta W/W$, which in addition takes into account the energy spread.
Beam currents of kA and normalised emittances of mm mrad scale are what today constitutes high-brightness beams.
Photocathode-based electron guns capable of generating such high-brightness beams were an enabling key development, which in turn allowed e.g. the demonstration of the first X-FEL, the Linear Coherent Light Source (LCLS)~\cite{Pellegrini2012,emma2010first} at the SLAC National Accelerator Laboratory.

Such electron guns utilize a laser pulse to release a burst of electrons from the solid
cathode material, typically via the photoelectric effect.
The laser spot size on the cathode material, the laser pulse duration and the residual transverse momenta of the emitted electrons are defining parameters for the initial (thermal) emittance of the released electron population \cite{Dowell2009PhysRevSTAB.12.074201,Eduard2015PhysRevSTAB.18.063401}.
To minimize subsequent space charge induced emittance growth, rapid acceleration immediately after emission is required~\cite{Eduard2015PhysRevSTAB.18.063401}.
However, the extraction and acceleration fields in electron guns and linacs based on radio frequency (RF) metallic cavities are typically constrained to some tens of \SI{}{MV/m}, due to material integrity limits of the accelerator building blocks. 

The initial picoseconds level duration of the electron beam at few Ampere current is a feasible trade-off aiming at thermal emittance optimization while operating at the photocathode gun limits \cite{Zhou:2015hsxOptimizationPhotocathode}.
Multi-stage compression techniques are required to reduce the duration of the produced electron beams to as short as a few tens of femtoseconds 
for kA beam currents.
However, electron beams are then subject to coherent synchrotron radiation (CSR) in the dispersion sections, known to induce energy and density modulations along the electron bunch. These microbunching instabilities have a detrimental impact on the obtainable final beam emittance and brightness \cite{PhysRevSTAB.5.064401,HuangBCMI2002,DiMitriPRAB2013}.
Summarizing, generation and acceleration of low emittance, short-pulse, high-brightness electron beams in conventional linacs faces a number of  fundamental challenges and limitations. 
Due to the importance of electron beam brightness and emittance for FELs~\cite{photonics2020317}, novel approaches such as those based on high-field cryogenic photo-guns are investigated \cite{rosenzweig2018ultra}. 

Plasma-based accelerators may provide an alternative route towards next generation light sources, and perhaps colliders \cite{USroadmapColbydoi:10.1142/S1793626816300012,hidding2019plasmaPWASC}. 
The chief attraction of plasma wakefield accelerators are the huge accelerating and focusing electric fields that they can support through collective oscillation 
of plasma electrons around the ions. 
Electron-beam driven plasma wakefield accelerators (PWFA) enable not only tens of GeV energy gains of electrons on metre-scale distances \cite{Blumenfeld2007,Litos2014Natureshort} in phase-constant  accelerators, but also provide a unique environment to realize plasma-based electron injector guns. 

The plasma photocathode \cite{HiddingPRL2012PhysRevLett.108.035001} has been invented \cite{hiddingpatent2011} to exploit the strong electric fields in plasma not only for acceleration, but also for high-quality bunch generation. In such a plasma photogun, a laser pulse is harnessed to release electrons -- just as in a classical photocathode.
It features a 
high-ionization threshold (HIT) component as underdense photocathode medium, sustained by a background low-ionisation threshold (LIT) medium-based plasma wave. The tens of GV/m extraction fields of the plasma wave at the same time act as bunch compressor, in a single ultracompact building block. 
However, key differences are that i) the electrons are produced directly inside the  accelerating structure, where they are immediately subject to tens of GV/m-scale accelerating and focusing electric fields, ii) the electrons are obtained from tunneling ionization of a higher ionization threshold component instead of single or multi-photon ionization, and iii) the releasing laser pulse propagates through underdense matter, in contrast to the  laser-solid interaction in conventional photocathodes.
The plasma photocathode is largely decoupled  from the plasma wakefield accelerator, and allows releasing plasma electrons at arbitrary positions  within the plasma wave, as well as various laser pulse geometries and angles \cite{hiddingpatent2011}.
The plasma photocathode injector offers an enormous degree of flexibility for electron beam production, but most fundamentally and importantly, it is a path towards ultra-high brightness electron beams.
This is because  i) the electrons, being released by laser intensities just above the tunnel ionization threshold, are born initially \enquote{ultra-cold} with very small residual momentum and thus minimized thermal emittance, ii) are released in a small transverse volume, iii) space-charge induced emittance growth is very limited due to the rapid  acceleration to relativistic energies, and iv) phase-mixing is very small due to the localized release volume.
At the same time, velocity bunching of the electrons to fs or even sub-fs duration in the accelerating plasma cavity with a size of typically a few hundreds of microns, is inherently suitable to produce bunches at kA-level currents.
Combined with nm rad-level emittances, this results in beam brightness many orders of magnitude larger than what is feasible with conventional photocathode gun linacs.
The plasma photocathode laser pulse properties and its intensity form factor convoluted with the HIT medium density and profile can be  tailored to control the rate of released electrons in space and time.
This facilitates precise control over the properties of the trapped electron beams, e.g. by using non-Gaussian laser pulse profiles, various focusing optics, laser frequencies, or simultaneous spatial and temporal focusing (SSTF) laser pulses for confined ionization volumes \cite{Manahan2019Royal}. 

Further, by taking advantage of multi-bunch injection \cite{hiddingarxiv2014} with multiple plasma photocathode lasers, electron beam dechirping can be performed  in the same plasma acceleration stage by the \enquote{escort bunch} beam-loading approach.
The relative energy spread of these electron beams can be potentially as low as $\simeq0.01\%$ at a couple of GeV energies, or even lower, thus resulting in unprecedented ultra-high 6D brightness \cite{ManahanHabib6d}.  

In this letter, we first discuss and provide new insights into the results obtained  within the \enquote{E-210: Trojan Horse} experimental campaign at SLAC FACET, which for the first time demonstrated a plasma photocathode injector 
\cite{Deng2019Trojan}, in  perpendicular injector geometry.
Then, the \enquote{E-310: Trojan Horse-II} successor program at SLAC FACET-II and related experiments are discussed.
These aim at demonstrating the full potential of the capabilities of the plasma photocathode injector towards ultrabright beams.
Finally,  witness beam parameters producible at SLAC FACET-II and parameter scans, which highlight the prospects for remarkable stability of the plasma photocathode output, are presented.

\section{\label{sec:E-210}E-210: Trojan Horse at FACET}
\noindent
The \enquote{E-210: Trojan Horse} experiment was realized at SLAC FACET in $90^{\circ}$ geometry between plasma wave and injector laser.
This choice of geometry was a balanced outcome of experimental boundary conditions, and the strategy that enabled this first proof-of-concept realization. It involved the innovation, development and exploitation of the plasma afterglow diagnostics \cite{Scherklarxiv2019} and the  so-called plasma torch downramp injection method \cite{WittigPlasmaTorchPhysRevSTAB.18.081304,Ullmann2020Torch} as stepping stones. 

\begin{figure*}
\includegraphics[width=1.0\textwidth]{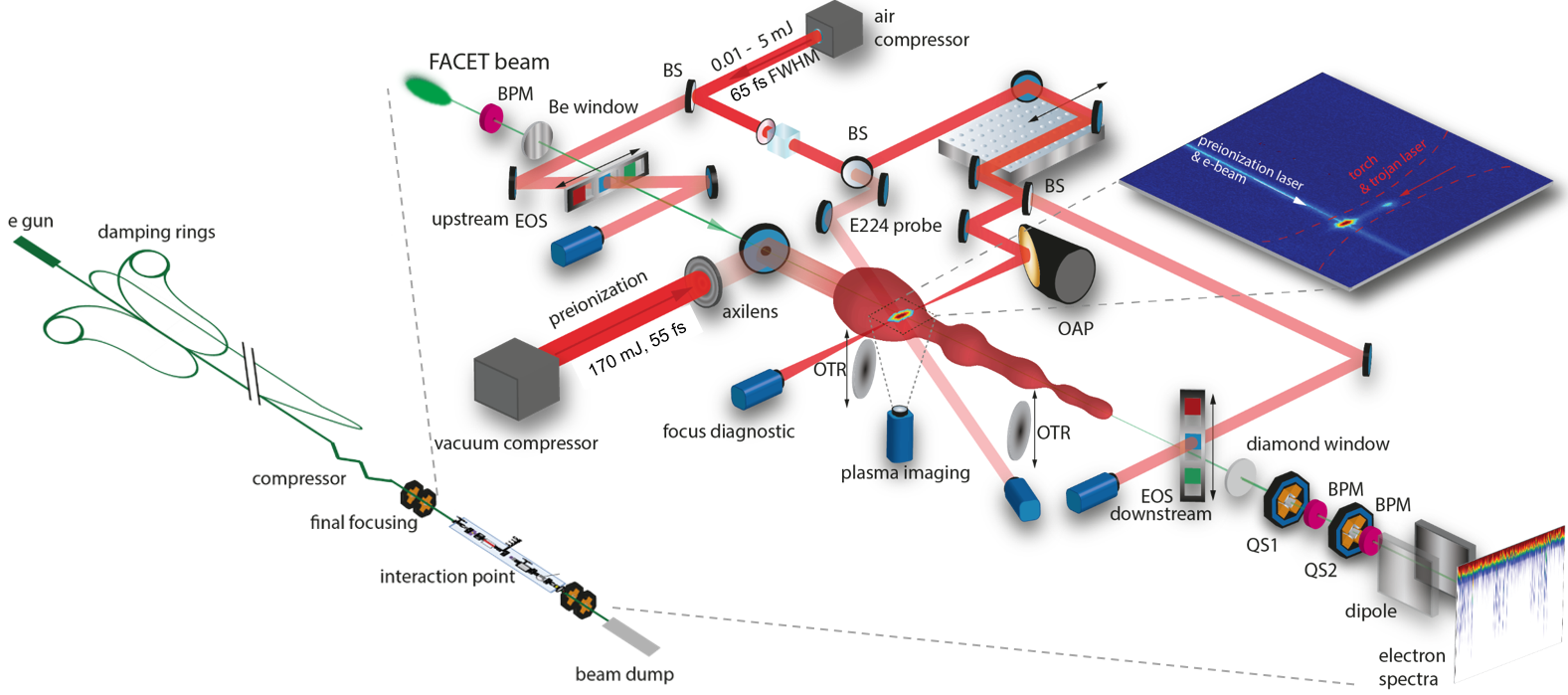}
\caption{\label{fig:E210setup}Setup of the E-210: Trojan Horse experiment at SLAC FACET. Key building blocks are FACET's electron driver beam, the preionization laser pulse that generates a hydrogen plasma channel around the electron beam axis, and the perpendicular injector laser pulse that ionized helium locally on electron beam propagation axis, thus enabling plasma afterglow, plasma torch and Trojan Horse experiments.}
\end{figure*}

\subsection{Experimental setup}
\begin{figure*}[h]
\includegraphics[width=1.0\textwidth]{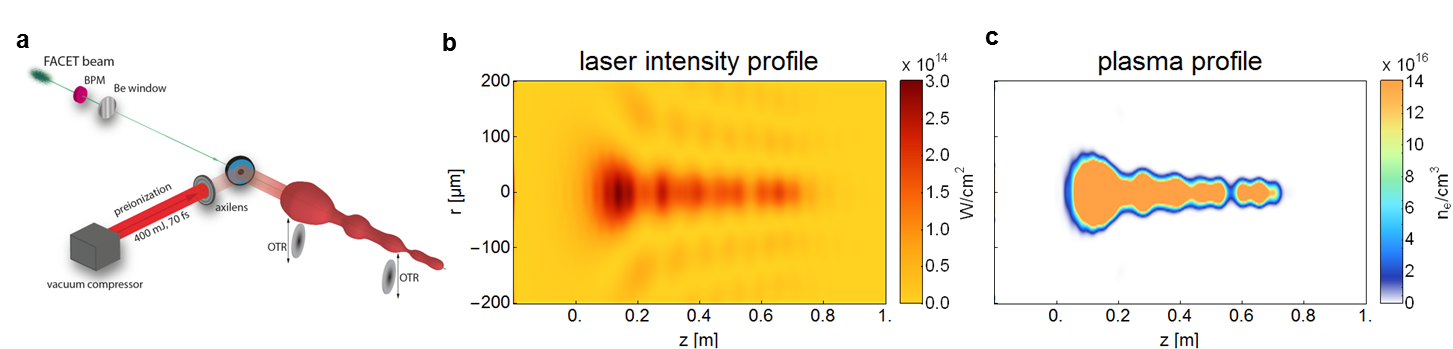}
\caption{\label{fig:E-210channelfromLaserIntensity}Relevant E-210 setup building block which produced the preionized plasma channel (a); projection of the produced approximate laser intensity profile (b); and corresponding hydrogen plasma profile produced by this intensity profile via tunneling ionization (c).}
\end{figure*}

\noindent
The FACET experiments took place in Sector 20 of the SLAC linac. Here, the high-current electron beam, generated by a thermionic cathode at the origin of the linac, compressed by a magnetic chicane and focused by a final focussing quadrupole system, interacted with a preionized plasma, as shown in  figure \ref{fig:E210setup}. The core experimental setup was assembled in the \enquote{picnic basket} interaction chamber, which was designed collaboratively by Radiabeam Technologies for the  \enquote{Plasma Photocathode Beam Brightness Transformer for Laser-Plasma Accelerators} project \cite{SBIRbrightnessTrafo} to host the E-210 experiment, and to assist many others. 
The whole region between FACET's upstream beryllium window and downstream diamond window was filled with a 50:50 mixture of hydrogen and helium gas. 
A preionized plasma channel with plasma electron density of $n_\mathrm{e} \approx 1.4\times 10^{17}\,\mathrm{cm^{-3}}$ was generated from hydrogen by a high-power laser pulse. The laser pulse propagated through an axilens, was folded on the electron beam propagation axis by a holed mirror and  then exhibited an axial intensity profile such that its corresponding electric field exceeded the hydrogen tunneling ionization threshold around the propagation axis.  
As indicated in figure \ref{fig:E210setup}, the resulting hydrogen plasma profile had a non-uniform width along the electron driver beam propagation axis. 
Figure \ref{fig:E-210channelfromLaserIntensity} concentrates on the relevant plasma preionization building block optics (a), provides a representation of the projected (calculated) laser intensity profile produced from the axilens (b) and shows the resulting (calculated) hydrogen plasma profile in projection (c). An important feature of laser-based preionization methods is that once the threshold of full ionization is reached and locally 100$\%$ of the gas is ionized,
shot-to-shot local intensity jitter does not matter as long as the full ionization intensity threshold is exceeded.  
This inherent levelling feature due to the intensity threshold is clear from comparing figures \ref{fig:E-210channelfromLaserIntensity} b) and c).
The selective ionization of hydrogen, without ionizing helium in the mix, in a large volume, was an enabling success achieved within E-210. 
However, the extent of the volume where this threshold of full ionization is reached, is critical.      
Very consequentially, the width of the channel obtained in the experiments was limited to approximately \SI{100}{\micro\metre}. For the most time during the channel propagation over its maximum length of $\Delta z \approx 65$ cm, the channel radius $r_\mathrm{c}$ was significantly narrower than the nominal blowout radius $R_\mathrm{b}$. This non-uniformity was a bottleneck and limitation in the experiment. 

Timing of the preionization laser with respect to the electron beam arrival could be tuned.  The laser generated plasma is comparatively cold and long-lasting due to the comparatively low power and intensity of the laser pulse. Hence, hydrodynamic effects impact plasma channel shape and density significantly only over extended  timescales, towards the ns-range. Plasma heating effects are an important field of study with regard e.g. to high repetition rate challenges of plasma-based accelerators \cite{hooker2018hofi,GilljohannPRXPhysRevX.9.011046,zgadzaj2020dissipation}. However, systematic studies of plasma heating effects were outside the core scope of E-210. Here, we were content with the preionization laser arriving $\simeq$ \SI{20}{ps} before the electron beam, which ensured steady operation, unaffected by shot-to-shot preionization timing jitter.    

In contrast, the timing of a second laser pulse which was strongly focused perpendicular to the electron beam path, was important on the fs-ps timescale. This laser pulse was crucial both for diagnostics as well as for injection of electrons to produce witness beams in the plasma wave. Hence, we installed  electro-optical sampling (EOS) units upstream and downstream of the main plasma interaction, based on further split-off laser pulses. The EOS provided time-stamping of shots, and benchmarking for the newly developed concept of plasma afterglow metrology \cite{Scherklarxiv2019}. 
The energy, transverse pointing, and delay of the plasma photocathode injector laser pulse could be varied, such that the laser pulse would ionize varying amounts of helium in a local filament across the electron driver beam axis. The laser energy budget was up to \SI{5.3}{mJ}, the transverse tuning range with respect to the electron driver beam axis covered  a few hundred \SI{}{\micro\metre}, and the temporal tuning range covered a range of few ps around the electron beam arrival time.

\subsection{Preionized plasma channel limitations}
\begin{figure*}
\centering
\includegraphics[width=1.0\textwidth]{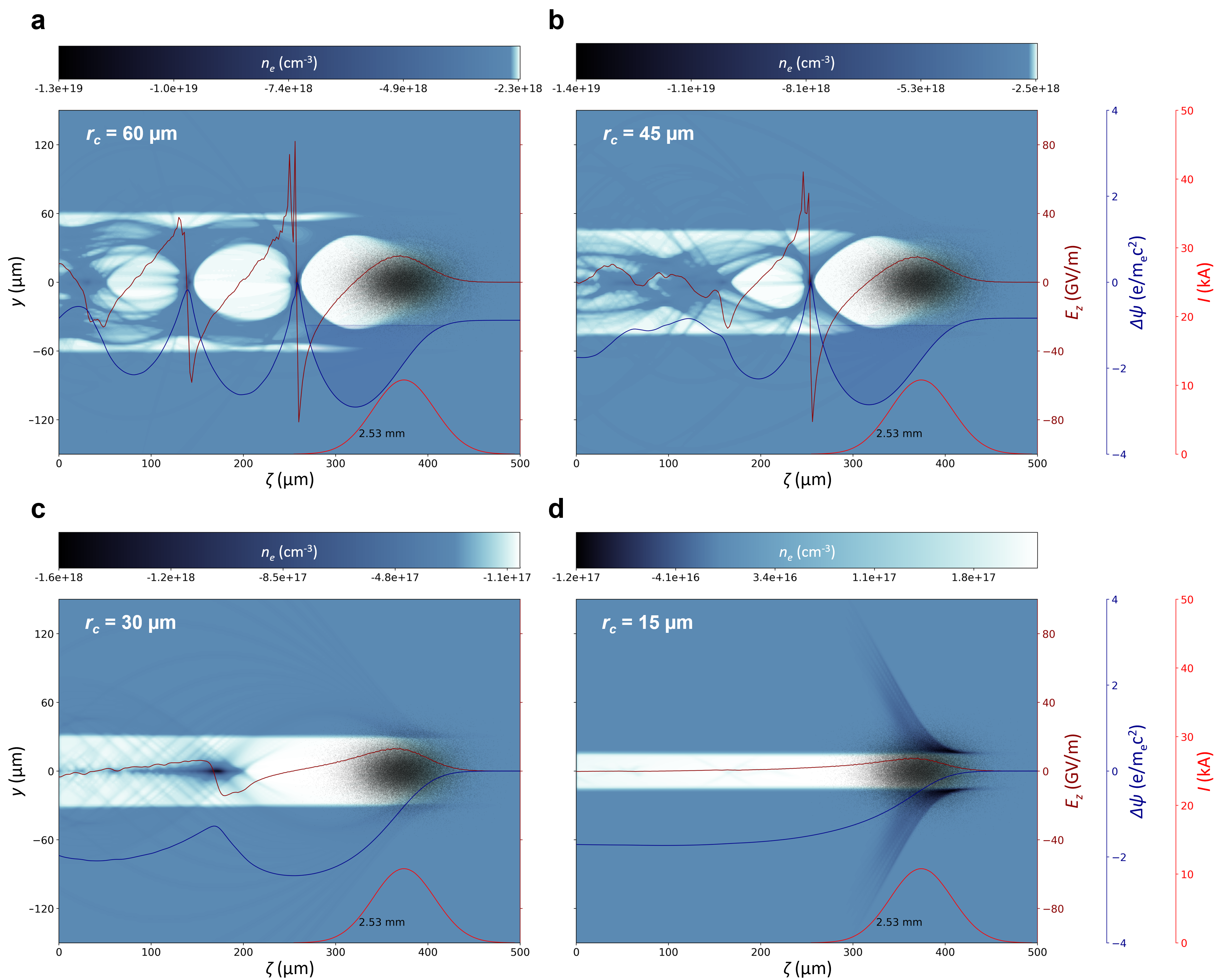}
\caption{3D PIC-simulations (VSim) of intense electron beam interaction with a preionized plasma channel of different radii $r_\mathrm{c}$. The FACET electron driver beam (black) propagates to the right, expels plasma electrons and sets up a nonlinear PWFA blowout as in a) and b), or for a thinner channel generates a wakeless ion channel as in c) and d) that could be used e.g. for light source applications.}
\label{fig:channel-inducedWakedynamics}
\end{figure*}

\noindent
This narrowing plasma channel shape has profound impact on the effective blowout shape, and hence on the corresponding electrostatic potential and electric field profile  of the wake. At FACET in E-210, this meant that several regimes of PWFA were realized along the plasma channel during one and the same shot. 
This plasma channel-induced blowout deformation and the behaviour of the plasma wake as it undergoes propagation in an increasingly narrower plasma channel is exemplified in figure \ref{fig:channel-inducedWakedynamics}, which shows the plasma electrons charge density, based on 3D particle-in-cell (PIC) simulations with VSim \cite{Nieter2004448}.

In figure \ref{fig:channel-inducedWakedynamics} a), the cylindrical plasma channel radius $r_\mathrm{c} =$~\SI{60}{\micro \metre}, and the expelled plasma blowout electrons see a re-attractive hydrogen ion background everywhere on their trajectories as they form the blowout shape, even at the point of maximal displacement from the axis, which is determined by the plasma density and the driver beam (shown in black) current profile $I$ (red solid line, shown at the bottom of the figure as projection of the longitudinal current profile of the Gaussian electron beam). 
The longitudinal electric field profile $E_\mathrm{acc}$ on axis is shown as dark red plot, and the underlying corresponding electrostatic trapping potential $\Delta \Psi$ is shown in blue. This situation represents the textbook case of full non-linear plasma wake in the blowout regime. 

When the plasma channel narrows to $r_\mathrm{c} =$~\SI{45}{\micro\metre} as shown in figure \ref{fig:channel-inducedWakedynamics} b), it is still just as wide as the nominal blowout radius $R_\mathrm{b}$, and hence the blowout shape in the first bucket and its electric field profile is in first approximation similar as in the case shown in figure \ref{fig:channel-inducedWakedynamics} a). 

However, plasma channel edge effects begin to impact the blowout structure. Some of the plasma electrons leave the ion channel and see a reduced re-attractive plasma potential. This results  in plasma frequency \enquote{redshifting} and an elongated plasma wave. In particular the second (and third) buckets are impacted substantially from the narrow plasma channel.

When further narrowing the plasma channel radius to $r_\mathrm{c} =$~\SI{30}{\micro\metre}, the restoring force of the ions is reduced significantly for electrons around their turning point as they reach a region where the ion density decreases sharply in the transverse direction. Here, a qualitative threshold is exceeded and the blowout breaks down, forming a much weaker plasma wave, as displayed in figure \ref{fig:channel-inducedWakedynamics} c). 
The constraints of the plasma channel width are reflected by the onset of 'snow-ploughed' plasma electrons \cite{BarovEnergyLossQtildePRST2004}, which do not return to axis on the time scale of $1/\omega_{\rm p}$ but are simply expelled outwards by the electron driver beam. 

When the channel radius is trimmed to $r_\mathrm{c} =$~\SI{15}{\micro\metre} in the simulation, most plasma electrons are snow-ploughed away, and the driver beam leaves behind an evacuated ion channel with constant and uniform density.
Practically no plasma blowout structure remains despite a decelerating field at the driver beam position: the longitudinal electric field is approximately zero behind the driver beam, while the ion channel still exhibits linear focusing forces. This, in effect,  represents a \textit{wakeless regime}, in which the electron driver beam expels plasma electrons to leave behind a pure focusing ion channel as shown in \ref{fig:channel-inducedWakedynamics} d). 

We emphasize that these highly complex channel-induced plasma wake dynamics, plasma blowout lengthening and the wakeless regime in the pure ion channel are not theoretical scenarios, but indeed have been encountered during realization of E-210. 
In fact, the scenarios described above have been repeatedly realized during each shot, as a result of the plasma channel radius decreasing and increasing repeatedly 
in the range from zero to $\approx$~\SI{100}{\micro\metre}. 

As described in \cite{Deng2019Trojan}, the limited channel width did impose a minimum hydrogen plasma density to be used, which forced to operate with a smaller-than-optimal blowout size,
and in turn put increased demands on spatiotemporal alignment and synchronization for plasma photocathode injection. 
This working point was in fact close to a ceiling of employable plasma densities, that arises from unwanted ionisation by  the  wake and the electron driver beam fields \cite{ManahanDarkCurrentPhysRevAccelBeams.19.011303}. In section III, it will be addressed how a wider plasma channel and operation at lower plasma densities stabilizes the PWFA and plasma photocathode combination substantially.  

\subsection{Energy gain limitations}
\noindent
In addition to the maximum channel width bottleneck and the need to squeeze the plasma wave through it, the quasi-periodically narrowing and expanding plasma channel had a profound effect on the energy gain of injected electrons in the plasma wakefields. 

\begin{figure*}
\begin{tabular}{cc} 
\includegraphics[width=.5\textwidth]{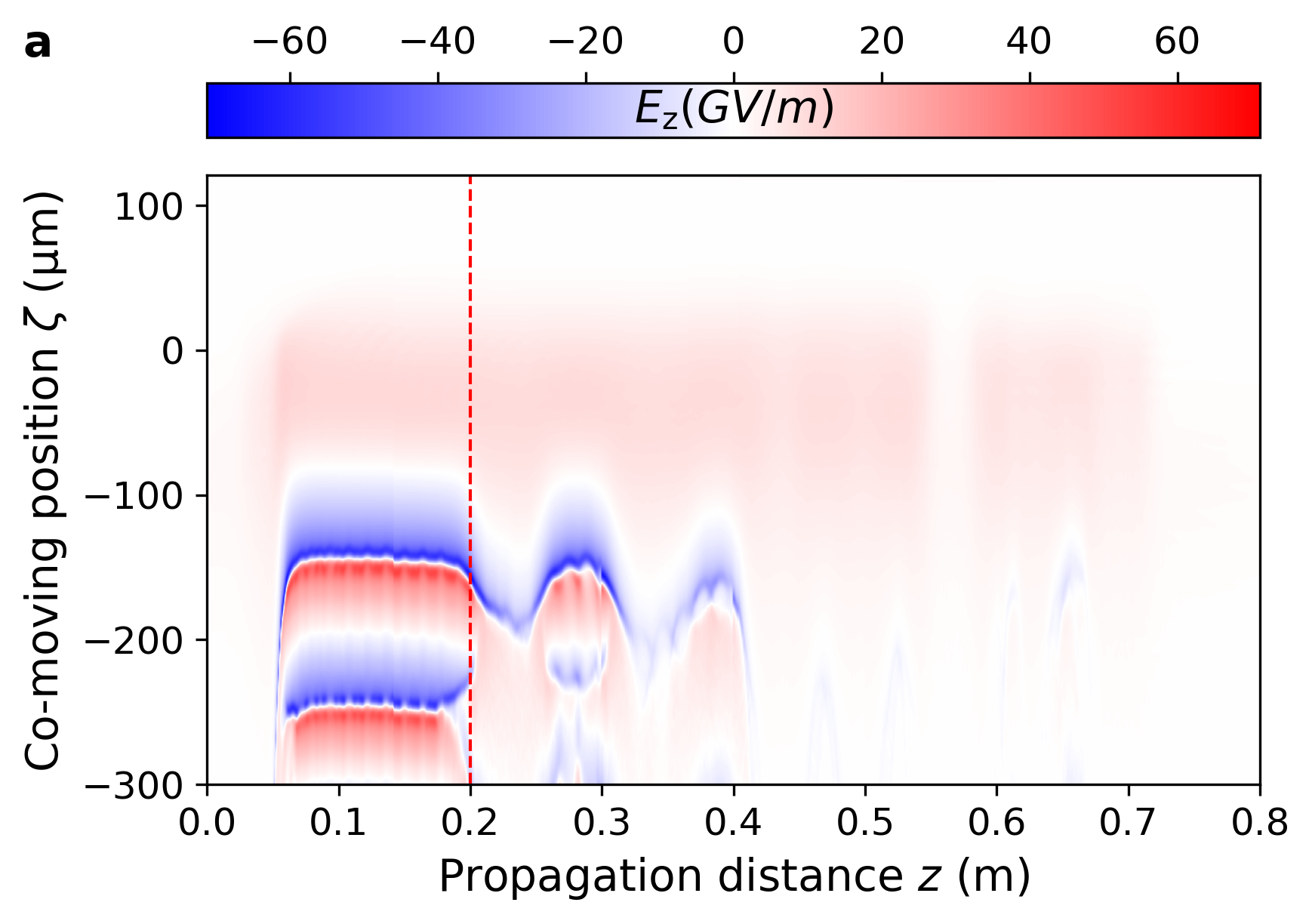}
    
\includegraphics[width=0.5\textwidth]{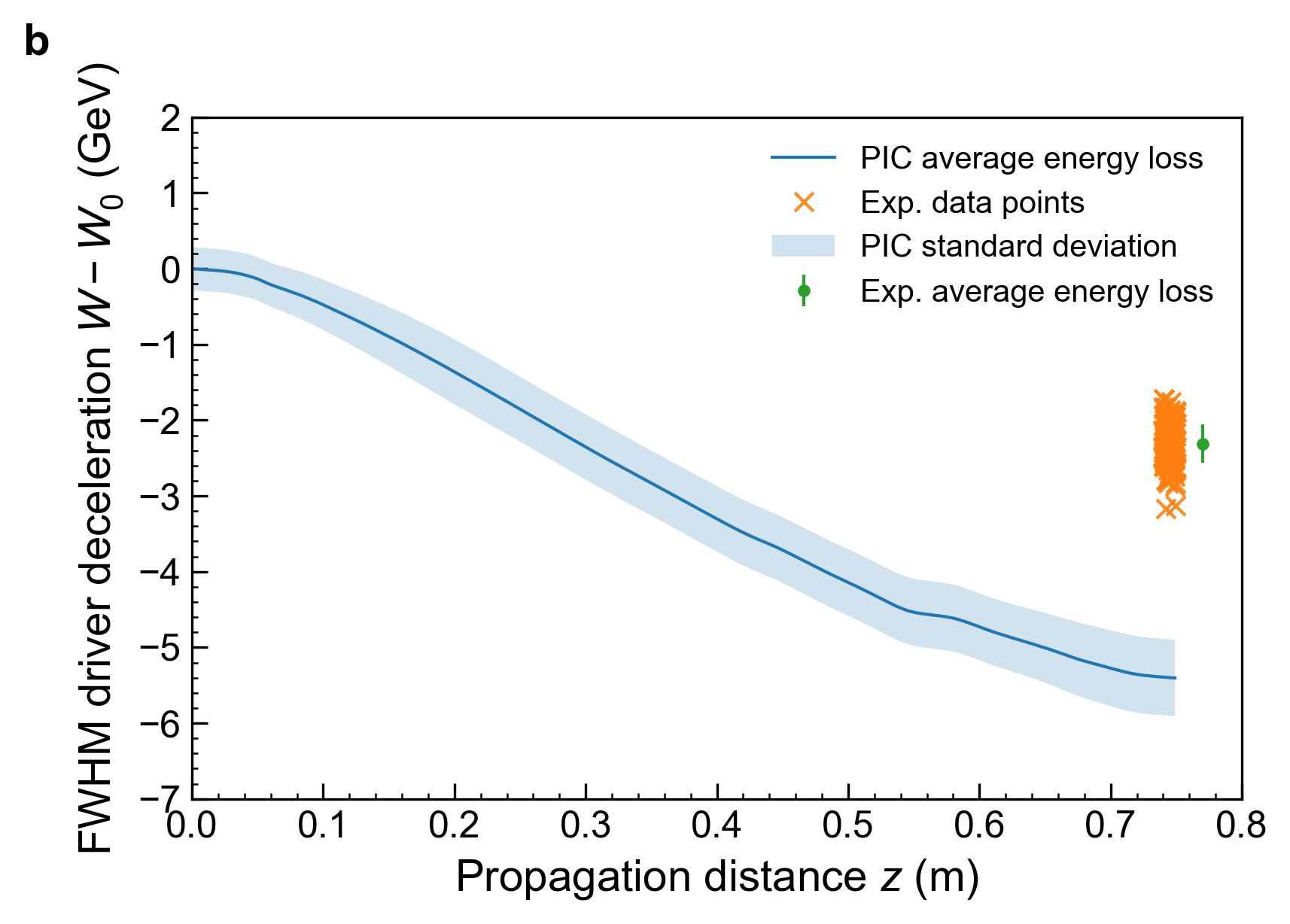}
\end{tabular} 

\caption{\label{fig:E210onaxisfieldevolution} a) shows the on-axis longitudinal wakefield evolution of blowout when propagating through preionized plasma channel shown in figure \ref{fig:E-210channelfromLaserIntensity} c), and b) shows estimates of driver beam deceleration obtained from PIC simulation and measured at the experiment. 
In a),  the vertical red dashed line denotes the experimental injection position of the witness beam. In b), the blue solid line shows simulated driver beam deceleration of  $\Delta W_{\mathrm{sim}}\approx$~\SI{5.4}{GeV} FWHM. The transparent tubes represent the standard deviation interval. The orange crosses represent measured data points of FWHM driver beam deceleration and the green point shows the average over 200 consecutive shots at $\Delta W_{\mathrm{exp}} \approx$~\SI{2.3}{GeV} with corresponding standard deviation error bars.} 
\end{figure*}

The unfavorable topology after the injection point at $z\approx$ 20 cm (dashed red line in figure \ref{fig:E210onaxisfieldevolution} a)) impacted the blowout shape and size along the propagation distance, resulting in a substantial variation of the effective wakefield phase at the witness beam trapping position.

Simulations show that the witness beam actually underwent a quasi-periodic transition from accelerating to decelerating phase of the wakefield.
In Ref. \cite{Deng2019Trojan} (supplementary figure 2 therein) we estimated projected energy gain outcomes for realistic trapping positions considering this wakefield evolution over the full plasma interaction distance of $\Delta z \approx$ 65 cm, while in the present work figure \ref{fig:E210onaxisfieldevolution} a) shows a waterfall plot of the corresponding on-axis accelerating electric field evolution during propagation along the plasma channel. 
The figure  highlights the shortcomings of the varying plasma channel and its significant effect on the longitudinal wakefield, and explains the effective energy gain limitation encountered in the E-210 experiment.
   
Experimental boundary conditions were responsible for restriction of the injection positionat  $z\approx 20$ cm. 
Numerical simulations indicate that,  for example, an injection position at around $z\approx 10$ cm, where the plasma channel reaches maximum width, would have allowed harnessing the full accelerating field of 50-60 GV/m over an extended distance. Estimates of this scenario show potential witness beam energy gains of multi-GeVs, instead of $\approx 1$ GeV as in our proof-of-concept experiments \cite{Deng2019Trojan}. This is supported by simulations of driver beam interaction with the preionized plasma and its deceleration to an average energy of  $\Delta W_{\mathrm{sim}}\approx$~\SI{5.4}{GeV} FWHM shown in figure \ref{fig:E210onaxisfieldevolution} b). The simulation data represents the strongest deceleration scenario for a shot with optimal alignment and maximum plasma channel size.
Corresponding measurements of the driver beam deceleration (orange crosses) are consistent with the simulation but  show a somewhat reduced average driver beam deceleration of FWHM $\Delta W_{\mathrm{exp}} \approx$~\SI{2.3}{GeV} in a range from $\Delta W_{\mathrm{min,max}} \approx$~\SI{1.7}{}-\SI{3.1}{GeV}. Reduced driver beam deceleration can be attributed to sub-optimal alignment and/or plasma channel size, and the large variation of experimentally observed driver beam deceleration is further evidence of strong shot-to-shot fluctuations.  

While wakefield dynamics induced by the plasma channel topology were a limiting factor during the experimental campaign, they are an interesting subject in its own right. For example, the wakeless regime can be an attractive operation point for betatron radiation generation and/or ion-channel lasers \cite{litos2018experimental,habib2019plasma}. 

\subsection{Injection considerations and procedure}
\begin{figure*}
\includegraphics[width=1.0\textwidth]{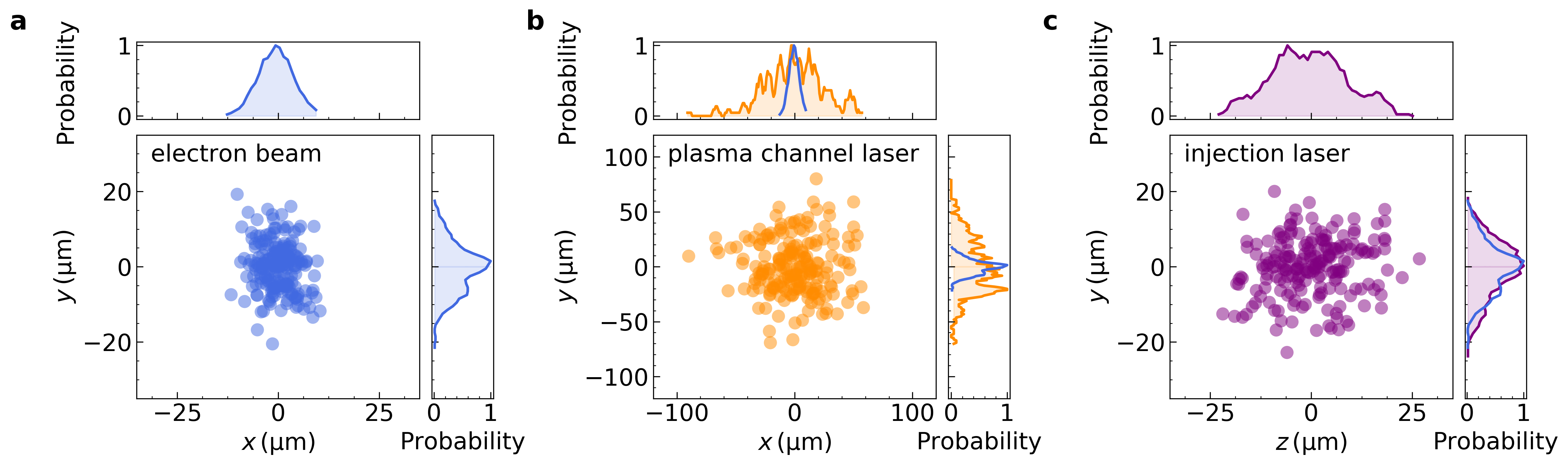}
\caption{\label{fig:E210jitters} Typical pointing jitters on target of electron driver beam (a), preionization plasma channel laser (b) and plasma photocathode injection laser (c) encountered during the E-210: Trojan Horse experiment at FACET. In subset plots the sold lines show the jitter distributions in the corresponding plane normalized to the miximum value. In b) and c) subset plots, additionally, electron driver beam jitter distributions (solid blue line) are presented for direct comparison.}
\end{figure*}

\noindent
At the core, the injection stability is a function of plasma blowout size on the one hand, and spatiotemporal alignment and synchronization precision with respect to the plasma photocathode injection laser on the other hand. 
Increasing the plasma blowout size by operating at reduced plasma densities therefore would increase the relative injection precision for a given absolute shot-to-shot jitter e.g. with regards to the pointing of the injector laser. 
However, in case of E-210, a larger blowout would also have meant that any shot-to-shot jitter of the preionization laser with respect to the driver electron beam axis would have brought the blowout closer to  the boundaries of the plasma channel or even would mean partial, asymmetric destruction of the blowout when it touches the plasma channel boundaries. 
Figure \ref{fig:E210jitters} summarizes typical experimentally encountered shot-to-shot spatial jitter of the electron driver beam (a), the preionization plasma channel laser (b), and the injection laser (c): the experimental shot-to-shot spatial jitter of the plasma channel preionization laser was substantially larger than that of the electron beam, or the injection laser.  This constellation further emphasizes the large impact of the preionization laser configuration on the E-210 experiment injection studies.           

\begin{figure*}
\includegraphics[width=1.0\textwidth]{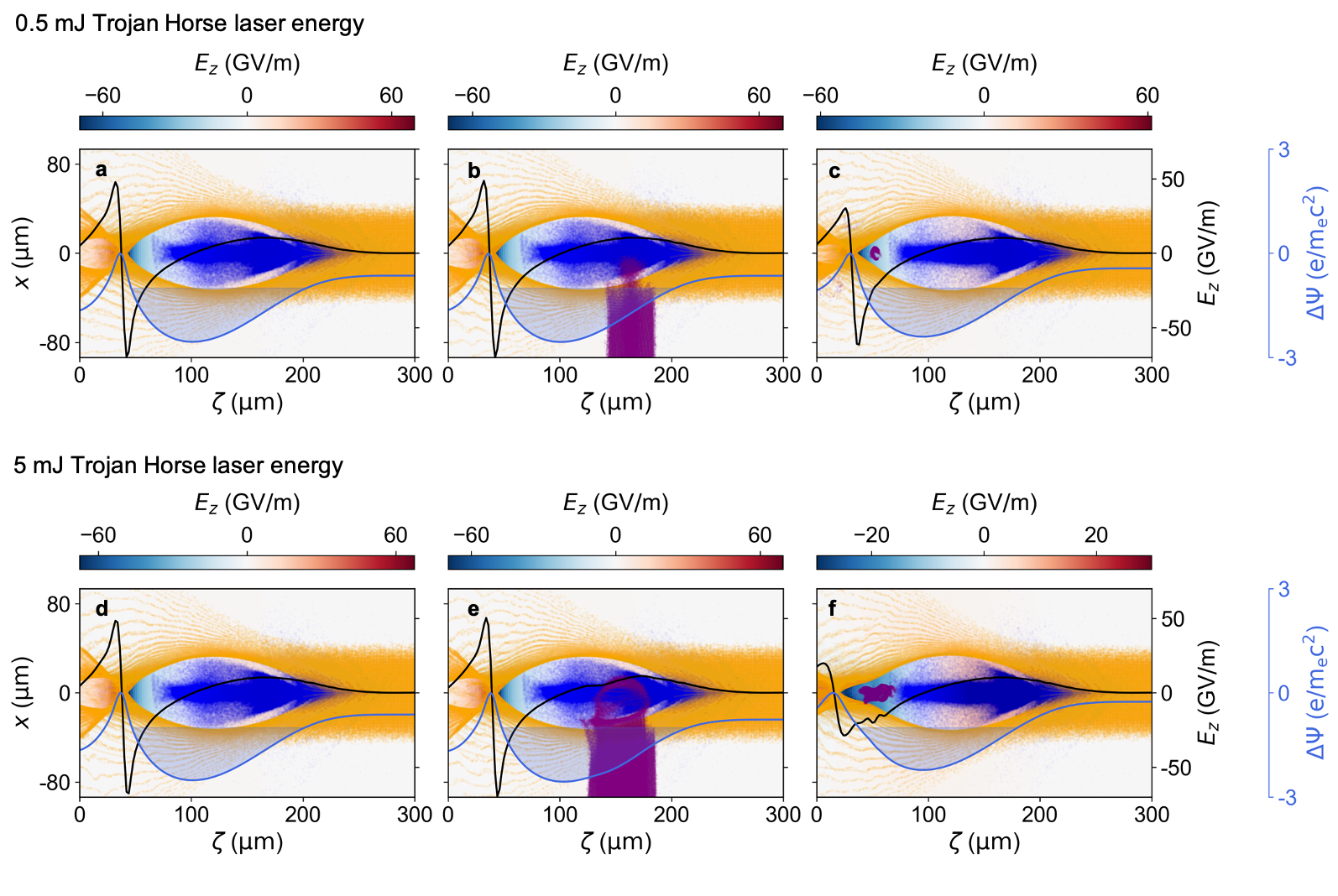}
\caption{\label{fig:E210trojanOldInjectionSims}Snapshots from PIC-simulations with VSim for the E-210 scenario. The top panel shows the situation before (a) at $ t=$~\SI{0}{}, during (b) at $ t\approx$~\SI{667}{fs} and after (c) injection at $ t\approx$~\SI{9.3}{ps} with an injector laser pulse at \SI{0.5}{mJ}; the bottom panel shows the corresponding situations when using \SI{5}{mJ} injection laser energy.}
\end{figure*}

Despite those jitters, using the spatiotemporal afterglow response \cite{Scherklarxiv2019} and the plasma torch injection mechanism  \cite{WittigPlasmaTorchPhysRevSTAB.18.081304,Ullmann2020Torch} enabled finding the suitable pointing and timing of the injector laser with respect to the plasma wave at the interaction point, and thus to access the plasma photocathode injection regime. 
Figure \ref{fig:E210trojanOldInjectionSims} visualizes 3D PIC simulations of the E-210 experiment. The electron beam driver (blue) propagates to the right, and drives a blowout that only just fits into the hydrogen plasma channel (orange dots) at the injection position $z \approx $~\SI{20}{cm}.
The top panel shows the situation for the injector laser pulse energy of \SI{0.5}{mJ}, while the bottom panel represents the \SI{5}{mJ} case.
The plasma photocathode injector laser pulse (not shown directly) with pulse duration of \SI{65}{fs} (FWHM), and vacuum spot size of $w_{0}=$~\SI{20}{\micro\metre} (r.m.s.) is propagating from bottom to top,  ionizes  helium, and thereby releases initially "ultra-cold"  helium electrons (purple) inside, but also outside of the wake due to its rather long Rayleigh length $Z_{\mathrm{R}}=\pi w_{0}^2/\SI{0.8}{\micro\metre} \approx$ \SI{1.57}{mm} compared to the blowout diameter of few tens of \SI{}{\micro\metre}. 
The solid black and blue profiles highlight on-axis longitudinal wakefield $E_\mathrm{z}$ and trapping potential $\Delta \Psi$, respectively. 
When the plasma wakefield swipes through, only those helium  electrons that are released within the electrostatic potential region of the wake that is capable of trapping electrons from rest  (indicated by the blue transparent region) are captured by the plasma wave, while all other helium electrons are lost to the background plasma. 
Frames a) and d) represent the situation before the laser pulse enters, and frames b) and e) show the He electrons (purple) appearing as result of the plasma photocathode laser pulse ionizing helium. 
Frames c) and f) then present the formed and trapped witness electron bunch with low charge (c) and high charge (f) as result of the different laser energies.  The higher injected charge for the 5 mJ case is a result of the large ionization volume inside the blowout. This can be seen by comparing the ionisation tracks in figure \ref{fig:E210trojanOldInjectionSims} b) and e), showing that for the 5 mJ case the ionisation track is substantially wider compared to the 0.5 mJ laser energy case.

Realization of first proof-of-concept demonstration of the plasma photocathode, and the many other scientific firsts realized during E-210, represent major experimental milestones towards production of ultracold electron beams. However, so far only the tip of the iceberg has been revealed. To reach the full potential of the plasma photocathode, several aspects are important to recognize. First, in E-210, the large ionization volume of the injection laser pulse even in the 0.5 mJ case, with comparatively long Rayleigh length, fills a large proportion of the comparatively small plasma blowout, which ultimately results in a large initial phase space volume and increases the obtainable emittance of the trapped witness beam. 
Secondly, because of the rather long driver beam compared to the blowout size, helium electrons are released in its immediate space charge field. 
They therefore are kicked out transversely to some extent and obtain significant transverse momentum, which likewise increases the emittance. 
In the E-210 scenario and its  boundary conditions,  the obtainable normalized emittance minimum therefore is at the single \SI{}{\micro\metre}-rad scale for the plane in the laser propagation axis, and  slightly better in the other transverse plane, since the electron release is not spread out across the entire extent of the blowout in this plane. 
This is predicted by simulations and is consistent with the experimentally derived emittance \cite{Deng2019Trojan}. 
This shows that  the plasma photocathode principle works \cite{Deng2019Trojan} exactly as anticipated. At the same time,  it can be realized even under sub-optimal boundary conditions, which is very encouraging for future, improved implementations of the scheme. 

Summarizing, the FACET E-210 experimental campaign was highly successful in demonstrating key milestones such as the feasibility of plasma photocathode injector \cite{Deng2019Trojan}, realization of the first density-downramp injection in PWFA \cite{Ullmann2020Torch} and by demonstrating  novel plasma-based diagnostics with large potential for non-interceptive precision metrology \cite{Scherklarxiv2019}. The experimental results, lessons learned, and modelling and understanding, also are crucial to design and prepare  the next generation of experiments.
Wider plasma channels for more stable injector and accelerator operation and larger energy gains, reduction of the ionization volume of the injection laser for full charge capture and lower emittances and/or different injection geometries, specifically collinear, are amongst the top priorities for future installations of plasma photocathodes at FACET-II and elsewhere.

\section{\label{sec:E-310}E-310: Trojan Horse-II at FACET-II}
\begin{figure}
\includegraphics[width=1.0\textwidth]{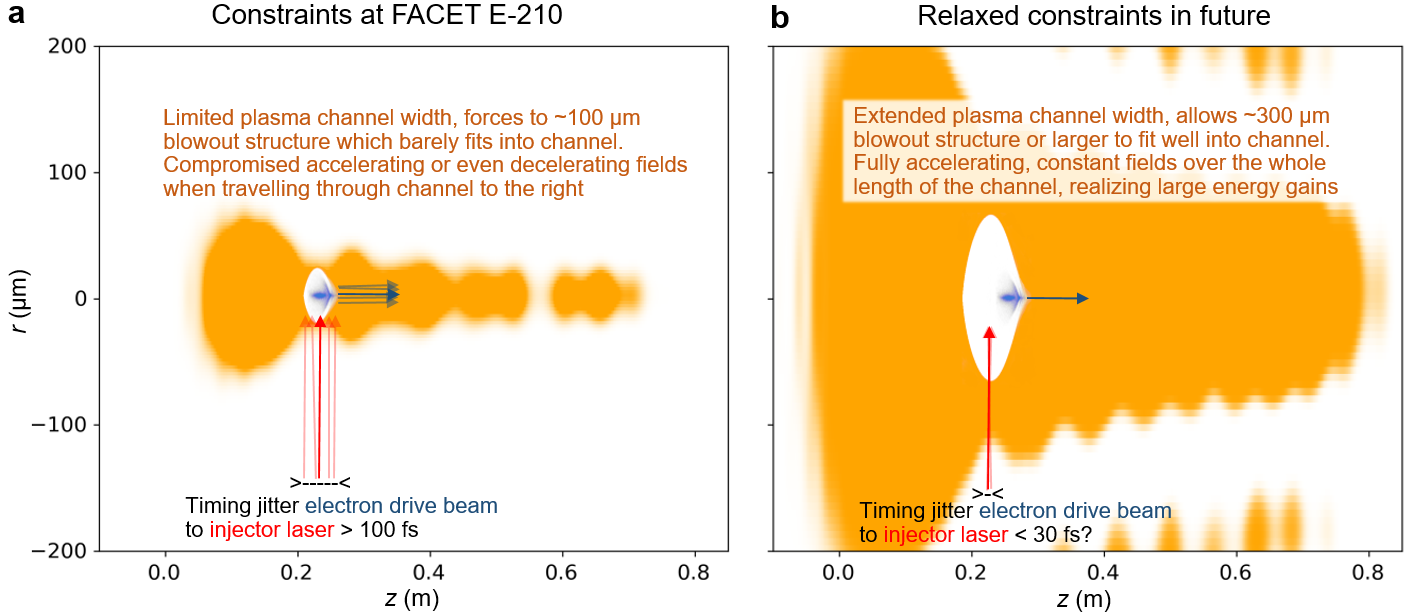}
\caption{\label{fig:SchematicFACET-FACET-II}Schematic key improvement at FACET-II: a broader preionized plasma channel will allow operation at reduced plasma density, which in turn allows higher stability and quality of the output electron bunch production.}
\end{figure}

\noindent
The \enquote{E-310: Trojan Horse-II} program at SLAC FACET-II aims at investigating various plasma phothocathode configurations, for example the realization of collinear and near-collinear geometry, and/or innovative approaches for reduced effective Rayleigh length of the plasma photocathode laser \cite{Manahan2019Royal}. 
 A wider plasma channel with larger blowout sizes combined with improved incoming  stability of driver electron beam and laser beams   will help to make headway towards improved witness beam quality, tunability, and stability. 
Even in 90$^\circ$ geometry these improvements promise much better stability and output beam quality. 
Figure \ref{fig:SchematicFACET-FACET-II} summarizes many of the the aspired improvements visually.  
The next sections will discuss and investigate some of these aspects.     

\subsection{\label{sec:FutureCapabilities}Future capabilities at FACET-II}

\noindent
The implementation of collinear geometry and/or confined laser release volumes and mitigation of driver beam kick to the released electrons is suitable to allow production of witness beams with increasingly improved emittance and brightness. 
Operation at reduced plasma densities, which requires wider preionization channels, can decrease the residual and correlated energy spread of the witness beam \cite{ManahanHabib6d}. Further, reduced plasma density does also relax the demands on the driver beam charge density, since the blowout regime can then be achieved easier. A thus reduced required driver beam density (and transverse matching) also elegantly avoids hot spots that may otherwise produce dark current, and the decreased plasma density naturally decreases potential wakefield vertex hot spots \cite{ManahanDarkCurrentPhysRevAccelBeams.19.011303}. 

\begin{figure*}
\includegraphics[width=1.0\textwidth]{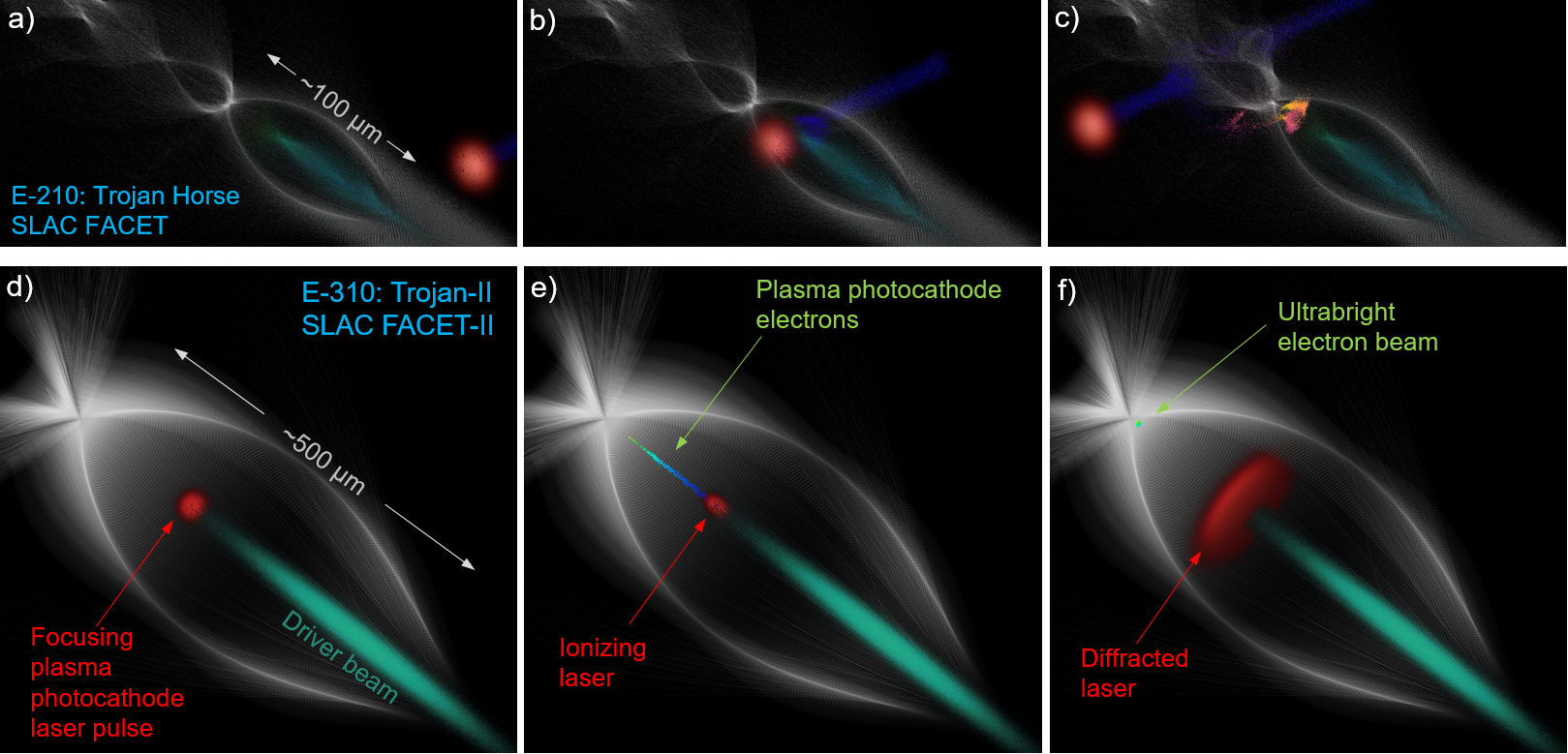}
\caption{\label{fig:E210vsE310-3DconceptualNew}The top panel shows a 3D visualization of the E-210 scenario before (a), during (b) and after (c) He electron release in a comparably small blowout. The bottom panel shows the corresponding situation for E-310 in collinear and co-propagating geometry.}
\end{figure*}

We emphasize again that if  the plasma channel can be made  wide enough by a sufficient margin,  shot-to-shot variations of pointing, energy, wavefronts of the preionization laser pulse may not have any influence at all on the plasma blowout shape: even if jitter in these parameters leads to variation of the channel width from shot-to-shot, it may not impact the acceleration process as long as full ionization is realized in the overlap region with the plasma wave, and the channel fully encompasses  the passing blowout structure, see figure \ref{fig:channel-inducedWakedynamics} a) and b).      

When the accelerating and focusing electric field profile is thereby constant, and the plasma blowout is larger and/or the electron driver beam is shorter as in E-210 so that there is no driver beam kick that could increase the transverse momentum of released HIT electrons, realization of witness bunches with normalized emittances down to nm rad scale, and hence ultrahigh brightness can be achieved in E-310. This is consistent with earlier works and  estimations of emittance and brightness in plasma photocathode scenarios \cite{HiddingPRL2012PhysRevLett.108.035001,hidding:570beyond,LiPhysRevLett.111.015003PRL2013,YunfengPhysRevSTAB.16.031303,BourgeoisHookerPhysRevLett.111.155004,Hidding5thgeneration2014,ThermalEmittancePhysRevSTAB.17.101301Schroeder2014,PhysRevLett.112.035003Xu2014,YuTwo-colrPhysRevLett.112.125001,ManahanHabib6d,Moon2019doi:10.1063/1.5108928,Manahan2019Royal}. 
The central importance of emittance and beam quality of electron beams for various applications is highlighted e.g. in US \cite{USroadmapColbydoi:10.1142/S1793626816300012} and UK ~\cite{hidding2019plasmaPWASC} roadmaps. 
Figure \ref{fig:E210vsE310-3DconceptualNew} visualizes and contrasts the E-210 scenario (top panel) with a potential  E-310 scenario (bottom panel). The electron driver beam (green) propagates from top left to bottom right and excites the plasma wave. 
The plasma photocathode release laser pulse (red) releases He electrons, which then form the trapped witness bunch. Key differences of these scenarios are the injection geometry (perpendicular vs. collinear) and the blowout size. 

\subsection{\label{sec:Stability}Beam parameter stability and tuneablity }

\noindent

While generation and acceleration of high quality electron beams  
is a supreme focus for plasma-based accelerator research,  in addition the reproducibility and tunability of the output  beam parameters is crucial towards realization of key applications such as free-electron lasers and, prospectively, high energy physics applications and perhaps even linear colliders. In conventional accelerators, a detailed statistical analysis is performed to identify prime sources of jitter, and subsequently a systematic approach is taken to eliminate or to minimize these jitter sources at the origin \cite{turner2016fel}. Similar strategies are required for plasma-based accelerators. Here, witness beam output parameter variation from shot-to-shot can be attributed to two major sources: the jitter of the plasma \textit{accelerator} on the one hand, and jitter of the witness beam \textit{injector} process on the other. In the Trojan Horse approach, accelerator and injector are largely decoupled, in contrast to  other plasma accelerator injection schemes, where the injection rate depends  crucially on the wake excitation and gas or plasma density profile encountered in a specific shot. 

With regard to the \textit{acceleration}, variations of plasma source and the incoming electron driver beam determine the size, strength and evolution of the plasma wakefield accelerator. As described in section II, main sources of jitter in the E-210 experiment originated from shot-to-shot variations of the preionization laser pulse as shown in figure \ref{fig:E210jitters} b), and periodic narrowing of the plasma channel as presented in figure \ref{fig:E-210channelfromLaserIntensity}. This accelerator building block can be substantially improved and impact of preionization laser jitter may  even be fully eliminated as discussed earlier.  Driver beam parameter stability can be significantly improved, at FACET-II e.g. by beam generation from a state-of-the-art photocathode. 

With regard to the witness beam \textit{injection}, primary factors that affect the plasma photocathode process are spatiotemporal alignment and synchronization of the injection laser with respect to the plasma wakefield accelerator, and the laser pulse intensity. 
Therefore, investigating, controlling and minimizing impact of jitter sources of incoming beams 
is important for optimized performance of the plasma photocathode. 

The plasma photocathode injection method offers control and stabilization advantages resulting from the inherently decoupled nature of this method. 
In order to explore how the identified plasma photocathode injector jitter sources impact the witness beam parameter range and stability, we have carried out extensive 3D PIC-simulation studies to explore the effect of  i) temporal jitter, ii) transverse spatial jitter, and iii) intensity jitter of the  plasma photocathode injector laser pulse inside a suitable plasma wakefield accelerator. 

The simulation parameter space is informed by the results and discussions from sections \ref{sec:E-210}-\ref{sec:E-310}. 
The electron driver beam reflects a possible working point within the FACET-II parameter space range, such that its energy is set to  
$W= $ \SI{10}{GeV} and 
its charge to $Q_{\rm d}= $ \SI{1.5}{nC}. 
The hydrogen plasma and helium gas density are set to $n_{\rm p}\approx$ \SI{1.78e+16}{cm}$^{-3}$ and $n_{\rm He}\approx$ \SI{2.27e+18}{cm}$^{-3}$, respectively. The hydrogen plasma density corresponds to a plasma  wavelength $\lambda_{\mathrm{p}} \approx$~\SI{250}{\micro\metre}, and the driver beam produces an elliptical blowout of similar length $L_{\rm b}\approx$~\SI{250}{\micro\metre} and radius of $R_{\rm b}\approx$~\SI{65}{\micro\metre}.  
The transverse normalized emittance of the driver beam is matched to the hydrogen plasma density $n_{\rm p}$, using  $\sigma_{\rm x,y}=\sqrt{\epsilon_{\rm n}/ \gamma k_{\rm \beta}}$, where $\epsilon_{\rm n}$ is the driver beam normalized emittance, $k_{\rm \beta}=\omega_{\rm p}/c\sqrt{2\gamma}$ is the betatron wavenumber, $\omega_{\rm p}$ is the plasma frequency, $c$ is the speed of light and $\gamma$ is the relativistic Lorentz factor of the FACET-II electron driver beam. The longitudinal size of the driver beam is optimized according to $\sigma_{\rm z}= \sqrt{2}/k_{\rm p}$ to satisfy the PWFA resonance condition, where $k_{\rm p}=\omega_{\rm p}/c$ is the plasma wavenumber.
The normalized emittance $\epsilon_{\rm n,x,y}$= \SI{50}{mm} mrad of the drive beam determines $\sigma_{\rm (x,y),rms}\approx $ \SI{4.5}{\micro\metre} and the resonance condition yields $\sigma_{\rm z,rms}\approx$ \SI{32}{\micro\metre}.

\begin{figure}
\includegraphics[width=0.5\textwidth]{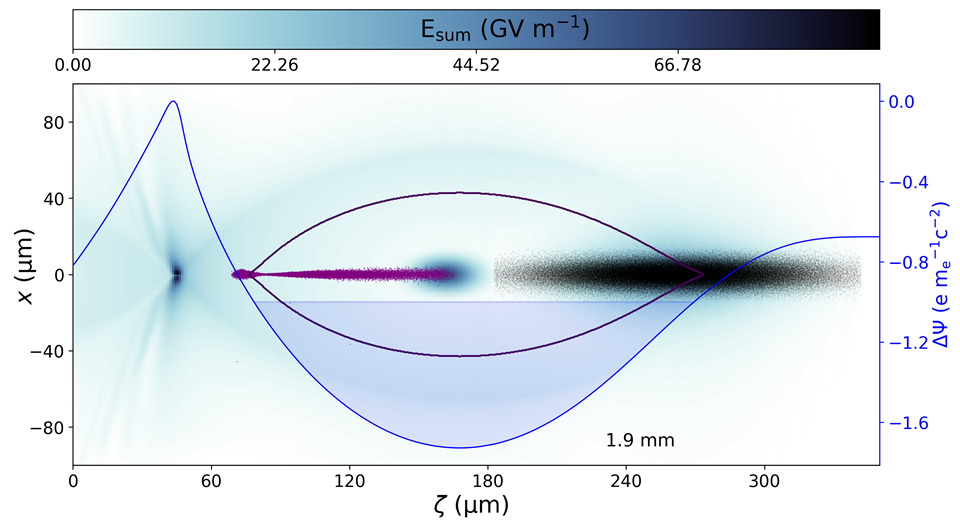}
\caption{\label{fig:LargerBlowout}Visualization of the plasma wakefield scenario used for the plasma photocathode parameter sensitivity studies. The driver beam (black) propagates to the right and the plasma photocathode laser pulse releases He electrons (purple) in the centre of the blowout.}
\end{figure}

Figure \ref{fig:LargerBlowout} visualizes the underlying scenario. The electron driver beam (black) propagates to the right and sets up the plasma blowout in its wake. The plasma photocathode laser pulse is currently in the process of releasing He electrons (purple) via tunneling ionization of the background He gas, with some of those electrons that have been released at the beginning of the injection process at $\zeta_{\rm i}=z-ct\approx$~\SI{161}{\micro\metre} already piling up at the trapping position $\zeta_{\rm f}$ within the blowout, which is defined by the electrostatic potential $\phi(\zeta)\propto\int E_{\rm z}(\zeta) d\zeta$ at each of the release slices.
The corresponding on-axis electrostatic trapping potential $\Delta \Psi= \Psi(\zeta_{\rm i})-\Psi(\zeta_{\rm f})=-m_{\rm e}c^{2}e^{-1} $ of the wake, with $\Psi(\zeta)=e m_{\rm e}^{-1}c^{-2}\phi(\zeta)$, is shown in blue, and the region in which plasma photocathode-released electrons would be trapped, corresponding to $\Delta \Psi \leq -1$, is represented by the blue shaded area, just as previously in figure 6. The default release position $\zeta_{\rm i}$ of He electrons is in the center of the hydrogen-based blowout, where the corresponding trapping potential has its minimum $\Delta \Psi_{\rm min}\approx-1.7$. The He electrons are born at rest, and hence are slipping backwards towards the blowout vertex while being quickly accelerated to relativistic energies due to the multi-\SI{}{GV/m} accelerating gradient. The purple solid ellipse approximates the
trapping volume, i.e. electrons released approximately within this volume from rest will be trapped by the electrostatic potential of the wave. The sum of the combined electric field is plotted, thereby showing the focused plasma photocathode laser pulse in the centre of the blowout.  The collinear plasma photocathode laser pulse has a FWHM pulse duration $\tau = 50$ fs, an r.m.s spot size $w_0 =$~\SI{7}{\micro\metre} and a default focus intensity in terms of the dimensionless laser amplitude $a_0 = e E/(\omega m_e c) = 0.018$, where  $E$ is the electric field amplitude and $\omega$ is the laser frequency.  We note that plasma photocathode injector parameters are purposefully optimized for a low witness charge regime to minimize beam loading effects, in order to focus on the impact of incoming injector laser pulse jitter contributions. Much higher witness charge values are possible to be released, for example straightforwardly by increasing the laser pulse intensity, or by increasing the He density. At elevated witness charge and current levels, advantageous effects of beamloading can be harnessed, while at even higher charge and current levels, the beam and its emittance becomes space charge dominated.   

Importantly, because of the parabolic shape of the trapping potential, its slope around the potential minimum is shallow. Therefore, the deviation in initial $\Delta \Psi(\zeta_{\rm i})$ around the release position is small, and consequently, even when electrons are released across an extended co-moving range  by the laser pulse, this provides a strongly reduced spread in final trapping position $\delta\zeta_{\rm f}$, and thus enables fs to sub-fs formed bunch duration. This does not  only constitute an automatic bunch compression of the injected electrons, but furthermore releasing electrons at this prominent position in the centre of the blowout makes it intrinsically resilient against relative timing variation:  even when electrons are released at different longitudinal co-moving positions around the default release position from shot-to-shot, they are trapped at rather similar accelerating phase positions in the wake. Because of the phase-locked feature in beam-driven acceleration this manifests itself in significantly reduced witness beam energy variation from shot-to-shot.

The final trapping position can be expressed as function of initial release position and plasma density as
\begin{equation} \label{eq:trapping}
\zeta_{\rm f} = -\left(\zeta_{\rm i}^2 +\frac{4\alpha_{\rm t}}{n_{\rm p}}\right)^{1/2},
\end{equation}
where $\alpha_{t}=m_{e}c^2\epsilon_{0}e^{-2}$ and $\zeta_{\rm i}$ is the initial release position within the trapping potential.  
Note that in this representation the potential minimum is at $\zeta_{\rm i}=0$.
Series expansion of equation \ref{eq:trapping} at $\zeta_{\rm i}=0$ yields $\zeta_{\rm f,t}\approx -2\sqrt{\alpha_{\rm t}/n_{\rm p}}-\zeta_{\rm i}^2/4\sqrt{\alpha_{\rm t}/n_{\rm p}}+\mathcal{O}(\zeta_{\rm i}^4)$.
From this we can immediately see that due to the quadratic $\zeta_{\rm i}$ term, releasing electrons at the trapping potential minimum, meaning  $\zeta_{\rm i}\approx0$ here, is an optimum that results in maximized stability of the  trapping position $\zeta_{\rm f}$.
Further, we can see again that lower plasma densities improve injection stability.

These simulation settings are used as reference case, and guided by the experimental spatiotemporal jitter of the plasma photocathode laser encountered in E210 (compare figure \ref{fig:E210jitters}), and taking into account improved performance possible with state-of-the-art photocathode based linacs, we perform injector parameter scans. 
This allows better understanding of the impact of laser jitters on witness beam parameters in FACET-II-like experimental settings, and will guide the injection and trapping process optimization in a realistic scenario. In this study we followed a conservative approach in terms of jitter values and statistical treatment of the simulation data, since laser pulse timing, alignment and intensity variation can be significantly better than in our study, and the simulation data with equidistant variation is treated as uniform distribution, while experimentally rather a Gaussian distribution of jitters (see Figure \ref{fig:E210jitters}) would be expected. 

\subsection{\label{sec:Laserenergyjitter}Injector laser timing jitter}
\begin{figure}
\includegraphics[width=0.5\textwidth]{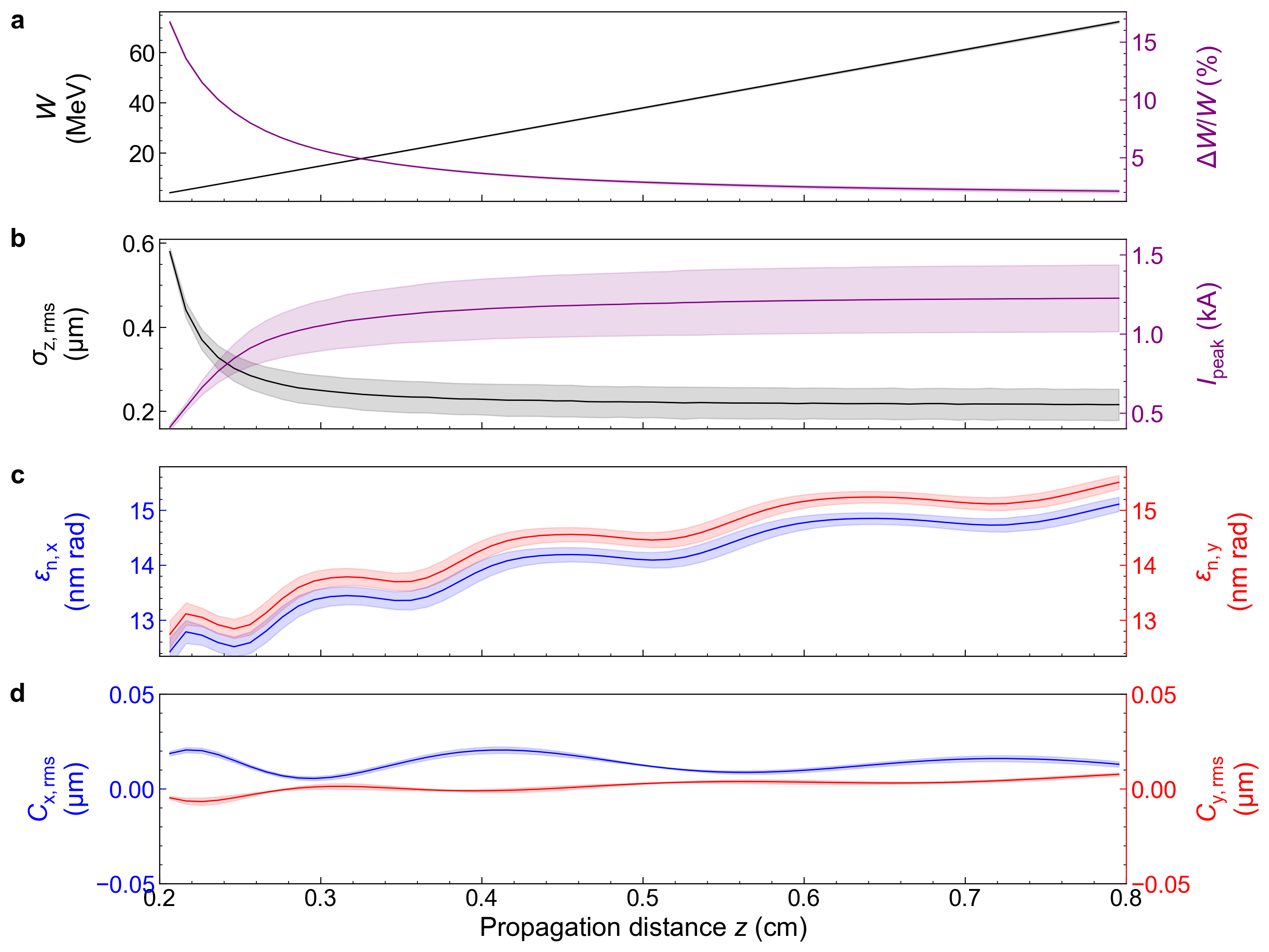}
\caption{\label{fig:Jitter-Detail-Timming}Evolution of witness bunch parameters vs. temporal jitter of \SI{30}{fs} of collinear plasma photocathode laser pulse release position.}
\end{figure}

\noindent
First, we vary the longitudinal release position by shifting the plasma photocathode laser longitudinally in the range of $\Delta \tau = 30$ fs while keeping all other settings constant at the above discussed default parameters. 
The choice of this range is informed by the typical level of synchronization that can be achieved in state-of-the-art linacs e.g. used for X-FEL machines. 
Figure \ref{fig:Jitter-Detail-Timming} shows the 3D PIC simulation results obtained by VSim over a propagation distance of \SI{0.8}{cm}. In order to capture the large blowout structure in its entirety and at the same time resolve relevant physics of PWFA, the co-moving simulation box consists of  $N_{\rm z}\times N_{\rm x}\times N_{\rm y}=358\times217\times217\approx 16.8$ million  cells with a spatial resolution of \SI{1}{\micro\metre} in each direction and an integration time step of $\Delta t \approx $ \SI{2}{fs}. The background hydrogen plasma is modeled with one macro-particle per cell (PPC). Absorption boundary layers are utilized to minimize field reflections.  The neutral helium is implemented as a fluid gas with a PPC of 1000, which increases the number of macro-particles in the witness beam and  improves simulation fidelity. The solid lines represent the average value over all simulations performed  in \SI{5}{fs}-steps up to the maximum delay of $\pm 30$ fs. The shading shows the standard deviation interval around the baseline.

From top to bottom, the evolution of energy gain $W$ (left $y$-axis), relative energy spread $\Delta W/W$ (right $y$-axis), the resulting witness bunch length $\sigma_\mathrm{z,rms}$ (left $y$-axis), peak current $I_\mathrm{peak}$ (right $y$-axis), normalized emittance $\epsilon_{\mathrm{n,x}}$ in $x$-direction and the other transverse plane  $\epsilon_{\mathrm{n,y}}$, and witness beam centroid $C_{\rm x,y,rms}$ in both planes are plotted as a function of propagation distance.

The witness beam energy, energy spread, normalized emittance, centroids and charge are particularly unaffected by temporal injection laser shifts (also see summary table \ref{tab:JitterDataSummaryTable}). The bunch duration and peak current are slightly more impacted by the timing jitter. 
As anticipated in the previous section, the excellent output beam parameter stability arises from the fact that the trapping position that corresponds to the release position in the wake's potential minimum effectively acts as attractor: due to the parabolic shape of the electrostatic potential, final trapping positions  $\zeta_{\rm f}$ of individual slices outside the potential minimum are clustered close behind the trapping position corresponding to the potential minimum. 
When designing plasma photocathodes, one may take the parabolic profile of the trapping potential into account in more detail: for example, a symmetric release volume around the trapping minimum will lead to a folding of two regions around the minimum onto the same trapping positions as a bijective projection, whereas releasing slices only on one side of the minimum results in a simpler injective projection.  Releasing farther away from the minimum means increasingly larger spread of trapping positions, and therefore larger energy spread, longer bunch duration and reduced current.     
With regard to the witness  beam charge, in this scan, excellent stability is seen. 
First, it is worthwhile to mention that all released witness electrons are trapped and form the witness bunch in this scenario, corresponding to a $100\%$ charge capture efficiency. 
This is not trivial, since high charge efficiency during injection or staging is a significant challenge in other approaches e.g. in LWFA \cite{SteinkeNature2016} or via external injection from a linac \cite{WuStagingNature2021}. 
In absolute numbers, the injected charge here amounts to   $Q\approx 2.375\pm 0.006$ pC across the parameter sweep. We note again that much higher charge levels up to the nC range are possible to be released, likewise with $100\%$ capture efficiency.
The  temporal jitter does not have a significant impact on the released charge because the electric wakefields close to the blowout centre are approximately zero; they hence do not contribute significantly to the tunneling ionization yield \cite{YunfengPhysRevSTAB.16.031303} of the laser.  
The excellent charge stability with jitters at the sub-percent level is a result of the decoupling of wakefield excitation and injection, enabled and controlled  by the plasma photocathode.

At the same time, the bunch length $\sigma_\mathrm{z,rms} \approx 0.22 \pm$\SI{0.04}{\micro\metre} (r.m.s.) and the corresponding peak current $I_\mathrm{peak} \approx 1.23 \pm$\SI{0.21}{kA} in this configuration reflects the auto-compression features as discussed: the substantial longitudinal release position variation is compensated by the inherently forgiving trapping mechanism.   

With regard to emittance, excellent average values and excellent stability is obtained in this scan, amounting to $\epsilon_\mathrm{n,x} \approx 15.11  \pm$\SI{0.13}{nm~rad} and $\epsilon_\mathrm{n,y} \approx 15.51 \pm$\SI{0.12}{nm~rad}, respectively (see figure \ref{fig:Jitter-Detail-Timming} c).
Same holds for the centroid variation as shown in figure \ref{fig:Jitter-Detail-Timming} d) and the magnitude of the centroid amplitude is of the order of sub-\SI{0.1}{\micro\metre}.

\subsection{\label{sec:mislaignment}Transverse injector laser shift}
\begin{figure}
\includegraphics[width=0.45\textwidth]{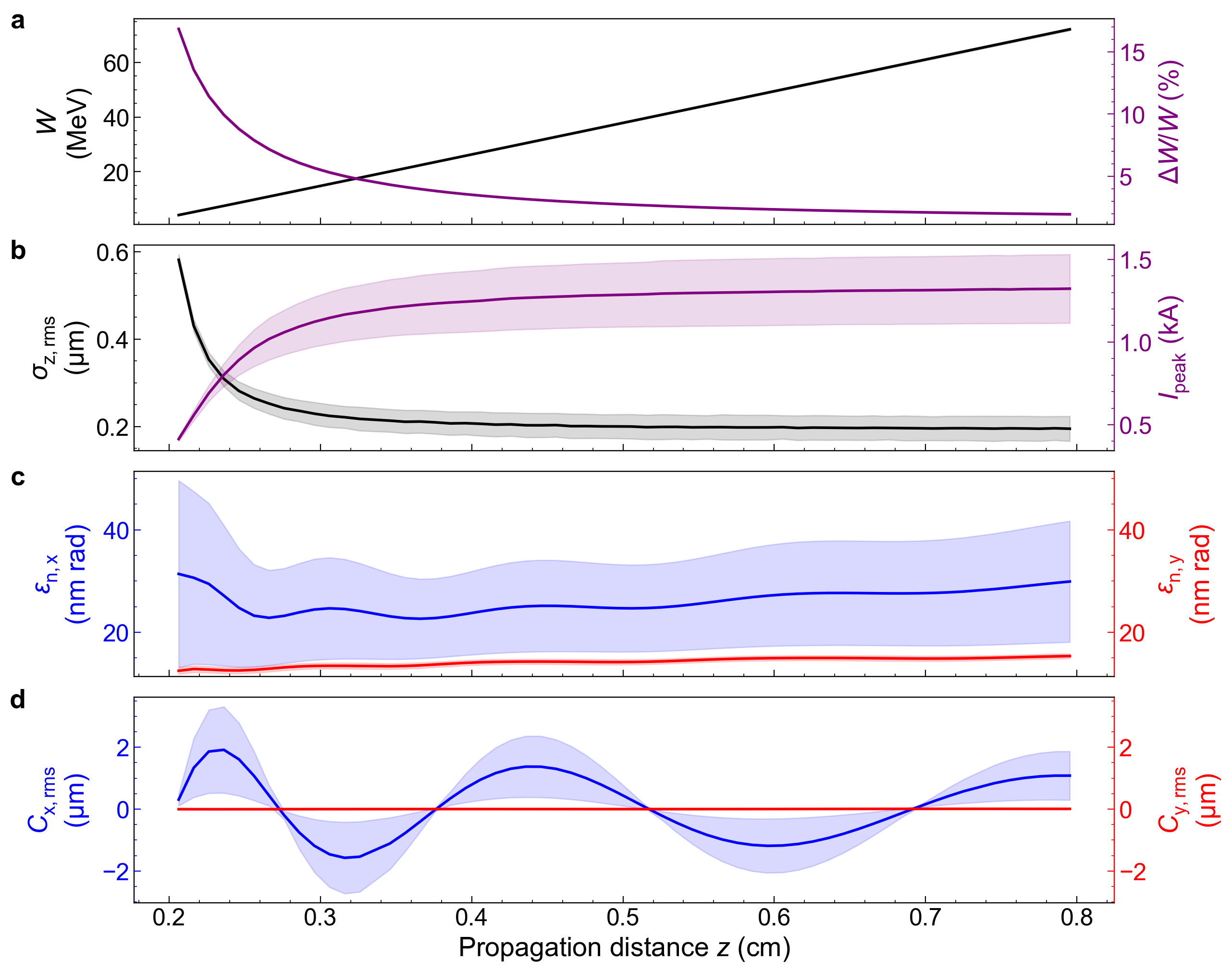}
\caption{\label{fig:Jitter-Detail-Xmisalignment}Evolution of witness bunch parameters vs. transverse jitter of collinear plasma photocathode laser pulse release position.}
\end{figure}

\noindent
Next, we study the impact of a transverse injector laser jitter based on the same default simulation setting as in the above section.  
Figure~\ref{fig:Jitter-Detail-Xmisalignment} shows details of the parameter evolution when the laser pulse is shifted by up to $\Delta X_\mathrm{Laser} =$\SI{10}{\micro\metre}. The solid lines represent the average value over all simulations performed  in \SI{2}{\micro\metre}-steps up to the maximum $\Delta X_\mathrm{Laser}$ shift, and the shading again shows the standard deviation interval around the baseline.
The maximum shift $X_\mathrm{Laser}$ corresponds to $\approx 15\%$ of the maximum plasma blowout radius $R_{\rm b}=$ \SI{65}{\micro\metre}. Plot axes have the same meaning as in figure \ref{fig:Jitter-Detail-Timming}.

While the absolute parameter scan range of \SI{10}{\micro\metre} is very similar to the longitudinal (temporal) scan range, the relative change of release position within the blowout is much larger in this scan, due to the elliptic shape of the plasma wave blowout. Nevertheless,   
similar or better stability level as for the temporal jitter scan is seen for witness beam energy, energy spread, charge, and peak current 
(also see summary table \ref{tab:JitterDataSummaryTable}). 
For example, the witness beam charge stability amounts to $Q\approx 2.371\pm$\SI{0.005}{pC} across the parameter sweep. As for the longitudinal jitter, the transverse wakefields do not contribute significantly to the charge yield jitter. 

The variation of emittance in $x$-direction is larger than in $y$-direction, as expected from an off-axis release position in $x$-direction because of the larger  transverse momenta of the electrons. Nevertheless, it amounts to $\approx 29.91 \pm$\SI{11.8}{nm~rad} -- these are values that are even in a worst-case scenario orders of magnitude better than from state-of-the-art linacs used e.g. for X-FELs.  
At the LCLS linac, for example, the (simulated) normalized emittance of the electron beam is of $\sim$\SI{}{mm \, mrad}-scale, and its shot-to-shot variation is of $\sim 0.5\,$mm mrad-scale \cite{Emma2010fyaPAC,987880LCLSjitterPAC2001}. This suggests that not only the average emittance can be by a factor of ~100 better than state-of-the-art, but also the emittance stability of the plasma photocathode may be 10 times better than at the best X-FEL linacs today. 
Of course, one may argue that a comparison between simulations with variations of only a few parameters with full-scale experimental results would be inherently unfair and that many experimental milestones are to be reached yet; however on the other hand, as explained earlier the combination of plasma photocathode, bunch compressor and accelerator within a single stage conceptually simplifies the setup substantially when compared to a state-of-the-art linac. The inherent robustness of the plasma photocathode with regard to physical principles, and the overall simplicity of the setup, may allow to bring the stability and controllability prospects to fruition. 

Electrons released off-axis experience the restoring force of the ion background in the $x$-direction. This excites collective transverse oscillations of the witness beam electrons only in the $x$-direction, apparent from the centroid evolution plotted in Figure \ref{fig:Jitter-Detail-Xmisalignment} d).   
However, it can be seen that the oscillation amplitudes quickly decrease with increasing beam energy.  Already at a witness beam energy of $W \approx$~\SI{70}{MeV} reached at the end of the \SI{0.8}{cm} propagation distance, the witness beam centroid amplitude reaches the \SI{}{\micro\metre} to sub-\SI{}{\micro\metre}-level, and will be further reduced with increasing beam energy. Again, this is a direct important advantage 
of releasing electrons at rest inside the wake.  Inherently, off-axis injected beams from plasma photocathodes will rapidly self-align to the driver beam propagation axis with increasing beam energy and therefore, alignment of the driver beam to the desired orbit is the sole challenge in minimizing centroid jitter of the witness beam at the plasma stage exit. This rapid reduction of the betatron oscillation is of multi-faceted advantage. 
Perhaps most strikingly, one may compare the situation with external injection of pre-accelerated electron beams. In such a scenario, where the electron beam may have an energy of already tens or hundreds of MeV, a transverse,  or even worse angular pointing jitter can be catastrophic. In the plasma wake, the relativistic electron beam will perform betatron oscillations with large amplitude much longer, and may even not be captured by the plasma wake. It will also not move to a favourable accelerating phase automatically, in contrast to electrons released by a plasma photocathode laser.
 A centroid offset and oscillation of the witness beam is not only problematic inside a plasma accelerator stage, but imposes fundamental challenges for beam transport post plasma, including complete beam loss due to a pointing exit angle outside the acceptance of the beam transport line.  
 Next to partial or complete witness beam loss,  hosing instabilities  \cite{PhysRevLett.118.174801,alberto2018stability} and beam energy spread and bunch duration growth during the acceleration \cite{pousa2019intrinsic} are also unwanted.
 
It shall be noted that the rapid inherent self-alignment feature  of  plasma photocathode injection makes this method an excellent candidate for test beams for emittance preservation studies in multi-stage plasma-based accelerators towards plasma-based linear collider efforts.   

\subsection{\label{sec:Laserenergyjitter}Injector laser intensity fluctuations}
Laser pulse energy or power fluctuations will result in variation of the plasma photocathode laser pulse intensity at the injection position. Other factors such as spot size, wavefront flatness etc. also can vary from shot-to-shot, and will also effectively result in an intensity variation. In turn, laser pulse intensity fluctuations will result in variation of the effective tunneling ionisation yield. 
Again, we performed a series of 3D PIC simulations using our baseline interaction parameter set and varied the dimensionless laser amplitude $a_0$ by up to $\pm 2\%$ around the baseline value of $a_0 = 0.018$, while keeping other laser parameters constant. 

\begin{figure}[ht]
\includegraphics[width=0.75\textwidth]{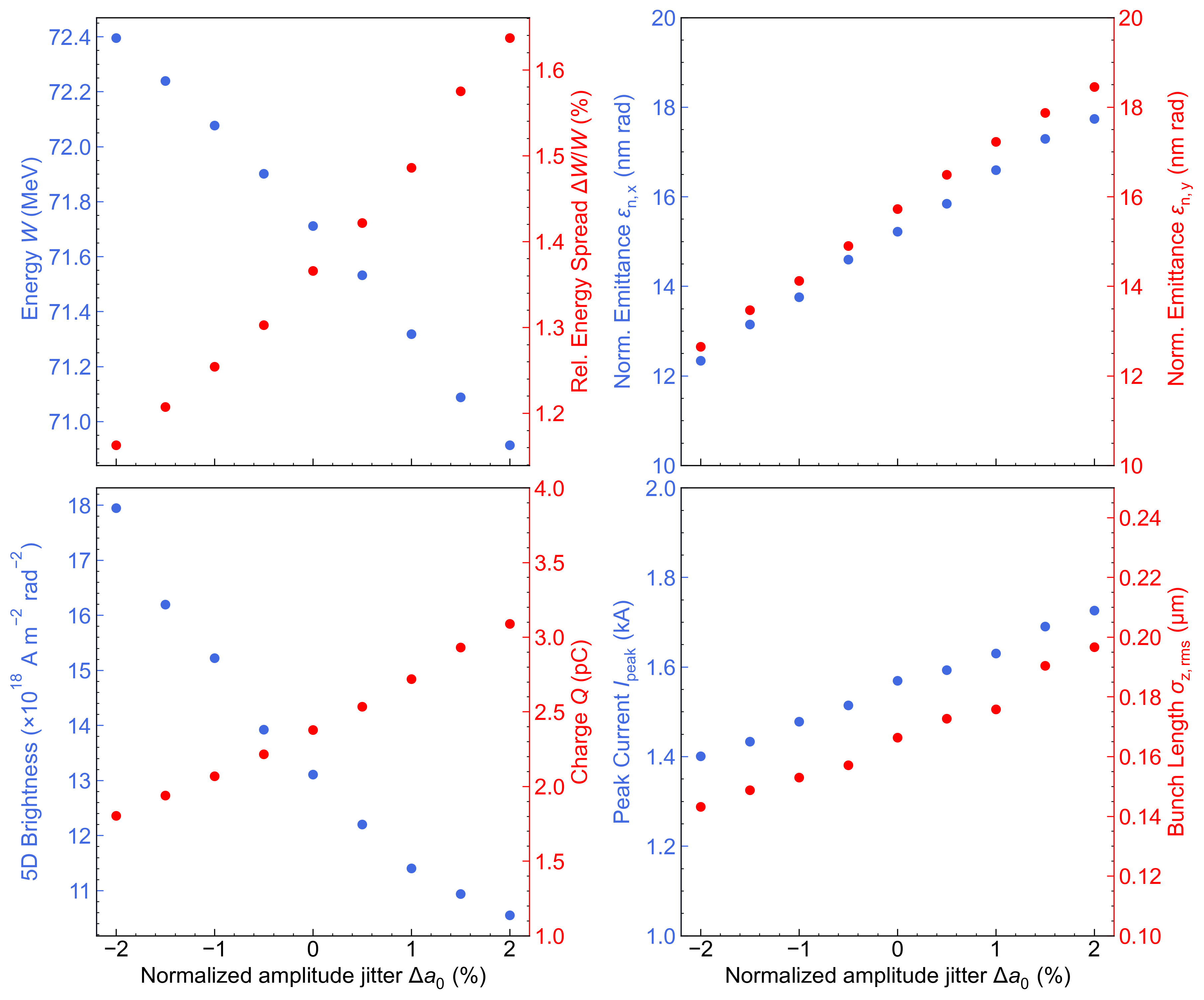}
\caption{\label{fig:Jitter-Summarya0}Condensed summary plot of key witness bunch parameters when scanning the effect of laser pulse intensity variations by varying $a_0$.}
\end{figure}

Figure \ref{fig:Jitter-Summarya0} summarizes the results of the laser intensity scan. Again, very high output parameter stability around outstanding average parameter values are obtained. If we analyse the individual plots, then the following observations and interpretations apply: With regard to output witness beam energy, the energy slightly decreases monotonically as the plasma photocathode laser intensity increases. The relative energy spread, in contrast, slightly increases monotonically as laser intensity increases. These trends can be attributed to beam loading as more charge is released and trapped when the release laser intensity increases (the released charge $Q$ is given in the bottom left panel), and the longer release volume, which results in a longer beam and larger energy gain differences between head and tail of the beam \cite{ManahanHabib6d}.  This systematic behavior indicates that  the energy spread can be adjusted by tuning the laser pulse energy at the percent-level, which is experimentally straightforward. Again, it may be worthwhile to highlight that even in a scenario where the laser energy and intensity may not be fully controllable to $a_0$ better than $\pm 2\%$, the energy stability of the output witness beam $W \approx71.69 \pm$\SI{0.68}{MeV}, and likewise the relative energy spread of $\Delta W/W \approx 1.38 \pm$\SI{0.15}{\%} are very promising, with slice energy spreads far below this level. 

The normalized emittance in both planes (top right panel in figure \ref{fig:Jitter-Summarya0}) increases monotonically with increasing laser intensity. Various  factors contribute to this: first, a higher $a_0$ means that a larger volume of He gas is ionized, as a larger laser pulse electric field means higher ionization rates, and hence He electron release also occurs farther away from axis, which increases the initial transverse phase space of the witness beam. Second, larger $a_0$ also means electrons are released over a longer spread in longitudinal direction, which increases the range of betatron oscillation phases which contribute to the final trapped bunch (phase mixing). Third, higher bunch charge implies larger intra-bunch space charge forces, which also slightly increase transverse electron momenta. And finally, larger $a_0$ will also increase the residual transverse momentum slightly, which also contributes to the finally obtained emittance. This thermal emittance contribution is estimated to be typically  negligible when compared to the other sources of emittance, which justifies modelling the photocathode laser pulse with an envelope function instead of fully resolving it. A comparative analysis of the different sources of emittance in relevant scenarios, and balanced optimization pathways will be undertaken in another study.

The charge yield $Q$ (bottom left panel) shows significant changes over the full range of $a_0$ variation, as expected. It shall be noted that energy stability of sub-mJ class lasers can be substantially better than the range considered here, which in turn allows much smaller charge jitter as considered here. 
For many applications, including key ones such as plasma-based X-FEL's, current is more important than charge. It is therefore of considerable interest that at the same time as the charge yield may increase due to higher $a_0$, also the bunch duration increases (bottom right panel). The associated current, therefore by far does not increase as much as the charge, because the witness bunch duration  increases with increasing charge. This auto-current-stabilization feature is then also inherited by the obtained brightness, another key performance parameter e.g. for FEL and other light source applications.    

\subsection{\label{sec:Laserenergyjitter}Overview of impact of  timing, transverse, and intensity jitter}

\begin{figure*}
\includegraphics[width=1.0\textwidth]{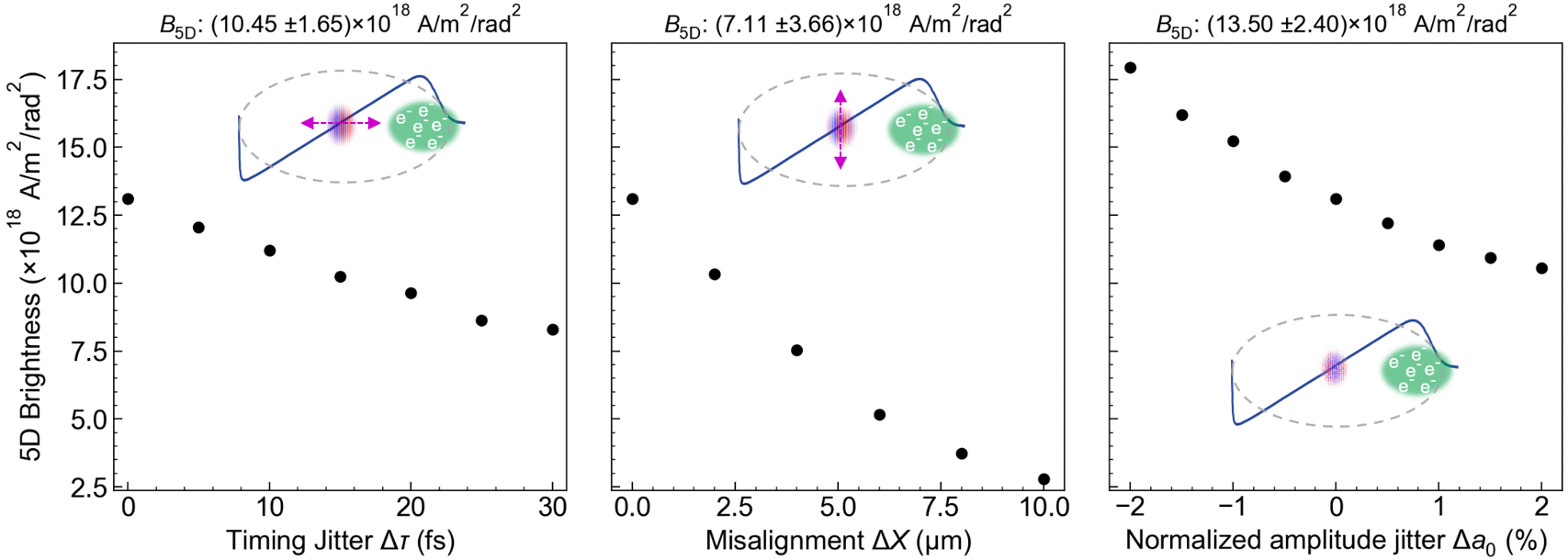}
\caption{\label{fig:Jitter-5Dbrightness}5D-brightness dependency of plasma photocathode timing release variation (left), transverse plasma photocathode release position offset $\Delta X$ (middle),  and of normalized amplitude $a_0$ of the plasma photocathode laser pulse.}
\end{figure*}

As a composite parameter, the witness beam brightness not only reflects key beam parameters, but can as well be used for quantifying overall beam stability. 
We compare the impact of the three main investigated jitter sources on the beam brightness in figure \ref{fig:Jitter-5Dbrightness}. 

In the left panel, impact of timing jitter  between release laser pulse and blowout structure is presented, whereas the middle panel shows the impact of transverse jitter. Although \SI{30}{fs} is approximately equivalent to the transverse jitter range of \SI{10}{\micro\metre}, the resulting 5D-brightness values (both mean and range) from the timing jitter study are even better than those obtained from the transverse jitter study.  
The modest impact on beam brightness compared to the other jitter sources is expected, because of the quadratic contribution of emittance $B_\mathrm{5D}\propto \epsilon_{\rm n}^{-2}$ and the previously described outstanding resilience of emittance vs. timing jitter. 
In summary, three factors are responsible for this. 
First, the elliptically shaped blowout has its principal axis in longitudinal direction, which as shown in figure \ref{fig:LargerBlowout}, means that the same absolute offset amounts to a relative offset which is smaller in the longitudinal direction of the elliptical blowout structure than in the transverse direction. Second, and more fundamentally consequential, the longitudinal  electrostatic potential of the wake has a parabolic profile and a local minimum around the blowout centre (see figure \ref{fig:LargerBlowout}). Therefore, this release region is particular resilient against longitudinal release position jitter as discussed above. Finally, the transverse momentum of electrons released at slightly different longitudinal positions is very similar. All these factors contribute to a substantially better emittance as obtained for the transverse offset scenario. This is a fortunate constellation, as at linac-driven systems pointing stability of a laser pulse (the transverse jitter) is typically better controllable (e.g., to the few \SI{}{\micro\metre}-scale \cite{PhysRevSTABCinquegranaPumpProbeJitter2014}) 
than timing. The inherent larger resilience of the plasma photoinjector to timing than to transverse offset is therefore a complementary advantageous fit to the poorer timing precision when compared to the transverse precision in linac-based systems. 
For completeness, in the right panel of figure \ref{fig:Jitter-5Dbrightness} we present again the $a_0$-dependency of the 5D-brightness, a plot corresponding to the one in figure \ref{fig:Jitter-Summarya0}, bottom left panel.

The beam parameters presented in this study are projected quantities, however, the witness beams' slice brightness can easily exceed  $B_\mathrm{5D}> 10^{20} \rm Am^{-2} \rm rad^{-2}$-levels. 
Electron beams with such unprecedented 5D-brightness are a central capability of the Trojan Horse plasma photocathode technique. The above considerations and simulations indicate that at the same time there are also extremely attractive prospects with regard to the stability and tunability of such output electron beams. 
This may be perceived as counter-intuitive, since conventional wisdom is rather that improvements in beam quality can be obtained mainly through higher complexity of the setup, which in turn puts much higher demands on technology control on aggregate.  

\begin{figure*}
\includegraphics[width=1.0\textwidth]{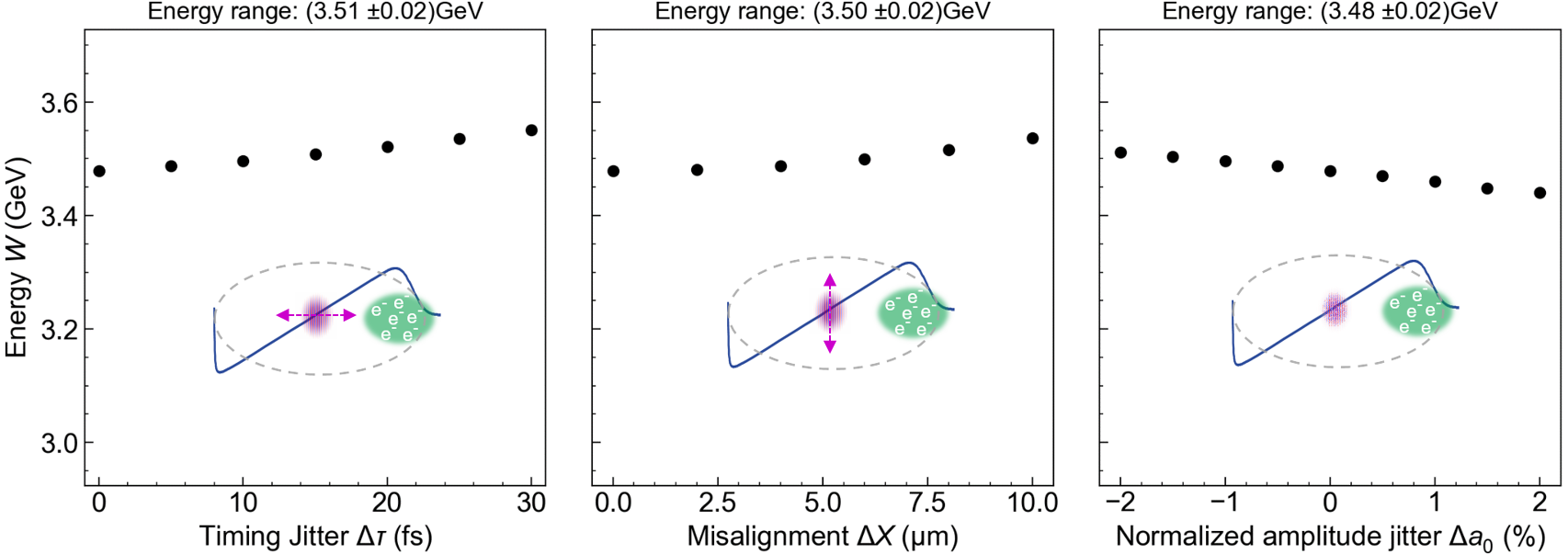}
\caption{\label{fig:Jitter-Energy}Electron energy dependency of plasma photocathode timing release variation (left), of  transverse plasma photocathode release position offset $\Delta X$ (middle), and of normalized amplitude $a_0$ of the plasma photocathode laser pulse.}
\end{figure*}

Next to stability e.g. of the emittance and 5D-brightness, a further crucial aspect is the output energy stability. This is important for applications, for example when aiming at an FEL, where the electron beam energy defines the resonant wavelength of the produced radiation. Energy stability is also crucial for electron beam transport from the end of the plasma stage towards the applications. We therefore have examined the energy stability of output witness beams with respect to the three jitter sources discussed above. Again, PIC-simulations were carried through with VSim over the initial, and defining phase of acceleration over \SI{8}{mm}. We then assumed a non-evolving, constant wakefield, which is justified in the case of highly relativistic, high current driver beams such as available at SLAC, and projected the central energy gain up to $W \approx$~\SI{3.5}{GeV}.

Figure \ref{fig:Jitter-Energy} summarizes the outcome of the jitter studies. 
When the timing is varied in longitudinal jitter scans (left panel of figure \ref{fig:Jitter-Energy}), a release position longitudinally outside the electrostatic potential minimum means a trapping position further behind in the wakefield, hence a slightly larger energy gain. 

When the transverse release position is shifted, the obtained energy actually also shows a slightly larger energy gain than when electrons are released on axis. When electrons are released off axis, they are performing said betatron oscillations around the axis, leading to reduction of the longitudinal velocity during the trapping process due to the relativistic momentum conservation. Therefore, electrons require longer acceleration distances to catch up with the plasma wave and are trapped further at the rear of the wakefield. The same point can be expressed by arguing with a smaller electrostatic wake potential outside the blowout center (see figure \ref{fig:LargerBlowout}).  

Hence, the accelerating longitudinal wakefields at the corresponding trapping positions (see the blue profile in the schematic insets in figure \ref{fig:Jitter-Energy}) for offset release are  slightly larger than at a trapping position earlier in the wake. 
One could aim at maximum energy gain by releasing at a position that ensures the latest possible trapping position, but here we have factored in a safety margin as regards the witness beam trapping position.
This is another advantage of the scheme: the electron driver beam will in practice have an energy and current jitter, and the plasma wakefield will evolve due to driver beam energy depletion, head erosion etc.  By allowing a sufficient safety margin and not aiming to trap at the very end of the initial wakefield distribution, one can effectively ensure that the electron witness beam is stably accelerated in a "safe zone" of the wakefield. 

Finally,  the $a_0$-dependency is scanned (right panel of figure \ref{fig:Jitter-Energy}). Here, it is observed that when the laser amplitude increases, the energy gain is decreasing -- as discussed before, this can be attributed to slight beam loading when a stronger laser pulse releases more charge (and current). 

In total, the resulting energy stability across all these scans is excellent: the energy variation amounts to the sub-1$\%$ level. This is a level similarly obtained at state-of-the-art linacs which drive X-FELs \cite{Emma2010fyaPAC,987880LCLSjitterPAC2001}.  

Table \ref{tab:JitterDataSummaryTable} summarizes the jitter of witness beam parameters with respect to spatiotemporal and intensity jitter of the injector laser around the baseline scenario. We note that the plasma photocathode spatiotemporal pointing jitter of incoming beams measured at FACET (see figure \ref{fig:E210jitters}) is of the same order of magnitude than the jitter assumed here in the sensitivity studies. However, today's technical capability for  jitter minimization of incoming beams is much better than that (and e.g. at FACET-II efforts are made to improve those), so we conclude that jitters as assumed here, even if e.g. longer propagation of the laser pulse in plasma in case of collinear injector geometry represents an additional challenge, are entirely possible, and likely can be much better. This further enhances prospects for improved stability of output beams.

\begin{center}
\begin{table*}[ht]
\caption{\label{tab:JitterDataSummaryTable}Witness beam parameter summary of plasma photocathode laser jitter analysis.}
 \begin{tabular}{l c c c} 
 \hline \hline
 Beam parameter & Timing jitter $\Delta \tau$  &  Pointing jitter $\Delta X$ & Laser amplitude jitter  $\Delta a_{0}$ \\ [0.5ex]
 \hline\hline
 Energy $W$ (MeV) & $72.38\pm0.69$ & $72.15\pm0.59$ & $71.69\pm0.68$ \\ 
 Energy spread ($\%$) & $1.52\pm0.11$ & $1.41\pm0.05$  & $1.38\pm0.15$ \\
 Charge (pC)& $2.375\pm0.006$ & $2.371\pm0.005$ & $2.41\pm0.42$ \\
 Peak current $I_\mathrm{p}$ (kA) & $1.23\pm0.21$ & $1.32\pm0.21$ & $1.56\pm0.11$ \\
 Bunch length ($\mu$m) & $0.22\pm0.04$ & $0.19\pm0.03$& $0.17\pm0.02$ \\
 Normalized emittance $\epsilon_\mathrm{n,x}$ (nm rad)  & $15.11\pm0.13$ & $29.91\pm11.80$ & $15.17\pm1.77$ \\
 Normalized mittance $\epsilon_\mathrm{n,y}$ (nm rad) & $15.51\pm0.12$ & $15.38\pm0.48$ & $15.66\pm1.90$ \\
 5D brightness ($\times 10^{18}$ A m$^{-2}$rad$^{-2}$) & $10.45\pm1.65$ & $7.11\pm3.66$ & $13.5\pm2.40$ \\
 \hline \hline
\end{tabular}
\end{table*}
\end{center}

\begin{figure*}
\includegraphics[width=1.0\textwidth]{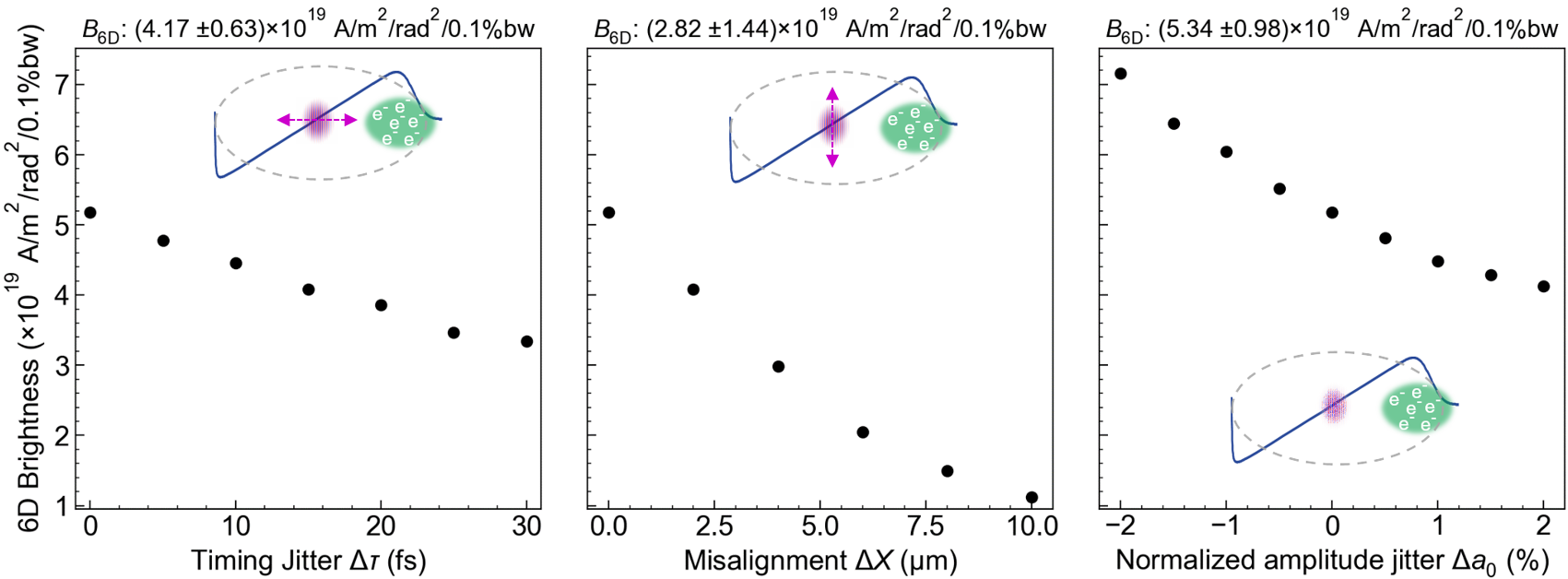}
\caption{\label{fig:Jitter-6Dbrightness}6D-brightness dependency of plasma photocathode timing release variation (left), of transverse plasma photocathode release position offset $\Delta X$ (middle),  and of normalized amplitude $a_0$ of the plasma photocathode laser pulse.}
\end{figure*}

Finally, although not the focus of this study, we comment on the energy spread and 6D brightness, defined as $B_\mathrm{6D} = B_\mathrm{5D}/(\Delta W_{\rm res}/W) 0.1\% \mathrm{bw}$ as introduced initially. The relative energy spread $\Delta W_{\rm res}/W$ included here is a further crucial parameter for beam transport, emittance and brightness preservation, and applications.  
 
In the following, we provide estimates for  $\Delta W_{\rm res}/W$  by considering 3D PIC simulation results in conjunction with the energy spread scaling law derived in
\cite{ManahanHabib6d} for the dechirping  obtainable by applying the escort beam-loading method. 

From the discussed PIC-simulations for the $\lambda_\mathrm{p} =  \SI{250}{\micro\metre}$ plasma wavelength case, we find that the acceleration gradient at the witness beam trapping position is $E_{\mathrm{z,trap}}\approx$ \SI{11.5}{\giga\volt\per\metre}.
For simplicity, we assume acceleration distances of $\Delta z\approx$ \SI{30}{cm} and obtain extrapolated witness beam energies $W\approx$ \SI{3.5}{GeV}.
These considerations are valid because of the phase-locked acceleration in beam-driven plasma wakefield acceleration, and viability of long acceleration distances  even in a single stage.
Considering the scaling law for the residual energy spread $\Delta W_{\rm res}\approx2\pi E_{\rm z,trap} w_{0}^2/5\lambda$ from \cite{ManahanHabib6d} at this plasma wavelength, we obtain a residual energy spread of $\Delta W_{\rm res}\approx$ \SI{0.88}{MeV}, which in case of full dechirping at an energy of \SI{3.5}{GeV} yields a relative energy spread of $\Delta W_{\rm res}/W\approx$ \SI{0.025}{\%}.
These are highly attractive values, in particular because they can be obtained in a single plasma accelerator stage in combination with the nm rad scale normalized emittances, which could prove an enabling feature for emittance and brightness preservation during transport.
The corresponding summary plots of the 6D brightness for the $\lambda_\mathrm{p} = \SI{250}{\micro\metre}$ case are shown in figure \ref{fig:Jitter-6Dbrightness}.
Since the estimated residual energy spread is not much larger than those of electron beams used for X-FEL systems, and since as discussed the 5D brightness is by orders of magnitude higher than state-of-the-art, consequently also the 6D brightness is orders of magnitude higher than state-of-the-art obtained at the best X-FEL systems today. 
The 6D brightness trend naturally also represents  the same trends as the 5D brightness in terms of variation of timing (left panel), transverse offset (middle panel), and laser intensity (right panel).

Working at even lower plasma densities can further reduce the residual and therefore also the relative energy spread according to  the $\Delta W_{\rm res}\propto n_{\rm p}^{1/2}$ scaling \cite{ManahanHabib6d} (albeit this is not computationally verified here due to computational costs). 
With regard to all jitter studies discussed here, it should be emphasized that we have assumed no variation of the plasma wakefield driver beam. In reality, one will have to take into account shot-to-shot jitters of the incoming electron driver beam, such as its current.
We will perform jitter studies, taking into account real jitter values of the FACET-II electron beam, when they are known and measured as they develop.
Again, the decoupled nature of the plasma photocathode injection process is expected to be highly beneficial for the obtainable witness beam output stability.

The obtainable values of 6D brightness, taking into account the improvement of relative energy spread $\Delta W/W$ at higher electron witness energies, are in agreement with the conceptional overview figure \ref{fig:6DbrightnessPlasmaPhotocathodevsElectronEnergy}. Here, the reach of 6D brightness values obtainable from the techniques and scenarios as described above are summarized, and contrasted with current state-of-the-art of other types of plasma accelerators, as well as with traditional linac-based X-FELs.    

\section{\label{sec:Conclusion}Conclusion}
The plasma photocathode approach extends the exploitation of plasma 'merely' as an accelerator module with superior gradients to also being a source of electron beams with superior characteristics. 
In summary, in this work we present details of the E-210: Trojan Horse experiment at FACET, thus amending pertinent reports on this aimed at a broader readership such as \cite{Deng2019Trojan}. We concentrate on limitations of this experiment, and outline improvement  measures and  techniques that we aim to develop and implement at FACET-II.

 Step-like tunneling ionization thresholds of gaseous media with low and high ionization thresholds have been harnessed in E-210. The preionized plasma channel is an important bottleneck that has to be widened to allow  unconstrained PWFA to be performed. The preionization laser pulse has been used to ionize hydrogen to support the PWFA, and the plasma photocathode laser pulse  was used to ionize helium, and to release and inject electrons inside the plasma wave. Once full ionization of the low ionization threshold medium is reached, excess preionization laser intensity does not change the local plasma density, as long as it stays below the tunneling ionization threshold of the high ionization threshold medium.  This 'peak limiter' provides an ionization intensity corridor feature that offers resilience towards shot-to-shot preionization laser pulse jitters in terms of power and effective intensity, and alignment. If a wide enough preionization channel can be generated in a stable manner, jitter sources resulting from the preionization channel, which had dominating impact on the witness output beam stability and quality for E-210, can be eliminated. 

\begin{figure}
\includegraphics[width=.48\textwidth]{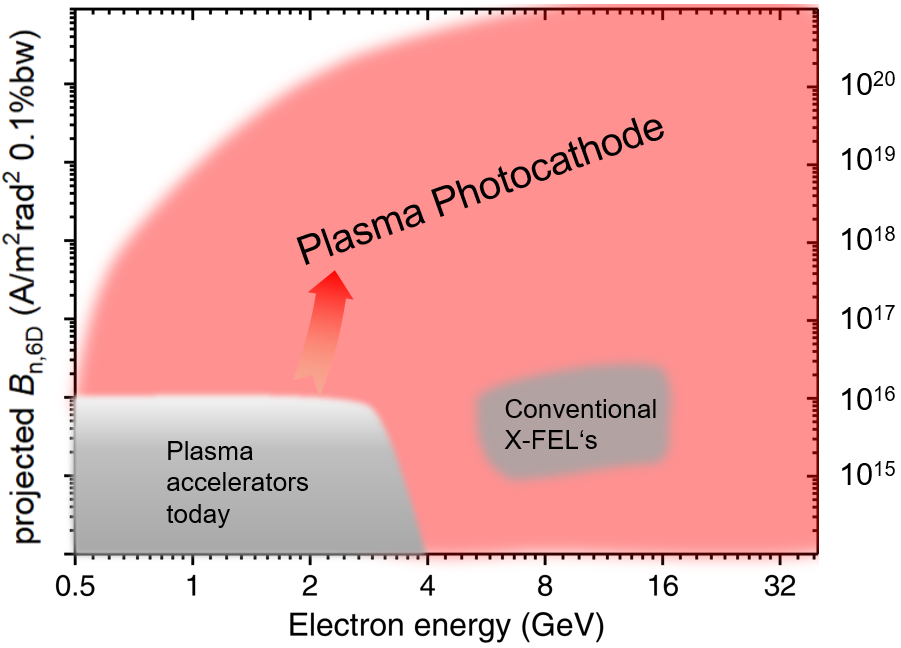}
\caption{\label{fig:6DbrightnessPlasmaPhotocathodevsElectronEnergy}Reach of the plasma photocathode as regards obtainable 6D-brightness in comparison to state-of-the-art.}
\end{figure}
A wider channel also will allow for operation at lower plasma density and correspondingly larger blowouts. This can solve a series of issues. For example, unwanted hot spots can be avoided, relative plasma photocathode spatiotemporal injection precision is improved, residual energy spread of injected electrons is decreased, and possible benefits for emittance can be harnessed. Additionally, a better stability of incoming electron driver beam and injector laser pulse, and electron to laser pulse synchronization will contribute to improved absolute injection precision.

We explore the impact of  spatiotemporal injection precision, and plasma photocathode injector laser intensity variation,  on witness bunch production in collinear geometry by simulation-based jitter studies. These indicate a remarkable potential for stability around target values of emittance and brightness which go beyond the state-of-the-art by many orders of magnitude. 

Such ultrabright electron beams would have fundamental impact on experiments and applications e.g. in photon science and high field and energy physics. The current theoretical reach of plasma photocathodes in terms of 6D brightness, based on technology that is today available, and on techniques developed so far, is depicted in figure \ref{fig:6DbrightnessPlasmaPhotocathodevsElectronEnergy}. 

While the applications of ultralow emittance and ultrabright electron beams are too numerous and wide-reaching to discuss them adequately here, we highlight two major thrusts: First, both the ultralow emittance, and ultrahigh brightness potential of plasma photocathodes makes them ideal candidates at the heart of advanced photon sources such as soft X-ray coherent synchrotron radiation sources \cite{alotaibi2020plasma,emma2021terawatt}, hard X-ray FEL, betatron radiation and ion channel lasers \cite{litos2018experimental,habib2019plasma}, and $\gamma$-ray sources \cite{habib2019plasma}. Second, such beams could be spearheading R\&D towards high energy physics colliders. The paramount importance of emittance and brightness for both thrusts is well-known  \cite{ANLbrightnessWhitePaper2003}, and hence e.g. plasma-based X-FELs are seen as a major milestone and stepping stone towards high energy physics R\&D \cite{hidding2019plasmaPWASC,European:2720131}. This strategic connection is  true for various types of accelerator  R$\&$D, but it is epitomized for plasma photocathode wakefield acceleration R\&D: ultralow emittance and ultrahigh brightness -- the chief attraction of plasma photocathodes -- is the key physics requirement for X-FELs \cite{ANLbrightnessWhitePaper2003,HiddingPRL2012PhysRevLett.108.035001} e.g. with regard to photon energy and gain, but also for HEP because of luminosity considerations. 

The plasma photocathode based X-FEL thrust is currently investigated in the UK STFC funded PWFA-FEL \cite{PWFA-FEL} design effort as a UK-US collaboration. Due to the far-reaching prospects of improved brightness, plasma photocathode beam brightness transformers are already considered as an addition to the UK XFEL in its Science Case   \cite{UKXFEL-ScienceCase2020} -- the first time  plasma-based X-FEL is part of a new X-FEL plan right from its conception.

For high energy physics and colliders in particular, ultrabright beams have several short- to long term applications. As mentioned already in \cite{ANLbrightnessWhitePaper2003}, one long-term prospect would be to open up the possibility of obviating the electron beam damping ring.  But there are also many short-term applications, such as using ultralow emittance beams from plasma photoguns as test beams for emittance preservation during staging. For a future TeV-class linear collider with many stages, for example, even a few nmrad-scale emittance growth per stage could be prohibitive for reaching luminosity goals -- therefore nmrad-scale test beams are required. This is coupled with the task of nmrad-scale emittance diagnostics, and other ultrabright electron beam diagnostics.     
Exploration of the production of spin-polarized electrons e.g. by using pre-polarized targets and/or ionization via (circularly) polarized \cite{Fano1969PhysRev.178.131} plasma photocathode laser(s) \cite{JoshiSpin2021} is a further R$\&$D topic that could further enhance the attractivity of plasma photoguns.

The ultralow emittance combined with femtosecond-level bunch duration -- corresponding to multi kA currents and linac-level energy spreads \cite{habib2019plasma} -- in principle also allows for extreme charge densities. The resulting collective, Lorentz-boosted unipolar electric field distribution is a unique modality, which makes them attractive e.g. for QED studies \cite{Yakimenko2019PhysRevLett.122.190404}. 
Further, the potential availability of intense hard x-ray or $\gamma$-ray beams, derived from ultrabright electrons produced by integrated plasma photocathode wakefield accelerators via novel and/or improved mechanisms \cite{habib2019plasma}, could enable novel constellations for particle and photon colliders as outlined in the UK-XFEL Science Case \cite{UKXFEL-ScienceCase2020}.

Finally, efforts to use plasma also as collective diagnostics of low emittance and/or high brightness beams e.g. via the plasma afterglow \cite{Scherklarxiv2019} mechanism, and for symmetric focusing of such beams via plasma lenses \cite{chen:1989prd,chen:1990prl,Doss2019} are highly synergistic with plasma  photocathodes.

The plasma afterglow technique \cite{Scherklarxiv2019} goes another conceptual step and aims to use plasma also as a highly sensitive detector medium. The overarching aim here is to retrieve important characteristics of electron and laser beams by harnessing the collective response of plasma with high sensitivity, but non-intrusively. The use of plasma as a high-sensitivity detector would then complete the trinity of plasma-based photoguns, accelerators and detectors. Jointly with plasma lenses and other plasma-based beam manipulation techniques \cite{WITTIG201683} we are therefore on the path to  an emerging, mutually reinforcing ecosystem of plasma- and laser-based building blocks. 

At FACET-II, an interconnected set of experiments will be used to explore and develop these approaches. This includes the E-310 to E-313 experiments series for electron beam generation and acceleration, and the E-315 and E-316 experiments for diagnostics. 

\begin{acknowledgments}
The FACET ‘E210: Trojan Horse’ plasma wakefield acceleration experiment was built and operated with support from  RadiaBeam Technologies (DOE contract no. DE-SC0009533), UCLA (US Department of Energy (DOE) contract no. DESC0009914), the FACET E200 team and DOE under contract no. DE-AC02-76SF00515, H2020 EuPRAXIA (grant no. 653782), Helmholtz VH-VI-503, EPSRC (grant no. EP/N028694/1) and the Research Council of Norway (grant no. 230450). B.H. acknowledges support from the DFG Emmy–Noether programme. For part of the work presented here, B.H., F.A.H., T.H., P.S., A.N., L.R. and D.C. were supported by the European Research Council (ERC) under the European Union’s Horizon 2020 research and innovation programme (NeXource: Next-generation Plasma-based Electron Beam Sources for High-brightness Photon Science, ERC Grant agreement No. 865877) and by the STFC PWFA-FEL programme ST/S006214/1. This work used computational resources of the National Energy Research Scientific Computing Center, which is supported by DOE DE-AC02-05CH11231, and Shaheen (project k1191). D.L.B. acknowledges support from the US DOE Office of High Energy Physics under award no. DE-SC0013855. J.R.C. acknowledges support from the National Science Foundation under award no. PHY 1734281.
\end{acknowledgments}

\bibliography{bib-prab}

\begin{thebibliography}{63}%
\makeatletter
\providecommand \@ifxundefined [1]{%
 \@ifx{#1\undefined}
}%
\providecommand \@ifnum [1]{%
 \ifnum #1\expandafter \@firstoftwo
 \else \expandafter \@secondoftwo
 \fi
}%
\providecommand \@ifx [1]{%
 \ifx #1\expandafter \@firstoftwo
 \else \expandafter \@secondoftwo
 \fi
}%
\providecommand \natexlab [1]{#1}%
\providecommand \enquote  [1]{``#1''}%
\providecommand \bibnamefont  [1]{#1}%
\providecommand \bibfnamefont [1]{#1}%
\providecommand \citenamefont [1]{#1}%
\providecommand \href@noop [0]{\@secondoftwo}%
\providecommand \href [0]{\begingroup \@sanitize@url \@href}%
\providecommand \@href[1]{\@@startlink{#1}\@@href}%
\providecommand \@@href[1]{\endgroup#1\@@endlink}%
\providecommand \@sanitize@url [0]{\catcode `\\12\catcode `\$12\catcode
  `\&12\catcode `\#12\catcode `\^12\catcode `\_12\catcode `\%12\relax}%
\providecommand \@@startlink[1]{}%
\providecommand \@@endlink[0]{}%
\providecommand \url  [0]{\begingroup\@sanitize@url \@url }%
\providecommand \@url [1]{\endgroup\@href {#1}{\urlprefix }}%
\providecommand \urlprefix  [0]{URL }%
\providecommand \Eprint [0]{\href }%
\providecommand \doibase [0]{https://doi.org/}%
\providecommand \selectlanguage [0]{\@gobble}%
\providecommand \bibinfo  [0]{\@secondoftwo}%
\providecommand \bibfield  [0]{\@secondoftwo}%
\providecommand \translation [1]{[#1]}%
\providecommand \BibitemOpen [0]{}%
\providecommand \bibitemStop [0]{}%
\providecommand \bibitemNoStop [0]{.\EOS\space}%
\providecommand \EOS [0]{\spacefactor3000\relax}%
\providecommand \BibitemShut  [1]{\csname bibitem#1\endcsname}%
\let\auto@bib@innerbib\@empty
\bibitem [{\citenamefont {Pellegrini}(2012)}]{Pellegrini2012}%
  \BibitemOpen
  \bibfield  {author} {\bibinfo {author} {\bibfnamefont {C.}~\bibnamefont
  {Pellegrini}},\ }\bibfield  {title} {\bibinfo {title} {The history of x-ray
  free-electron lasers},\ }\href {https://doi.org/10.1140/epjh/e2012-20064-5}
  {\bibfield  {journal} {\bibinfo  {journal} {The European Physical Journal H}\
  }\textbf {\bibinfo {volume} {37}},\ \bibinfo {pages} {659} (\bibinfo {year}
  {2012})}\BibitemShut {NoStop}%
\bibitem [{\citenamefont {Emma}\ \emph {et~al.}(2010)\citenamefont {Emma},
  \citenamefont {Akre}, \citenamefont {Arthur}, \citenamefont {Bionta},
  \citenamefont {Bostedt}, \citenamefont {Bozek}, \citenamefont {Brachmann},
  \citenamefont {Bucksbaum}, \citenamefont {Coffee}, \citenamefont {Decker}
  \emph {et~al.}}]{emma2010first}%
  \BibitemOpen
  \bibfield  {author} {\bibinfo {author} {\bibfnamefont {P.}~\bibnamefont
  {Emma}}, \bibinfo {author} {\bibfnamefont {R.}~\bibnamefont {Akre}}, \bibinfo
  {author} {\bibfnamefont {J.}~\bibnamefont {Arthur}}, \bibinfo {author}
  {\bibfnamefont {R.}~\bibnamefont {Bionta}}, \bibinfo {author} {\bibfnamefont
  {C.}~\bibnamefont {Bostedt}}, \bibinfo {author} {\bibfnamefont
  {J.}~\bibnamefont {Bozek}}, \bibinfo {author} {\bibfnamefont
  {A.}~\bibnamefont {Brachmann}}, \bibinfo {author} {\bibfnamefont
  {P.}~\bibnamefont {Bucksbaum}}, \bibinfo {author} {\bibfnamefont
  {R.}~\bibnamefont {Coffee}}, \bibinfo {author} {\bibfnamefont {F.-J.}\
  \bibnamefont {Decker}}, \emph {et~al.},\ }\bibfield  {title} {\bibinfo
  {title} {First lasing and operation of an {\aa}ngstrom-wavelength
  free-electron laser},\ }\href@noop {} {\bibfield  {journal} {\bibinfo
  {journal} {nature photonics}\ }\textbf {\bibinfo {volume} {4}},\ \bibinfo
  {pages} {641} (\bibinfo {year} {2010})}\BibitemShut {NoStop}%
\bibitem [{\citenamefont {Dowell}\ and\ \citenamefont
  {Schmerge}(2009)}]{Dowell2009PhysRevSTAB.12.074201}%
  \BibitemOpen
  \bibfield  {author} {\bibinfo {author} {\bibfnamefont {D.~H.}\ \bibnamefont
  {Dowell}}\ and\ \bibinfo {author} {\bibfnamefont {J.~F.}\ \bibnamefont
  {Schmerge}},\ }\bibfield  {title} {\bibinfo {title} {Quantum efficiency and
  thermal emittance of metal photocathodes},\ }\href
  {https://doi.org/10.1103/PhysRevSTAB.12.074201} {\bibfield  {journal}
  {\bibinfo  {journal} {Phys. Rev. ST Accel. Beams}\ }\textbf {\bibinfo
  {volume} {12}},\ \bibinfo {pages} {074201} (\bibinfo {year}
  {2009})}\BibitemShut {NoStop}%
\bibitem [{\citenamefont {Prat}\ \emph {et~al.}(2015)\citenamefont {Prat},
  \citenamefont {Bettoni}, \citenamefont {Braun}, \citenamefont {Divall},\ and\
  \citenamefont {Schietinger}}]{Eduard2015PhysRevSTAB.18.063401}%
  \BibitemOpen
  \bibfield  {author} {\bibinfo {author} {\bibfnamefont {E.}~\bibnamefont
  {Prat}}, \bibinfo {author} {\bibfnamefont {S.}~\bibnamefont {Bettoni}},
  \bibinfo {author} {\bibfnamefont {H.-H.}\ \bibnamefont {Braun}}, \bibinfo
  {author} {\bibfnamefont {M.~C.}\ \bibnamefont {Divall}},\ and\ \bibinfo
  {author} {\bibfnamefont {T.}~\bibnamefont {Schietinger}},\ }\bibfield
  {title} {\bibinfo {title} {Measurements of intrinsic emittance dependence on
  rf field for copper photocathodes},\ }\href
  {https://doi.org/10.1103/PhysRevSTAB.18.063401} {\bibfield  {journal}
  {\bibinfo  {journal} {Phys. Rev. ST Accel. Beams}\ }\textbf {\bibinfo
  {volume} {18}},\ \bibinfo {pages} {063401} (\bibinfo {year}
  {2015})}\BibitemShut {NoStop}%
\bibitem [{\citenamefont {Zhou}\ \emph {et~al.}(2015)\citenamefont {Zhou},
  \citenamefont {Bohler}, \citenamefont {Ding}, \citenamefont {Gilevich},
  \citenamefont {Huang}, \citenamefont {Loos}, \citenamefont {Ratner},\ and\
  \citenamefont {Vetter}}]{Zhou:2015hsxOptimizationPhotocathode}%
  \BibitemOpen
  \bibfield  {author} {\bibinfo {author} {\bibfnamefont {F.}~\bibnamefont
  {Zhou}}, \bibinfo {author} {\bibfnamefont {D.}~\bibnamefont {Bohler}},
  \bibinfo {author} {\bibfnamefont {Y.}~\bibnamefont {Ding}}, \bibinfo {author}
  {\bibfnamefont {S.}~\bibnamefont {Gilevich}}, \bibinfo {author}
  {\bibfnamefont {Z.}~\bibnamefont {Huang}}, \bibinfo {author} {\bibfnamefont
  {H.}~\bibnamefont {Loos}}, \bibinfo {author} {\bibfnamefont {D.}~\bibnamefont
  {Ratner}},\ and\ \bibinfo {author} {\bibfnamefont {S.}~\bibnamefont
  {Vetter}},\ }\bibfield  {title} {\bibinfo {title} {{Characterizing and
  Optimizing Photocathode Laser Distributions for Ultra-low Emittance Electron
  Beam Operations}},\ }\href@noop {} {\  (\bibinfo {year} {2015})}\BibitemShut
  {NoStop}%
\bibitem [{\citenamefont {Heifets}\ \emph {et~al.}(2002)\citenamefont
  {Heifets}, \citenamefont {Stupakov},\ and\ \citenamefont
  {Krinsky}}]{PhysRevSTAB.5.064401}%
  \BibitemOpen
  \bibfield  {author} {\bibinfo {author} {\bibfnamefont {S.}~\bibnamefont
  {Heifets}}, \bibinfo {author} {\bibfnamefont {G.}~\bibnamefont {Stupakov}},\
  and\ \bibinfo {author} {\bibfnamefont {S.}~\bibnamefont {Krinsky}},\
  }\bibfield  {title} {\bibinfo {title} {Coherent synchrotron radiation
  instability in a bunch compressor},\ }\href
  {https://doi.org/10.1103/PhysRevSTAB.5.064401} {\bibfield  {journal}
  {\bibinfo  {journal} {Phys. Rev. ST Accel. Beams}\ }\textbf {\bibinfo
  {volume} {5}},\ \bibinfo {pages} {064401} (\bibinfo {year}
  {2002})}\BibitemShut {NoStop}%
\bibitem [{\citenamefont {Huang}\ and\ \citenamefont
  {Kim}(2002)}]{HuangBCMI2002}%
  \BibitemOpen
  \bibfield  {author} {\bibinfo {author} {\bibfnamefont {Z.}~\bibnamefont
  {Huang}}\ and\ \bibinfo {author} {\bibfnamefont {K.-J.}\ \bibnamefont
  {Kim}},\ }\bibfield  {title} {\bibinfo {title} {Formulas for coherent
  synchrotron radiation microbunching in a bunch compressor chicane},\ }\href
  {https://doi.org/10.1103/PhysRevSTAB.5.074401} {\bibfield  {journal}
  {\bibinfo  {journal} {Phys. Rev. ST Accel. Beams}\ }\textbf {\bibinfo
  {volume} {5}},\ \bibinfo {pages} {074401} (\bibinfo {year}
  {2002})}\BibitemShut {NoStop}%
\bibitem [{\citenamefont {Di~Mitri}(2013)}]{DiMitriPRAB2013}%
  \BibitemOpen
  \bibfield  {author} {\bibinfo {author} {\bibfnamefont {S.}~\bibnamefont
  {Di~Mitri}},\ }\bibfield  {title} {\bibinfo {title} {Maximum brightness of
  linac-driven electron beams in the presence of collective effects},\ }\href
  {https://doi.org/10.1103/PhysRevSTAB.16.050701} {\bibfield  {journal}
  {\bibinfo  {journal} {Phys. Rev. ST Accel. Beams}\ }\textbf {\bibinfo
  {volume} {16}},\ \bibinfo {pages} {050701} (\bibinfo {year}
  {2013})}\BibitemShut {NoStop}%
\bibitem [{\citenamefont {Di~Mitri}(2015)}]{photonics2020317}%
  \BibitemOpen
  \bibfield  {author} {\bibinfo {author} {\bibfnamefont {S.}~\bibnamefont
  {Di~Mitri}},\ }\bibfield  {title} {\bibinfo {title} {On the importance of
  electron beam brightness in high gain free electron lasers},\ }\href
  {https://doi.org/10.3390/photonics2020317} {\bibfield  {journal} {\bibinfo
  {journal} {Photonics}\ }\textbf {\bibinfo {volume} {2}},\ \bibinfo {pages}
  {317} (\bibinfo {year} {2015})}\BibitemShut {NoStop}%
\bibitem [{\citenamefont {Rosenzweig}\ \emph {et~al.}(2018)\citenamefont
  {Rosenzweig}, \citenamefont {Cahill}, \citenamefont {Carlsten}, \citenamefont
  {Castorina}, \citenamefont {Croia}, \citenamefont {Emma}, \citenamefont
  {Fukusawa}, \citenamefont {Spataro}, \citenamefont {Alesini}, \citenamefont
  {Dolgashev} \emph {et~al.}}]{rosenzweig2018ultra}%
  \BibitemOpen
  \bibfield  {author} {\bibinfo {author} {\bibfnamefont {J.~B.}\ \bibnamefont
  {Rosenzweig}}, \bibinfo {author} {\bibfnamefont {A.}~\bibnamefont {Cahill}},
  \bibinfo {author} {\bibfnamefont {B.}~\bibnamefont {Carlsten}}, \bibinfo
  {author} {\bibfnamefont {G.}~\bibnamefont {Castorina}}, \bibinfo {author}
  {\bibfnamefont {M.}~\bibnamefont {Croia}}, \bibinfo {author} {\bibfnamefont
  {C.}~\bibnamefont {Emma}}, \bibinfo {author} {\bibfnamefont {A.}~\bibnamefont
  {Fukusawa}}, \bibinfo {author} {\bibfnamefont {B.}~\bibnamefont {Spataro}},
  \bibinfo {author} {\bibfnamefont {D.}~\bibnamefont {Alesini}}, \bibinfo
  {author} {\bibfnamefont {V.}~\bibnamefont {Dolgashev}}, \emph {et~al.},\
  }\bibfield  {title} {\bibinfo {title} {Ultra-high brightness electron beams
  from very-high field cryogenic radiofrequency photocathode sources},\
  }\href@noop {} {\bibfield  {journal} {\bibinfo  {journal} {Nuclear
  Instruments and Methods in Physics Research Section A: Accelerators,
  Spectrometers, Detectors and Associated Equipment}\ }\textbf {\bibinfo
  {volume} {909}},\ \bibinfo {pages} {224} (\bibinfo {year}
  {2018})}\BibitemShut {NoStop}%
\bibitem [{\citenamefont {Colby}\ and\ \citenamefont
  {Len}(2016)}]{USroadmapColbydoi:10.1142/S1793626816300012}%
  \BibitemOpen
  \bibfield  {author} {\bibinfo {author} {\bibfnamefont {E.~R.}\ \bibnamefont
  {Colby}}\ and\ \bibinfo {author} {\bibfnamefont {L.~K.}\ \bibnamefont
  {Len}},\ }\bibfield  {title} {\bibinfo {title} {Roadmap to the future},\
  }\href {https://doi.org/10.1142/S1793626816300012} {\bibfield  {journal}
  {\bibinfo  {journal} {Reviews of Accelerator Science and Technology}\
  }\textbf {\bibinfo {volume} {09}},\ \bibinfo {pages} {1} (\bibinfo {year}
  {2016})},\ \Eprint
  {https://arxiv.org/abs/https://doi.org/10.1142/S1793626816300012}
  {https://doi.org/10.1142/S1793626816300012} \BibitemShut {NoStop}%
\bibitem [{\citenamefont {Hidding}\ \emph
  {et~al.}(2019{\natexlab{a}})\citenamefont {Hidding}, \citenamefont {Hooker},
  \citenamefont {Jamison}, \citenamefont {Muratori}, \citenamefont {Murphy},
  \citenamefont {Najmudin}, \citenamefont {Pattathil}, \citenamefont {Sarri},
  \citenamefont {Streeter}, \citenamefont {Welsch} \emph
  {et~al.}}]{hidding2019plasmaPWASC}%
  \BibitemOpen
  \bibfield  {author} {\bibinfo {author} {\bibfnamefont {B.}~\bibnamefont
  {Hidding}}, \bibinfo {author} {\bibfnamefont {S.}~\bibnamefont {Hooker}},
  \bibinfo {author} {\bibfnamefont {S.}~\bibnamefont {Jamison}}, \bibinfo
  {author} {\bibfnamefont {B.}~\bibnamefont {Muratori}}, \bibinfo {author}
  {\bibfnamefont {C.}~\bibnamefont {Murphy}}, \bibinfo {author} {\bibfnamefont
  {Z.}~\bibnamefont {Najmudin}}, \bibinfo {author} {\bibfnamefont
  {R.}~\bibnamefont {Pattathil}}, \bibinfo {author} {\bibfnamefont
  {G.}~\bibnamefont {Sarri}}, \bibinfo {author} {\bibfnamefont
  {M.}~\bibnamefont {Streeter}}, \bibinfo {author} {\bibfnamefont
  {C.}~\bibnamefont {Welsch}}, \emph {et~al.},\ }\bibfield  {title} {\bibinfo
  {title} {Plasma wakefield accelerator research 2019-2040: A community-driven
  uk roadmap compiled by the plasma wakefield accelerator steering committee
  (pwasc)},\ }\href@noop {} {\bibfield  {journal} {\bibinfo  {journal} {arXiv
  preprint arXiv:1904.09205}\ } (\bibinfo {year}
  {2019}{\natexlab{a}})}\BibitemShut {NoStop}%
\bibitem [{\citenamefont {Blumenfeld}\ \emph {et~al.}(2007)\citenamefont
  {Blumenfeld} \emph {et~al.}}]{Blumenfeld2007}%
  \BibitemOpen
  \bibfield  {author} {\bibinfo {author} {\bibfnamefont {I.}~\bibnamefont
  {Blumenfeld}} \emph {et~al.},\ }\bibfield  {title} {\bibinfo {title} {Energy
  doubling of 42 gev electrons in a metre-scale plasma wakefield accelerator},\
  }\href@noop {} {\bibfield  {journal} {\bibinfo  {journal} {Nature}\ }\textbf
  {\bibinfo {volume} {445}},\ \bibinfo {pages} {741} (\bibinfo {year}
  {2007})}\BibitemShut {NoStop}%
\bibitem [{\citenamefont {Litos}\ \emph {et~al.}(2014)\citenamefont {Litos}
  \emph {et~al.}}]{Litos2014Natureshort}%
  \BibitemOpen
  \bibfield  {author} {\bibinfo {author} {\bibfnamefont {M.}~\bibnamefont
  {Litos}} \emph {et~al.},\ }\bibfield  {title} {\bibinfo {title}
  {High-efficiency acceleration of an electron beam in a plasma wakefield
  accelerator},\ }\href@noop {} {\bibfield  {journal} {\bibinfo  {journal}
  {Nature}\ }\textbf {\bibinfo {volume} {515}},\ \bibinfo {pages} {92}
  (\bibinfo {year} {2014})}\BibitemShut {NoStop}%
\bibitem [{\citenamefont {Hidding}\ \emph
  {et~al.}(2012{\natexlab{a}})\citenamefont {Hidding}, \citenamefont
  {Pretzler}, \citenamefont {Rosenzweig}, \citenamefont {K\"onigstein},
  \citenamefont {Schiller},\ and\ \citenamefont
  {Bruhwiler}}]{HiddingPRL2012PhysRevLett.108.035001}%
  \BibitemOpen
  \bibfield  {author} {\bibinfo {author} {\bibfnamefont {B.}~\bibnamefont
  {Hidding}}, \bibinfo {author} {\bibfnamefont {G.}~\bibnamefont {Pretzler}},
  \bibinfo {author} {\bibfnamefont {J.~B.}\ \bibnamefont {Rosenzweig}},
  \bibinfo {author} {\bibfnamefont {T.}~\bibnamefont {K\"onigstein}}, \bibinfo
  {author} {\bibfnamefont {D.}~\bibnamefont {Schiller}},\ and\ \bibinfo
  {author} {\bibfnamefont {D.~L.}\ \bibnamefont {Bruhwiler}},\ }\bibfield
  {title} {\bibinfo {title} {Ultracold electron bunch generation via plasma
  photocathode emission and acceleration in a beam-driven plasma blowout},\
  }\href {https://doi.org/10.1103/PhysRevLett.108.035001} {\bibfield  {journal}
  {\bibinfo  {journal} {Phys. Rev. Lett.}\ }\textbf {\bibinfo {volume} {108}},\
  \bibinfo {pages} {035001} (\bibinfo {year} {2012}{\natexlab{a}})}\BibitemShut
  {NoStop}%
\bibitem [{\citenamefont {Hidding}\ \emph {et~al.}(2011)\citenamefont
  {Hidding}, \citenamefont {Pretzler}, \citenamefont {Bruhwiler},\ and\
  \citenamefont {Rosenzweig}}]{hiddingpatent2011}%
  \BibitemOpen
  \bibfield  {author} {\bibinfo {author} {\bibfnamefont {B.}~\bibnamefont
  {Hidding}}, \bibinfo {author} {\bibfnamefont {G.}~\bibnamefont {Pretzler}},
  \bibinfo {author} {\bibfnamefont {D.}~\bibnamefont {Bruhwiler}},\ and\
  \bibinfo {author} {\bibfnamefont {J.}~\bibnamefont {Rosenzweig}},\
  }\href@noop {} {\bibinfo {title} {Method for generating electron beams in a
  hybrid plasma accelerator}} (\bibinfo {year} {2011}),\ \bibinfo {note}
  {german Patent DE 10 2011 104 858.1, US/PCT patent Ser. No.
  PCT/US12/043002}\BibitemShut {NoStop}%
\bibitem [{\citenamefont {Manahan}\ \emph {et~al.}(2019)\citenamefont
  {Manahan}, \citenamefont {Habib}, \citenamefont {Scherkl}, \citenamefont
  {Ullmann}, \citenamefont {Beaton}, \citenamefont {Sutherland}, \citenamefont
  {Kirwan}, \citenamefont {Delinikolas}, \citenamefont {Heinemann},
  \citenamefont {Altuijri}, \citenamefont {Knetsch}, \citenamefont {Karger},
  \citenamefont {Cook}, \citenamefont {Bruhwiler}, \citenamefont {Sheng},
  \citenamefont {Rosenzweig},\ and\ \citenamefont
  {Hidding}}]{Manahan2019Royal}%
  \BibitemOpen
  \bibfield  {author} {\bibinfo {author} {\bibfnamefont {G.~G.}\ \bibnamefont
  {Manahan}}, \bibinfo {author} {\bibfnamefont {A.~F.}\ \bibnamefont {Habib}},
  \bibinfo {author} {\bibfnamefont {P.}~\bibnamefont {Scherkl}}, \bibinfo
  {author} {\bibfnamefont {D.}~\bibnamefont {Ullmann}}, \bibinfo {author}
  {\bibfnamefont {A.}~\bibnamefont {Beaton}}, \bibinfo {author} {\bibfnamefont
  {A.}~\bibnamefont {Sutherland}}, \bibinfo {author} {\bibfnamefont
  {G.}~\bibnamefont {Kirwan}}, \bibinfo {author} {\bibfnamefont
  {P.}~\bibnamefont {Delinikolas}}, \bibinfo {author} {\bibfnamefont
  {T.}~\bibnamefont {Heinemann}}, \bibinfo {author} {\bibfnamefont
  {R.}~\bibnamefont {Altuijri}}, \bibinfo {author} {\bibfnamefont
  {A.}~\bibnamefont {Knetsch}}, \bibinfo {author} {\bibfnamefont
  {O.}~\bibnamefont {Karger}}, \bibinfo {author} {\bibfnamefont {N.~M.}\
  \bibnamefont {Cook}}, \bibinfo {author} {\bibfnamefont {D.~L.}\ \bibnamefont
  {Bruhwiler}}, \bibinfo {author} {\bibfnamefont {Z.-M.}\ \bibnamefont
  {Sheng}}, \bibinfo {author} {\bibfnamefont {J.~B.}\ \bibnamefont
  {Rosenzweig}},\ and\ \bibinfo {author} {\bibfnamefont {B.}~\bibnamefont
  {Hidding}},\ }\bibfield  {title} {\bibinfo {title} {Advanced schemes for
  underdense plasma photocathode wakefield accelerators: pathways towards
  ultrahigh brightness electron beams},\ }\href
  {https://doi.org/10.1098/rsta.2018.0182} {\bibfield  {journal} {\bibinfo
  {journal} {Philosophical Transactions of the Royal Society A: Mathematical,
  Physical and Engineering Sciences}\ }\textbf {\bibinfo {volume} {377}},\
  \bibinfo {pages} {20180182} (\bibinfo {year} {2019})},\ \Eprint
  {https://arxiv.org/abs/https://royalsocietypublishing.org/doi/pdf/10.1098/rsta.2018.0182}
  {https://royalsocietypublishing.org/doi/pdf/10.1098/rsta.2018.0182}
  \BibitemShut {NoStop}%
\bibitem [{\citenamefont {Hidding}\ \emph
  {et~al.}(2014{\natexlab{a}})\citenamefont {Hidding} \emph
  {et~al.}}]{hiddingarxiv2014}%
  \BibitemOpen
  \bibfield  {author} {\bibinfo {author} {\bibfnamefont {B.}~\bibnamefont
  {Hidding}} \emph {et~al.},\ }\bibfield  {title} {\bibinfo {title} {Tunable
  electron multibunch production in plasma wakefield accelerators},\
  }\href@noop {} {\bibfield  {journal} {\bibinfo  {journal} {arXiv}\ ,\
  \bibinfo {pages} {1403.1109}} (\bibinfo {year}
  {2014}{\natexlab{a}})}\BibitemShut {NoStop}%
\bibitem [{\citenamefont {Manahan}\ \emph {et~al.}(2017)\citenamefont
  {Manahan}, \citenamefont {Habib}, \citenamefont {Scherkl}, \citenamefont
  {Delinikolas}, \citenamefont {Beaton}, \citenamefont {Knetsch}, \citenamefont
  {Karger}, \citenamefont {Wittig}, \citenamefont {Heinemann}, \citenamefont
  {Sheng}, \citenamefont {Cary}, \citenamefont {Bruhwiler}, \citenamefont
  {Rosenzweig},\ and\ \citenamefont {Hidding}}]{ManahanHabib6d}%
  \BibitemOpen
  \bibfield  {author} {\bibinfo {author} {\bibfnamefont {G.}~\bibnamefont
  {Manahan}}, \bibinfo {author} {\bibfnamefont {A.}~\bibnamefont {Habib}},
  \bibinfo {author} {\bibfnamefont {P.}~\bibnamefont {Scherkl}}, \bibinfo
  {author} {\bibfnamefont {P.}~\bibnamefont {Delinikolas}}, \bibinfo {author}
  {\bibfnamefont {A.}~\bibnamefont {Beaton}}, \bibinfo {author} {\bibfnamefont
  {A.}~\bibnamefont {Knetsch}}, \bibinfo {author} {\bibfnamefont
  {O.}~\bibnamefont {Karger}}, \bibinfo {author} {\bibfnamefont
  {G.}~\bibnamefont {Wittig}}, \bibinfo {author} {\bibfnamefont
  {T.}~\bibnamefont {Heinemann}}, \bibinfo {author} {\bibfnamefont
  {Z.}~\bibnamefont {Sheng}}, \bibinfo {author} {\bibfnamefont
  {J.}~\bibnamefont {Cary}}, \bibinfo {author} {\bibfnamefont {D.}~\bibnamefont
  {Bruhwiler}}, \bibinfo {author} {\bibfnamefont {J.}~\bibnamefont
  {Rosenzweig}},\ and\ \bibinfo {author} {\bibfnamefont {B.}~\bibnamefont
  {Hidding}},\ }\bibfield  {title} {\bibinfo {title} {Single-stage plasma-based
  correlated energy spread compensation for ultrahigh 6d brightness electron
  beams},\ }\href {https://strathprints.strath.ac.uk/60819/} {\bibfield
  {journal} {\bibinfo  {journal} {Nature Communications}\ }\textbf {\bibinfo
  {volume} {8}} (\bibinfo {year} {2017})}\BibitemShut {NoStop}%
\bibitem [{\citenamefont {Deng}\ \emph {et~al.}(2019)\citenamefont {Deng},
  \citenamefont {Karger}, \citenamefont {Heinemann}, \citenamefont {Knetsch},
  \citenamefont {Scherkl}, \citenamefont {Manahan}, \citenamefont {Beaton},
  \citenamefont {Ullmann}, \citenamefont {Wittig}, \citenamefont {Habib} \emph
  {et~al.}}]{Deng2019Trojan}%
  \BibitemOpen
  \bibfield  {author} {\bibinfo {author} {\bibfnamefont {A.}~\bibnamefont
  {Deng}}, \bibinfo {author} {\bibfnamefont {O.}~\bibnamefont {Karger}},
  \bibinfo {author} {\bibfnamefont {T.}~\bibnamefont {Heinemann}}, \bibinfo
  {author} {\bibfnamefont {A.}~\bibnamefont {Knetsch}}, \bibinfo {author}
  {\bibfnamefont {P.}~\bibnamefont {Scherkl}}, \bibinfo {author} {\bibfnamefont
  {G.}~\bibnamefont {Manahan}}, \bibinfo {author} {\bibfnamefont
  {A.}~\bibnamefont {Beaton}}, \bibinfo {author} {\bibfnamefont
  {D.}~\bibnamefont {Ullmann}}, \bibinfo {author} {\bibfnamefont
  {G.}~\bibnamefont {Wittig}}, \bibinfo {author} {\bibfnamefont
  {A.}~\bibnamefont {Habib}}, \emph {et~al.},\ }\bibfield  {title} {\bibinfo
  {title} {Generation and acceleration of electron bunches from a plasma
  photocathode},\ }\href@noop {} {\bibfield  {journal} {\bibinfo  {journal}
  {Nature Physics}\ ,\ \bibinfo {pages} {1}} (\bibinfo {year}
  {2019})}\BibitemShut {NoStop}%
\bibitem [{\citenamefont {Scherkl}\ \emph {et~al.}(2019)\citenamefont
  {Scherkl}, \citenamefont {Knetsch}, \citenamefont {Heinemann}, \citenamefont
  {Sutherland}, \citenamefont {Habib}, \citenamefont {Karger}, \citenamefont
  {Ullmann}, \citenamefont {Beaton}, \citenamefont {Kirwan}, \citenamefont
  {Manahan}, \citenamefont {Xi}, \citenamefont {Deng}, \citenamefont {Litos},
  \citenamefont {OShea}, \citenamefont {Green}, \citenamefont {Clarke},
  \citenamefont {Andonian}, \citenamefont {Assmann}, \citenamefont
  {Jaroszynski}, \citenamefont {Bruhwiler}, \citenamefont {Smith},
  \citenamefont {Cary}, \citenamefont {Hogan}, \citenamefont {Yakimenko},
  \citenamefont {Rosenzweig},\ and\ \citenamefont
  {Hidding}}]{Scherklarxiv2019}%
  \BibitemOpen
  \bibfield  {author} {\bibinfo {author} {\bibfnamefont {P.}~\bibnamefont
  {Scherkl}}, \bibinfo {author} {\bibfnamefont {A.}~\bibnamefont {Knetsch}},
  \bibinfo {author} {\bibfnamefont {T.}~\bibnamefont {Heinemann}}, \bibinfo
  {author} {\bibfnamefont {A.}~\bibnamefont {Sutherland}}, \bibinfo {author}
  {\bibfnamefont {A.~F.}\ \bibnamefont {Habib}}, \bibinfo {author}
  {\bibfnamefont {O.}~\bibnamefont {Karger}}, \bibinfo {author} {\bibfnamefont
  {D.}~\bibnamefont {Ullmann}}, \bibinfo {author} {\bibfnamefont
  {A.}~\bibnamefont {Beaton}}, \bibinfo {author} {\bibfnamefont
  {G.}~\bibnamefont {Kirwan}}, \bibinfo {author} {\bibfnamefont
  {G.}~\bibnamefont {Manahan}}, \bibinfo {author} {\bibfnamefont
  {Y.}~\bibnamefont {Xi}}, \bibinfo {author} {\bibfnamefont {A.}~\bibnamefont
  {Deng}}, \bibinfo {author} {\bibfnamefont {M.~D.}\ \bibnamefont {Litos}},
  \bibinfo {author} {\bibfnamefont {B.~D.}\ \bibnamefont {OShea}}, \bibinfo
  {author} {\bibfnamefont {S.~Z.}\ \bibnamefont {Green}}, \bibinfo {author}
  {\bibfnamefont {C.~I.}\ \bibnamefont {Clarke}}, \bibinfo {author}
  {\bibfnamefont {G.}~\bibnamefont {Andonian}}, \bibinfo {author}
  {\bibfnamefont {R.}~\bibnamefont {Assmann}}, \bibinfo {author} {\bibfnamefont
  {D.~A.}\ \bibnamefont {Jaroszynski}}, \bibinfo {author} {\bibfnamefont
  {D.~L.}\ \bibnamefont {Bruhwiler}}, \bibinfo {author} {\bibfnamefont
  {J.}~\bibnamefont {Smith}}, \bibinfo {author} {\bibfnamefont {J.~R.}\
  \bibnamefont {Cary}}, \bibinfo {author} {\bibfnamefont {M.~J.}\ \bibnamefont
  {Hogan}}, \bibinfo {author} {\bibfnamefont {V.}~\bibnamefont {Yakimenko}},
  \bibinfo {author} {\bibfnamefont {J.~B.}\ \bibnamefont {Rosenzweig}},\ and\
  \bibinfo {author} {\bibfnamefont {B.}~\bibnamefont {Hidding}},\ }\bibfield
  {title} {\bibinfo {title} {{Plasma-photonic spatiotemporal synchronization of
  relativistic electron and laser beams}},\ }\href
  {http://arxiv.org/abs/1908.09263} {\  (\bibinfo {year} {2019})},\ \Eprint
  {https://arxiv.org/abs/1908.09263} {arXiv:1908.09263} \BibitemShut {NoStop}%
\bibitem [{\citenamefont {Wittig}\ \emph {et~al.}(2015)\citenamefont {Wittig},
  \citenamefont {Karger}, \citenamefont {Knetsch}, \citenamefont {Xi},
  \citenamefont {Deng}, \citenamefont {Rosenzweig}, \citenamefont {Bruhwiler},
  \citenamefont {Smith}, \citenamefont {Manahan}, \citenamefont {Sheng},
  \citenamefont {Jaroszynski},\ and\ \citenamefont
  {Hidding}}]{WittigPlasmaTorchPhysRevSTAB.18.081304}%
  \BibitemOpen
  \bibfield  {author} {\bibinfo {author} {\bibfnamefont {G.}~\bibnamefont
  {Wittig}}, \bibinfo {author} {\bibfnamefont {O.}~\bibnamefont {Karger}},
  \bibinfo {author} {\bibfnamefont {A.}~\bibnamefont {Knetsch}}, \bibinfo
  {author} {\bibfnamefont {Y.}~\bibnamefont {Xi}}, \bibinfo {author}
  {\bibfnamefont {A.}~\bibnamefont {Deng}}, \bibinfo {author} {\bibfnamefont
  {J.~B.}\ \bibnamefont {Rosenzweig}}, \bibinfo {author} {\bibfnamefont
  {D.~L.}\ \bibnamefont {Bruhwiler}}, \bibinfo {author} {\bibfnamefont
  {J.}~\bibnamefont {Smith}}, \bibinfo {author} {\bibfnamefont {G.~G.}\
  \bibnamefont {Manahan}}, \bibinfo {author} {\bibfnamefont {Z.-M.}\
  \bibnamefont {Sheng}}, \bibinfo {author} {\bibfnamefont {D.~A.}\ \bibnamefont
  {Jaroszynski}},\ and\ \bibinfo {author} {\bibfnamefont {B.}~\bibnamefont
  {Hidding}},\ }\bibfield  {title} {\bibinfo {title} {Optical plasma torch
  electron bunch generation in plasma wakefield accelerators},\ }\href
  {https://doi.org/10.1103/PhysRevSTAB.18.081304} {\bibfield  {journal}
  {\bibinfo  {journal} {Phys. Rev. ST Accel. Beams}\ }\textbf {\bibinfo
  {volume} {18}},\ \bibinfo {pages} {081304} (\bibinfo {year}
  {2015})}\BibitemShut {NoStop}%
\bibitem [{\citenamefont {Ullmann}\ \emph {et~al.}(2020)\citenamefont
  {Ullmann}, \citenamefont {Scherkl}, \citenamefont {Knetsch}, \citenamefont
  {Heinemann}, \citenamefont {Sutherland}, \citenamefont {Habib}, \citenamefont
  {Karger}, \citenamefont {Beaton}, \citenamefont {Manahan}, \citenamefont
  {Deng}, \citenamefont {Andonian}, \citenamefont {Litos}, \citenamefont
  {OShea}, \citenamefont {Bruhwiler}, \citenamefont {Cary}, \citenamefont
  {Hogan}, \citenamefont {Rosenzweig},\ and\ \citenamefont
  {Hidding}}]{Ullmann2020Torch}%
  \BibitemOpen
  \bibfield  {author} {\bibinfo {author} {\bibfnamefont {D.}~\bibnamefont
  {Ullmann}}, \bibinfo {author} {\bibfnamefont {P.}~\bibnamefont {Scherkl}},
  \bibinfo {author} {\bibfnamefont {A.}~\bibnamefont {Knetsch}}, \bibinfo
  {author} {\bibfnamefont {T.}~\bibnamefont {Heinemann}}, \bibinfo {author}
  {\bibfnamefont {A.}~\bibnamefont {Sutherland}}, \bibinfo {author}
  {\bibfnamefont {A.~F.}\ \bibnamefont {Habib}}, \bibinfo {author}
  {\bibfnamefont {O.~S.}\ \bibnamefont {Karger}}, \bibinfo {author}
  {\bibfnamefont {A.}~\bibnamefont {Beaton}}, \bibinfo {author} {\bibfnamefont
  {G.~G.}\ \bibnamefont {Manahan}}, \bibinfo {author} {\bibfnamefont
  {A.}~\bibnamefont {Deng}}, \bibinfo {author} {\bibfnamefont {G.}~\bibnamefont
  {Andonian}}, \bibinfo {author} {\bibfnamefont {M.~D.}\ \bibnamefont {Litos}},
  \bibinfo {author} {\bibfnamefont {B.~D.}\ \bibnamefont {OShea}}, \bibinfo
  {author} {\bibfnamefont {D.~L.}\ \bibnamefont {Bruhwiler}}, \bibinfo {author}
  {\bibfnamefont {J.~R.}\ \bibnamefont {Cary}}, \bibinfo {author}
  {\bibfnamefont {V.}~\bibnamefont {Hogan}, \bibfnamefont {M.~J.~Yakimenko}},
  \bibinfo {author} {\bibfnamefont {J.~B.}\ \bibnamefont {Rosenzweig}},\ and\
  \bibinfo {author} {\bibfnamefont {B.}~\bibnamefont {Hidding}},\ }\bibfield
  {title} {\bibinfo {title} {All-optical density downramp injection in
  electron-driven plasma wakefield accelerators},\ }\href@noop {} {\bibfield
  {journal} {\bibinfo  {journal} {arXiv preprint arXiv:2007.12634}\ } (\bibinfo
  {year} {2020})}\BibitemShut {NoStop}%
\bibitem [{SBI()}]{SBIRbrightnessTrafo}%
  \BibitemOpen
  \href {https://www.sbir.gov/sbirsearch/detail/407701} {\bibinfo {title}
  {Radiabeam: Plasma photocathode beam brightness transformer for
  laser-plasma-wakefield accelerators, {DOE DESC0009533},
  2013-2016}}\BibitemShut {NoStop}%
\bibitem [{\citenamefont {Shalloo}\ \emph {et~al.}(2018)\citenamefont
  {Shalloo}, \citenamefont {Arran}, \citenamefont {Corner}, \citenamefont
  {Holloway}, \citenamefont {Jonnerby}, \citenamefont {Walczak}, \citenamefont
  {Milchberg},\ and\ \citenamefont {Hooker}}]{hooker2018hofi}%
  \BibitemOpen
  \bibfield  {author} {\bibinfo {author} {\bibfnamefont {R.~J.}\ \bibnamefont
  {Shalloo}}, \bibinfo {author} {\bibfnamefont {C.}~\bibnamefont {Arran}},
  \bibinfo {author} {\bibfnamefont {L.}~\bibnamefont {Corner}}, \bibinfo
  {author} {\bibfnamefont {J.}~\bibnamefont {Holloway}}, \bibinfo {author}
  {\bibfnamefont {J.}~\bibnamefont {Jonnerby}}, \bibinfo {author}
  {\bibfnamefont {R.}~\bibnamefont {Walczak}}, \bibinfo {author} {\bibfnamefont
  {H.~M.}\ \bibnamefont {Milchberg}},\ and\ \bibinfo {author} {\bibfnamefont
  {S.~M.}\ \bibnamefont {Hooker}},\ }\bibfield  {title} {\bibinfo {title}
  {Hydrodynamic optical-field-ionized plasma channels},\ }\href
  {https://doi.org/10.1103/PhysRevE.97.053203} {\bibfield  {journal} {\bibinfo
  {journal} {Phys. Rev. E}\ }\textbf {\bibinfo {volume} {97}},\ \bibinfo
  {pages} {053203} (\bibinfo {year} {2018})}\BibitemShut {NoStop}%
\bibitem [{\citenamefont {Gilljohann}\ \emph {et~al.}(2019)\citenamefont
  {Gilljohann}, \citenamefont {Ding}, \citenamefont {D\"opp}, \citenamefont
  {G\"otzfried}, \citenamefont {Schindler}, \citenamefont {Schilling},
  \citenamefont {Corde}, \citenamefont {Debus}, \citenamefont {Heinemann},
  \citenamefont {Hidding}, \citenamefont {Hooker}, \citenamefont {Irman},
  \citenamefont {Kononenko}, \citenamefont {Kurz}, \citenamefont {Martinez
  de~la Ossa}, \citenamefont {Schramm},\ and\ \citenamefont
  {Karsch}}]{GilljohannPRXPhysRevX.9.011046}%
  \BibitemOpen
  \bibfield  {author} {\bibinfo {author} {\bibfnamefont {M.~F.}\ \bibnamefont
  {Gilljohann}}, \bibinfo {author} {\bibfnamefont {H.}~\bibnamefont {Ding}},
  \bibinfo {author} {\bibfnamefont {A.}~\bibnamefont {D\"opp}}, \bibinfo
  {author} {\bibfnamefont {J.}~\bibnamefont {G\"otzfried}}, \bibinfo {author}
  {\bibfnamefont {S.}~\bibnamefont {Schindler}}, \bibinfo {author}
  {\bibfnamefont {G.}~\bibnamefont {Schilling}}, \bibinfo {author}
  {\bibfnamefont {S.}~\bibnamefont {Corde}}, \bibinfo {author} {\bibfnamefont
  {A.}~\bibnamefont {Debus}}, \bibinfo {author} {\bibfnamefont
  {T.}~\bibnamefont {Heinemann}}, \bibinfo {author} {\bibfnamefont
  {B.}~\bibnamefont {Hidding}}, \bibinfo {author} {\bibfnamefont {S.~M.}\
  \bibnamefont {Hooker}}, \bibinfo {author} {\bibfnamefont {A.}~\bibnamefont
  {Irman}}, \bibinfo {author} {\bibfnamefont {O.}~\bibnamefont {Kononenko}},
  \bibinfo {author} {\bibfnamefont {T.}~\bibnamefont {Kurz}}, \bibinfo {author}
  {\bibfnamefont {A.}~\bibnamefont {Martinez de~la Ossa}}, \bibinfo {author}
  {\bibfnamefont {U.}~\bibnamefont {Schramm}},\ and\ \bibinfo {author}
  {\bibfnamefont {S.}~\bibnamefont {Karsch}},\ }\bibfield  {title} {\bibinfo
  {title} {Direct observation of plasma waves and dynamics induced by
  laser-accelerated electron beams},\ }\href
  {https://doi.org/10.1103/PhysRevX.9.011046} {\bibfield  {journal} {\bibinfo
  {journal} {Phys. Rev. X}\ }\textbf {\bibinfo {volume} {9}},\ \bibinfo {pages}
  {011046} (\bibinfo {year} {2019})}\BibitemShut {NoStop}%
\bibitem [{\citenamefont {Zgadzaj}\ \emph {et~al.}(2020)\citenamefont
  {Zgadzaj}, \citenamefont {Li}, \citenamefont {Downer}, \citenamefont
  {Sosedkin}, \citenamefont {Khudyakov}, \citenamefont {Lotov}, \citenamefont
  {Silva}, \citenamefont {Vieira}, \citenamefont {Allen}, \citenamefont
  {Gessner} \emph {et~al.}}]{zgadzaj2020dissipation}%
  \BibitemOpen
  \bibfield  {author} {\bibinfo {author} {\bibfnamefont {R.}~\bibnamefont
  {Zgadzaj}}, \bibinfo {author} {\bibfnamefont {Z.}~\bibnamefont {Li}},
  \bibinfo {author} {\bibfnamefont {M.}~\bibnamefont {Downer}}, \bibinfo
  {author} {\bibfnamefont {A.}~\bibnamefont {Sosedkin}}, \bibinfo {author}
  {\bibfnamefont {V.}~\bibnamefont {Khudyakov}}, \bibinfo {author}
  {\bibfnamefont {K.}~\bibnamefont {Lotov}}, \bibinfo {author} {\bibfnamefont
  {T.}~\bibnamefont {Silva}}, \bibinfo {author} {\bibfnamefont
  {J.}~\bibnamefont {Vieira}}, \bibinfo {author} {\bibfnamefont
  {J.}~\bibnamefont {Allen}}, \bibinfo {author} {\bibfnamefont
  {S.}~\bibnamefont {Gessner}}, \emph {et~al.},\ }\bibfield  {title} {\bibinfo
  {title} {Dissipation of electron-beam-driven plasma wakes},\ }\href@noop {}
  {\bibfield  {journal} {\bibinfo  {journal} {arXiv preprint arXiv:2001.09401}\
  } (\bibinfo {year} {2020})}\BibitemShut {NoStop}%
\bibitem [{\citenamefont {Nieter}\ and\ \citenamefont
  {Cary}(2004)}]{Nieter2004448}%
  \BibitemOpen
  \bibfield  {author} {\bibinfo {author} {\bibfnamefont {C.}~\bibnamefont
  {Nieter}}\ and\ \bibinfo {author} {\bibfnamefont {J.~R.}\ \bibnamefont
  {Cary}},\ }\bibfield  {title} {\bibinfo {title} {Vorpal: a versatile plasma
  simulation code},\ }\href {https://doi.org/DOI: 10.1016/j.jcp.2003.11.004}
  {\bibfield  {journal} {\bibinfo  {journal} {Journal of Computational
  Physics}\ }\textbf {\bibinfo {volume} {196}},\ \bibinfo {pages} {448 }
  (\bibinfo {year} {2004})}\BibitemShut {NoStop}%
\bibitem [{\citenamefont {Barov}\ \emph {et~al.}(2004)\citenamefont {Barov},
  \citenamefont {Rosenzweig}, \citenamefont {Thompson},\ and\ \citenamefont
  {Yoder}}]{BarovEnergyLossQtildePRST2004}%
  \BibitemOpen
  \bibfield  {author} {\bibinfo {author} {\bibfnamefont {N.}~\bibnamefont
  {Barov}}, \bibinfo {author} {\bibfnamefont {J.~B.}\ \bibnamefont
  {Rosenzweig}}, \bibinfo {author} {\bibfnamefont {M.~C.}\ \bibnamefont
  {Thompson}},\ and\ \bibinfo {author} {\bibfnamefont {R.~B.}\ \bibnamefont
  {Yoder}},\ }\bibfield  {title} {\bibinfo {title} {Energy loss of a
  high-charge bunched electron beam in plasma: Analysis},\ }\href
  {https://doi.org/10.1103/PhysRevSTAB.7.061301} {\bibfield  {journal}
  {\bibinfo  {journal} {Phys. Rev. ST Accel. Beams}\ }\textbf {\bibinfo
  {volume} {7}},\ \bibinfo {pages} {061301} (\bibinfo {year}
  {2004})}\BibitemShut {NoStop}%
\bibitem [{\citenamefont {Manahan}\ \emph {et~al.}(2016)\citenamefont
  {Manahan}, \citenamefont {Deng}, \citenamefont {Karger}, \citenamefont {Xi},
  \citenamefont {Knetsch}, \citenamefont {Litos}, \citenamefont {Wittig},
  \citenamefont {Heinemann}, \citenamefont {Smith}, \citenamefont {Sheng},
  \citenamefont {Jaroszynski}, \citenamefont {Andonian}, \citenamefont
  {Bruhwiler}, \citenamefont {Rosenzweig},\ and\ \citenamefont
  {Hidding}}]{ManahanDarkCurrentPhysRevAccelBeams.19.011303}%
  \BibitemOpen
  \bibfield  {author} {\bibinfo {author} {\bibfnamefont {G.~G.}\ \bibnamefont
  {Manahan}}, \bibinfo {author} {\bibfnamefont {A.}~\bibnamefont {Deng}},
  \bibinfo {author} {\bibfnamefont {O.}~\bibnamefont {Karger}}, \bibinfo
  {author} {\bibfnamefont {Y.}~\bibnamefont {Xi}}, \bibinfo {author}
  {\bibfnamefont {A.}~\bibnamefont {Knetsch}}, \bibinfo {author} {\bibfnamefont
  {M.}~\bibnamefont {Litos}}, \bibinfo {author} {\bibfnamefont
  {G.}~\bibnamefont {Wittig}}, \bibinfo {author} {\bibfnamefont
  {T.}~\bibnamefont {Heinemann}}, \bibinfo {author} {\bibfnamefont
  {J.}~\bibnamefont {Smith}}, \bibinfo {author} {\bibfnamefont {Z.~M.}\
  \bibnamefont {Sheng}}, \bibinfo {author} {\bibfnamefont {D.~A.}\ \bibnamefont
  {Jaroszynski}}, \bibinfo {author} {\bibfnamefont {G.}~\bibnamefont
  {Andonian}}, \bibinfo {author} {\bibfnamefont {D.~L.}\ \bibnamefont
  {Bruhwiler}}, \bibinfo {author} {\bibfnamefont {J.~B.}\ \bibnamefont
  {Rosenzweig}},\ and\ \bibinfo {author} {\bibfnamefont {B.}~\bibnamefont
  {Hidding}},\ }\bibfield  {title} {\bibinfo {title} {Hot spots and dark
  current in advanced plasma wakefield accelerators},\ }\href
  {https://doi.org/10.1103/PhysRevAccelBeams.19.011303} {\bibfield  {journal}
  {\bibinfo  {journal} {Phys. Rev. Accel. Beams}\ }\textbf {\bibinfo {volume}
  {19}},\ \bibinfo {pages} {011303} (\bibinfo {year} {2016})}\BibitemShut
  {NoStop}%
\bibitem [{\citenamefont {Litos}\ \emph {et~al.}(2018)\citenamefont {Litos},
  \citenamefont {Ariniello}, \citenamefont {Doss}, \citenamefont {Hunt-Stone},\
  and\ \citenamefont {Cary}}]{litos2018experimental}%
  \BibitemOpen
  \bibfield  {author} {\bibinfo {author} {\bibfnamefont {M.}~\bibnamefont
  {Litos}}, \bibinfo {author} {\bibfnamefont {R.}~\bibnamefont {Ariniello}},
  \bibinfo {author} {\bibfnamefont {C.}~\bibnamefont {Doss}}, \bibinfo {author}
  {\bibfnamefont {K.}~\bibnamefont {Hunt-Stone}},\ and\ \bibinfo {author}
  {\bibfnamefont {J.~R.}\ \bibnamefont {Cary}},\ }\bibfield  {title} {\bibinfo
  {title} {Experimental opportunities for the ion channel laser},\ }in\
  \href@noop {} {\emph {\bibinfo {booktitle} {2018 IEEE Advanced Accelerator
  Concepts Workshop (AAC)}}}\ (\bibinfo {organization} {IEEE},\ \bibinfo {year}
  {2018})\ pp.\ \bibinfo {pages} {1--5}\BibitemShut {NoStop}%
\bibitem [{\citenamefont {Habib}\ \emph {et~al.}(2019)\citenamefont {Habib},
  \citenamefont {Scherkl}, \citenamefont {Manahan}, \citenamefont {Heinemann},
  \citenamefont {Ullmann}, \citenamefont {Sutherland}, \citenamefont {Knetsch},
  \citenamefont {Litos}, \citenamefont {Hogan}, \citenamefont {Rosenzweig}
  \emph {et~al.}}]{habib2019plasma}%
  \BibitemOpen
  \bibfield  {author} {\bibinfo {author} {\bibfnamefont {A.~F.}\ \bibnamefont
  {Habib}}, \bibinfo {author} {\bibfnamefont {P.}~\bibnamefont {Scherkl}},
  \bibinfo {author} {\bibfnamefont {G.~G.}\ \bibnamefont {Manahan}}, \bibinfo
  {author} {\bibfnamefont {T.}~\bibnamefont {Heinemann}}, \bibinfo {author}
  {\bibfnamefont {D.}~\bibnamefont {Ullmann}}, \bibinfo {author} {\bibfnamefont
  {A.}~\bibnamefont {Sutherland}}, \bibinfo {author} {\bibfnamefont
  {A.}~\bibnamefont {Knetsch}}, \bibinfo {author} {\bibfnamefont
  {M.}~\bibnamefont {Litos}}, \bibinfo {author} {\bibfnamefont
  {M.}~\bibnamefont {Hogan}}, \bibinfo {author} {\bibfnamefont
  {J.}~\bibnamefont {Rosenzweig}}, \emph {et~al.},\ }\bibfield  {title}
  {\bibinfo {title} {Plasma accelerator-based ultrabright x-ray beams from
  ultrabright electron beams},\ }in\ \href@noop {} {\emph {\bibinfo {booktitle}
  {Advances in Laboratory-based X-Ray Sources, Optics, and Applications
  VII}}},\ Vol.\ \bibinfo {volume} {11110}\ (\bibinfo {organization}
  {International Society for Optics and Photonics},\ \bibinfo {year} {2019})\
  p.\ \bibinfo {pages} {111100A}\BibitemShut {NoStop}%
\bibitem [{\citenamefont {Hidding}\ \emph
  {et~al.}(2012{\natexlab{b}})\citenamefont {Hidding}, \citenamefont
  {Rosenzweig}, \citenamefont {Xi}, \citenamefont {O'Shea}, \citenamefont
  {Andonian}, \citenamefont {Schiller}, \citenamefont {Barber}, \citenamefont
  {Williams}, \citenamefont {Pretzler}, \citenamefont {K\"{o}nigstein},
  \citenamefont {Kleeschulte}, \citenamefont {Hogan}, \citenamefont {Litos},
  \citenamefont {Corde}, \citenamefont {White}, \citenamefont {Muggli},
  \citenamefont {Bruhwiler},\ and\ \citenamefont {Lotov}}]{hidding:570beyond}%
  \BibitemOpen
  \bibfield  {author} {\bibinfo {author} {\bibfnamefont {B.}~\bibnamefont
  {Hidding}}, \bibinfo {author} {\bibfnamefont {J.~B.}\ \bibnamefont
  {Rosenzweig}}, \bibinfo {author} {\bibfnamefont {Y.}~\bibnamefont {Xi}},
  \bibinfo {author} {\bibfnamefont {B.}~\bibnamefont {O'Shea}}, \bibinfo
  {author} {\bibfnamefont {G.}~\bibnamefont {Andonian}}, \bibinfo {author}
  {\bibfnamefont {D.}~\bibnamefont {Schiller}}, \bibinfo {author}
  {\bibfnamefont {S.}~\bibnamefont {Barber}}, \bibinfo {author} {\bibfnamefont
  {O.}~\bibnamefont {Williams}}, \bibinfo {author} {\bibfnamefont
  {G.}~\bibnamefont {Pretzler}}, \bibinfo {author} {\bibfnamefont
  {T.}~\bibnamefont {K\"{o}nigstein}}, \bibinfo {author} {\bibfnamefont
  {F.}~\bibnamefont {Kleeschulte}}, \bibinfo {author} {\bibfnamefont {M.~J.}\
  \bibnamefont {Hogan}}, \bibinfo {author} {\bibfnamefont {M.}~\bibnamefont
  {Litos}}, \bibinfo {author} {\bibfnamefont {S.}~\bibnamefont {Corde}},
  \bibinfo {author} {\bibfnamefont {W.~W.}\ \bibnamefont {White}}, \bibinfo
  {author} {\bibfnamefont {P.}~\bibnamefont {Muggli}}, \bibinfo {author}
  {\bibfnamefont {D.~L.}\ \bibnamefont {Bruhwiler}},\ and\ \bibinfo {author}
  {\bibfnamefont {K.}~\bibnamefont {Lotov}},\ }\bibfield  {title} {\bibinfo
  {title} {Beyond injection: Trojan horse underdense photocathode plasma
  wakefield acceleration},\ }\href {https://doi.org/10.1063/1.4773760}
  {\bibfield  {journal} {\bibinfo  {journal} {AIP Conference Proceedings}\
  }\textbf {\bibinfo {volume} {1507}},\ \bibinfo {pages} {570} (\bibinfo {year}
  {2012}{\natexlab{b}})}\BibitemShut {NoStop}%
\bibitem [{\citenamefont {Li}\ \emph {et~al.}(2013)\citenamefont {Li},
  \citenamefont {Hua}, \citenamefont {Xu}, \citenamefont {Zhang}, \citenamefont
  {Yan}, \citenamefont {Du}, \citenamefont {Huang}, \citenamefont {Chen},
  \citenamefont {Tang}, \citenamefont {Lu}, \citenamefont {Joshi},
  \citenamefont {Mori},\ and\ \citenamefont
  {Gu}}]{LiPhysRevLett.111.015003PRL2013}%
  \BibitemOpen
  \bibfield  {author} {\bibinfo {author} {\bibfnamefont {F.}~\bibnamefont
  {Li}}, \bibinfo {author} {\bibfnamefont {J.~F.}\ \bibnamefont {Hua}},
  \bibinfo {author} {\bibfnamefont {X.~L.}\ \bibnamefont {Xu}}, \bibinfo
  {author} {\bibfnamefont {C.~J.}\ \bibnamefont {Zhang}}, \bibinfo {author}
  {\bibfnamefont {L.~X.}\ \bibnamefont {Yan}}, \bibinfo {author} {\bibfnamefont
  {Y.~C.}\ \bibnamefont {Du}}, \bibinfo {author} {\bibfnamefont {W.~H.}\
  \bibnamefont {Huang}}, \bibinfo {author} {\bibfnamefont {H.~B.}\ \bibnamefont
  {Chen}}, \bibinfo {author} {\bibfnamefont {C.~X.}\ \bibnamefont {Tang}},
  \bibinfo {author} {\bibfnamefont {W.}~\bibnamefont {Lu}}, \bibinfo {author}
  {\bibfnamefont {C.}~\bibnamefont {Joshi}}, \bibinfo {author} {\bibfnamefont
  {W.~B.}\ \bibnamefont {Mori}},\ and\ \bibinfo {author} {\bibfnamefont
  {Y.~Q.}\ \bibnamefont {Gu}},\ }\bibfield  {title} {\bibinfo {title}
  {Generating high-brightness electron beams via ionization injection by
  transverse colliding lasers in a plasma-wakefield accelerator},\ }\href
  {https://doi.org/10.1103/PhysRevLett.111.015003} {\bibfield  {journal}
  {\bibinfo  {journal} {Phys. Rev. Lett.}\ }\textbf {\bibinfo {volume} {111}},\
  \bibinfo {pages} {015003} (\bibinfo {year} {2013})}\BibitemShut {NoStop}%
\bibitem [{\citenamefont {Xi}\ \emph {et~al.}(2013)\citenamefont {Xi},
  \citenamefont {Hidding}, \citenamefont {Bruhwiler}, \citenamefont
  {Pretzler},\ and\ \citenamefont {Rosenzweig}}]{YunfengPhysRevSTAB.16.031303}%
  \BibitemOpen
  \bibfield  {author} {\bibinfo {author} {\bibfnamefont {Y.}~\bibnamefont
  {Xi}}, \bibinfo {author} {\bibfnamefont {B.}~\bibnamefont {Hidding}},
  \bibinfo {author} {\bibfnamefont {D.}~\bibnamefont {Bruhwiler}}, \bibinfo
  {author} {\bibfnamefont {G.}~\bibnamefont {Pretzler}},\ and\ \bibinfo
  {author} {\bibfnamefont {J.~B.}\ \bibnamefont {Rosenzweig}},\ }\bibfield
  {title} {\bibinfo {title} {Hybrid modeling of relativistic underdense plasma
  photocathode injectors},\ }\href
  {https://doi.org/10.1103/PhysRevSTAB.16.031303} {\bibfield  {journal}
  {\bibinfo  {journal} {Phys. Rev. ST Accel. Beams}\ }\textbf {\bibinfo
  {volume} {16}},\ \bibinfo {pages} {031303} (\bibinfo {year}
  {2013})}\BibitemShut {NoStop}%
\bibitem [{\citenamefont {Bourgeois}\ \emph {et~al.}(2013)\citenamefont
  {Bourgeois}, \citenamefont {Cowley},\ and\ \citenamefont
  {Hooker}}]{BourgeoisHookerPhysRevLett.111.155004}%
  \BibitemOpen
  \bibfield  {author} {\bibinfo {author} {\bibfnamefont {N.}~\bibnamefont
  {Bourgeois}}, \bibinfo {author} {\bibfnamefont {J.}~\bibnamefont {Cowley}},\
  and\ \bibinfo {author} {\bibfnamefont {S.~M.}\ \bibnamefont {Hooker}},\
  }\bibfield  {title} {\bibinfo {title} {Two-pulse ionization injection into
  quasilinear laser wakefields},\ }\href
  {https://doi.org/10.1103/PhysRevLett.111.155004} {\bibfield  {journal}
  {\bibinfo  {journal} {Phys. Rev. Lett.}\ }\textbf {\bibinfo {volume} {111}},\
  \bibinfo {pages} {155004} (\bibinfo {year} {2013})}\BibitemShut {NoStop}%
\bibitem [{\citenamefont {Hidding}\ \emph
  {et~al.}(2014{\natexlab{b}})\citenamefont {Hidding}, \citenamefont {Manahan},
  \citenamefont {Karger}, \citenamefont {Knetsch}, \citenamefont {Wittig},
  \citenamefont {Jaroszynski}, \citenamefont {Sheng}, \citenamefont {Xi},
  \citenamefont {Deng}, \citenamefont {Rosenzweig}, \citenamefont {Andonian},
  \citenamefont {Murokh}, \citenamefont {Pretzler}, \citenamefont {Bruhwiler},\
  and\ \citenamefont {Smith}}]{Hidding5thgeneration2014}%
  \BibitemOpen
  \bibfield  {author} {\bibinfo {author} {\bibfnamefont {B.}~\bibnamefont
  {Hidding}}, \bibinfo {author} {\bibfnamefont {G.~G.}\ \bibnamefont
  {Manahan}}, \bibinfo {author} {\bibfnamefont {O.}~\bibnamefont {Karger}},
  \bibinfo {author} {\bibfnamefont {A.}~\bibnamefont {Knetsch}}, \bibinfo
  {author} {\bibfnamefont {G.}~\bibnamefont {Wittig}}, \bibinfo {author}
  {\bibfnamefont {D.~A.}\ \bibnamefont {Jaroszynski}}, \bibinfo {author}
  {\bibfnamefont {Z.-M.}\ \bibnamefont {Sheng}}, \bibinfo {author}
  {\bibfnamefont {Y.}~\bibnamefont {Xi}}, \bibinfo {author} {\bibfnamefont
  {A.}~\bibnamefont {Deng}}, \bibinfo {author} {\bibfnamefont {J.~B.}\
  \bibnamefont {Rosenzweig}}, \bibinfo {author} {\bibfnamefont
  {G.}~\bibnamefont {Andonian}}, \bibinfo {author} {\bibfnamefont
  {A.}~\bibnamefont {Murokh}}, \bibinfo {author} {\bibfnamefont
  {G.}~\bibnamefont {Pretzler}}, \bibinfo {author} {\bibfnamefont {D.~L.}\
  \bibnamefont {Bruhwiler}},\ and\ \bibinfo {author} {\bibfnamefont
  {J.}~\bibnamefont {Smith}},\ }\bibfield  {title} {\bibinfo {title} {Ultrahigh
  brightness bunches from hybrid plasma accelerators as drivers of 5th
  generation light sources},\ }\href
  {http://stacks.iop.org/0953-4075/47/i=23/a=234010} {\bibfield  {journal}
  {\bibinfo  {journal} {Journal of Physics B: Atomic, Molecular and Optical
  Physics}\ }\textbf {\bibinfo {volume} {47}},\ \bibinfo {pages} {234010}
  (\bibinfo {year} {2014}{\natexlab{b}})}\BibitemShut {NoStop}%
\bibitem [{\citenamefont {Schroeder}\ \emph {et~al.}(2014)\citenamefont
  {Schroeder}, \citenamefont {Vay}, \citenamefont {Esarey}, \citenamefont
  {Bulanov}, \citenamefont {Benedetti}, \citenamefont {Yu}, \citenamefont
  {Chen}, \citenamefont {Geddes},\ and\ \citenamefont
  {Leemans}}]{ThermalEmittancePhysRevSTAB.17.101301Schroeder2014}%
  \BibitemOpen
  \bibfield  {author} {\bibinfo {author} {\bibfnamefont {C.~B.}\ \bibnamefont
  {Schroeder}}, \bibinfo {author} {\bibfnamefont {J.-L.}\ \bibnamefont {Vay}},
  \bibinfo {author} {\bibfnamefont {E.}~\bibnamefont {Esarey}}, \bibinfo
  {author} {\bibfnamefont {S.~S.}\ \bibnamefont {Bulanov}}, \bibinfo {author}
  {\bibfnamefont {C.}~\bibnamefont {Benedetti}}, \bibinfo {author}
  {\bibfnamefont {L.-L.}\ \bibnamefont {Yu}}, \bibinfo {author} {\bibfnamefont
  {M.}~\bibnamefont {Chen}}, \bibinfo {author} {\bibfnamefont {C.~G.~R.}\
  \bibnamefont {Geddes}},\ and\ \bibinfo {author} {\bibfnamefont {W.~P.}\
  \bibnamefont {Leemans}},\ }\bibfield  {title} {\bibinfo {title} {Thermal
  emittance from ionization-induced trapping in plasma accelerators},\ }\href
  {https://doi.org/10.1103/PhysRevSTAB.17.101301} {\bibfield  {journal}
  {\bibinfo  {journal} {Phys. Rev. ST Accel. Beams}\ }\textbf {\bibinfo
  {volume} {17}},\ \bibinfo {pages} {101301} (\bibinfo {year}
  {2014})}\BibitemShut {NoStop}%
\bibitem [{\citenamefont {Xu}\ \emph {et~al.}(2014)\citenamefont {Xu} \emph
  {et~al.}}]{PhysRevLett.112.035003Xu2014}%
  \BibitemOpen
  \bibfield  {author} {\bibinfo {author} {\bibfnamefont {X.~L.}\ \bibnamefont
  {Xu}} \emph {et~al.},\ }\bibfield  {title} {\bibinfo {title} {Phase-space
  dynamics of ionization injection in plasma-based accelerators},\ }\href
  {https://doi.org/10.1103/PhysRevLett.112.035003} {\bibfield  {journal}
  {\bibinfo  {journal} {Phys. Rev. Lett.}\ }\textbf {\bibinfo {volume} {112}},\
  \bibinfo {pages} {035003} (\bibinfo {year} {2014})}\BibitemShut {NoStop}%
\bibitem [{\citenamefont {Yu}\ \emph {et~al.}(2014)\citenamefont {Yu},
  \citenamefont {Esarey}, \citenamefont {Schroeder}, \citenamefont {Vay},
  \citenamefont {Benedetti}, \citenamefont {Geddes}, \citenamefont {Chen},\
  and\ \citenamefont {Leemans}}]{YuTwo-colrPhysRevLett.112.125001}%
  \BibitemOpen
  \bibfield  {author} {\bibinfo {author} {\bibfnamefont {L.-L.}\ \bibnamefont
  {Yu}}, \bibinfo {author} {\bibfnamefont {E.}~\bibnamefont {Esarey}}, \bibinfo
  {author} {\bibfnamefont {C.}~\bibnamefont {Schroeder}}, \bibinfo {author}
  {\bibfnamefont {J.-L.}\ \bibnamefont {Vay}}, \bibinfo {author} {\bibfnamefont
  {C.}~\bibnamefont {Benedetti}}, \bibinfo {author} {\bibfnamefont
  {C.}~\bibnamefont {Geddes}}, \bibinfo {author} {\bibfnamefont
  {M.}~\bibnamefont {Chen}},\ and\ \bibinfo {author} {\bibfnamefont
  {W.}~\bibnamefont {Leemans}},\ }\bibfield  {title} {\bibinfo {title}
  {Two-color laser-ionization injection},\ }\href
  {https://doi.org/10.1103/PhysRevLett.112.125001} {\bibfield  {journal}
  {\bibinfo  {journal} {Phys. Rev. Lett.}\ }\textbf {\bibinfo {volume} {112}},\
  \bibinfo {pages} {125001} (\bibinfo {year} {2014})}\BibitemShut {NoStop}%
\bibitem [{\citenamefont {Moon}\ \emph {et~al.}(2019)\citenamefont {Moon},
  \citenamefont {Kumar}, \citenamefont {Hur},\ and\ \citenamefont
  {Chung}}]{Moon2019doi:10.1063/1.5108928}%
  \BibitemOpen
  \bibfield  {author} {\bibinfo {author} {\bibfnamefont {K.}~\bibnamefont
  {Moon}}, \bibinfo {author} {\bibfnamefont {S.}~\bibnamefont {Kumar}},
  \bibinfo {author} {\bibfnamefont {M.}~\bibnamefont {Hur}},\ and\ \bibinfo
  {author} {\bibfnamefont {M.}~\bibnamefont {Chung}},\ }\bibfield  {title}
  {\bibinfo {title} {Longitudinal phase space dynamics of witness bunch during
  the trojan horse injection for plasma-based particle accelerators},\ }\href
  {https://doi.org/10.1063/1.5108928} {\bibfield  {journal} {\bibinfo
  {journal} {Physics of Plasmas}\ }\textbf {\bibinfo {volume} {26}},\ \bibinfo
  {pages} {073103} (\bibinfo {year} {2019})},\ \Eprint
  {https://arxiv.org/abs/https://doi.org/10.1063/1.5108928}
  {https://doi.org/10.1063/1.5108928} \BibitemShut {NoStop}%
\bibitem [{\citenamefont {Turner}\ \emph {et~al.}(2016)\citenamefont {Turner},
  \citenamefont {Akre}, \citenamefont {Brachmann}, \citenamefont {Decker},
  \citenamefont {Ding}, \citenamefont {Emma}, \citenamefont {Feng},
  \citenamefont {Fisher}, \citenamefont {Frisch}, \citenamefont {Gilevich}
  \emph {et~al.}}]{turner2016fel}%
  \BibitemOpen
  \bibfield  {author} {\bibinfo {author} {\bibfnamefont {J.}~\bibnamefont
  {Turner}}, \bibinfo {author} {\bibfnamefont {R.}~\bibnamefont {Akre}},
  \bibinfo {author} {\bibfnamefont {A.}~\bibnamefont {Brachmann}}, \bibinfo
  {author} {\bibfnamefont {F.-J.}\ \bibnamefont {Decker}}, \bibinfo {author}
  {\bibfnamefont {Y.}~\bibnamefont {Ding}}, \bibinfo {author} {\bibfnamefont
  {P.}~\bibnamefont {Emma}}, \bibinfo {author} {\bibfnamefont {Y.}~\bibnamefont
  {Feng}}, \bibinfo {author} {\bibfnamefont {A.}~\bibnamefont {Fisher}},
  \bibinfo {author} {\bibfnamefont {J.}~\bibnamefont {Frisch}}, \bibinfo
  {author} {\bibfnamefont {A.}~\bibnamefont {Gilevich}}, \emph {et~al.},\
  }\bibfield  {title} {\bibinfo {title} {Fel beam stability in the lcls},\ }in\
  \href@noop {} {\emph {\bibinfo {booktitle} {Conf. Proc. C110328: 2423-2425,
  2011}}},\ \bibinfo {series and number} {\bibinfo {number} {SLAC-PUB-16660}}\
  (\bibinfo {organization} {SLAC National Accelerator Lab., Menlo Park, CA
  (United States)},\ \bibinfo {year} {2016})\BibitemShut {NoStop}%
\bibitem [{\citenamefont {Steinke}\ \emph {et~al.}(2016)\citenamefont
  {Steinke}, \citenamefont {van Tilborg}, \citenamefont {Benedetti},
  \citenamefont {Geddes}, \citenamefont {Schroeder}, \citenamefont {Daniels},
  \citenamefont {Swanson}, \citenamefont {Gonsalves}, \citenamefont {Nakamura},
  \citenamefont {Matlis}, \citenamefont {Shaw}, \citenamefont {Esarey},\ and\
  \citenamefont {Leemans}}]{SteinkeNature2016}%
  \BibitemOpen
  \bibfield  {author} {\bibinfo {author} {\bibfnamefont {S.}~\bibnamefont
  {Steinke}}, \bibinfo {author} {\bibfnamefont {J.}~\bibnamefont {van
  Tilborg}}, \bibinfo {author} {\bibfnamefont {C.}~\bibnamefont {Benedetti}},
  \bibinfo {author} {\bibfnamefont {C.~G.~R.}\ \bibnamefont {Geddes}}, \bibinfo
  {author} {\bibfnamefont {C.~B.}\ \bibnamefont {Schroeder}}, \bibinfo {author}
  {\bibfnamefont {J.}~\bibnamefont {Daniels}}, \bibinfo {author} {\bibfnamefont
  {K.~K.}\ \bibnamefont {Swanson}}, \bibinfo {author} {\bibfnamefont {A.~J.}\
  \bibnamefont {Gonsalves}}, \bibinfo {author} {\bibfnamefont {K.}~\bibnamefont
  {Nakamura}}, \bibinfo {author} {\bibfnamefont {N.~H.}\ \bibnamefont
  {Matlis}}, \bibinfo {author} {\bibfnamefont {B.~H.}\ \bibnamefont {Shaw}},
  \bibinfo {author} {\bibfnamefont {E.}~\bibnamefont {Esarey}},\ and\ \bibinfo
  {author} {\bibfnamefont {W.~P.}\ \bibnamefont {Leemans}},\ }\bibfield
  {title} {\bibinfo {title} {Multistage coupling of independent laser-plasma
  accelerators},\ }\href {https://doi.org/10.1038/nature16525} {\bibfield
  {journal} {\bibinfo  {journal} {Nature}\ }\textbf {\bibinfo {volume} {530}},\
  \bibinfo {pages} {190} (\bibinfo {year} {2016})}\BibitemShut {NoStop}%
\bibitem [{\citenamefont {Wu}\ \emph {et~al.}(2021)\citenamefont {Wu},
  \citenamefont {Hua}, \citenamefont {Zhou}, \citenamefont {Zhang},
  \citenamefont {Liu}, \citenamefont {Peng}, \citenamefont {Fang},
  \citenamefont {Ning}, \citenamefont {Nie}, \citenamefont {Li}, \citenamefont
  {Zhang}, \citenamefont {Pai}, \citenamefont {Du}, \citenamefont {Lu},
  \citenamefont {Mori},\ and\ \citenamefont {Joshi}}]{WuStagingNature2021}%
  \BibitemOpen
  \bibfield  {author} {\bibinfo {author} {\bibfnamefont {Y.}~\bibnamefont
  {Wu}}, \bibinfo {author} {\bibfnamefont {J.}~\bibnamefont {Hua}}, \bibinfo
  {author} {\bibfnamefont {Z.}~\bibnamefont {Zhou}}, \bibinfo {author}
  {\bibfnamefont {J.}~\bibnamefont {Zhang}}, \bibinfo {author} {\bibfnamefont
  {S.}~\bibnamefont {Liu}}, \bibinfo {author} {\bibfnamefont {B.}~\bibnamefont
  {Peng}}, \bibinfo {author} {\bibfnamefont {Y.}~\bibnamefont {Fang}}, \bibinfo
  {author} {\bibfnamefont {X.}~\bibnamefont {Ning}}, \bibinfo {author}
  {\bibfnamefont {Z.}~\bibnamefont {Nie}}, \bibinfo {author} {\bibfnamefont
  {F.}~\bibnamefont {Li}}, \bibinfo {author} {\bibfnamefont {C.}~\bibnamefont
  {Zhang}}, \bibinfo {author} {\bibfnamefont {C.-H.}\ \bibnamefont {Pai}},
  \bibinfo {author} {\bibfnamefont {Y.}~\bibnamefont {Du}}, \bibinfo {author}
  {\bibfnamefont {W.}~\bibnamefont {Lu}}, \bibinfo {author} {\bibfnamefont
  {W.~B.}\ \bibnamefont {Mori}},\ and\ \bibinfo {author} {\bibfnamefont
  {C.}~\bibnamefont {Joshi}},\ }\bibfield  {title} {\bibinfo {title}
  {High-throughput injection--acceleration of electron bunches from a linear
  accelerator to a laser wakefield accelerator},\ }\href
  {https://doi.org/10.1038/s41567-021-01202-6} {\bibfield  {journal} {\bibinfo
  {journal} {Nature Physics}\ }\textbf {\bibinfo {volume} {17}},\ \bibinfo
  {pages} {801} (\bibinfo {year} {2021})}\BibitemShut {NoStop}%
\bibitem [{\citenamefont {Emma}(2010)}]{Emma2010fyaPAC}%
  \BibitemOpen
  \bibfield  {author} {\bibinfo {author} {\bibfnamefont {P.}~\bibnamefont
  {Emma}},\ }\bibfield  {title} {\bibinfo {title} {{First Lasing of the LCLS
  X-Ray FEL at 1.5 Å}},\ }in\ \href@noop {} {\emph {\bibinfo {booktitle}
  {{Particle Accelerator Conference (PAC 09)}}}}\ (\bibinfo {year} {2010})\ p.\
  \bibinfo {pages} {TH3PBI01}\BibitemShut {NoStop}%
\bibitem [{\citenamefont {{Borland}}\ \emph {et~al.}(2001)\citenamefont
  {{Borland}}, \citenamefont {{Chae}}, \citenamefont {{Milton}}, \citenamefont
  {{Soliday}}, \citenamefont {{Bharadwaj}}, \citenamefont {{Emma}},
  \citenamefont {{Krejcik}}, \citenamefont {{Limborg}}, \citenamefont
  {{Nuhn}},\ and\ \citenamefont {{Woodley}}}]{987880LCLSjitterPAC2001}%
  \BibitemOpen
  \bibfield  {author} {\bibinfo {author} {\bibfnamefont {M.}~\bibnamefont
  {{Borland}}}, \bibinfo {author} {\bibfnamefont {Y.~.}\ \bibnamefont
  {{Chae}}}, \bibinfo {author} {\bibfnamefont {S.}~\bibnamefont {{Milton}}},
  \bibinfo {author} {\bibfnamefont {R.}~\bibnamefont {{Soliday}}}, \bibinfo
  {author} {\bibfnamefont {V.}~\bibnamefont {{Bharadwaj}}}, \bibinfo {author}
  {\bibfnamefont {P.}~\bibnamefont {{Emma}}}, \bibinfo {author} {\bibfnamefont
  {P.}~\bibnamefont {{Krejcik}}}, \bibinfo {author} {\bibfnamefont
  {C.}~\bibnamefont {{Limborg}}}, \bibinfo {author} {\bibfnamefont {H.~.}\
  \bibnamefont {{Nuhn}}},\ and\ \bibinfo {author} {\bibfnamefont
  {M.}~\bibnamefont {{Woodley}}},\ }\bibfield  {title} {\bibinfo {title}
  {Start-to-end jitter simulations of the linac coherent light source},\ }in\
  \href {https://doi.org/10.1109/PAC.2001.987880} {\emph {\bibinfo {booktitle}
  {PACS2001. Proceedings of the 2001 Particle Accelerator Conference (Cat.
  No.01CH37268)}}},\ Vol.~\bibinfo {volume} {4}\ (\bibinfo {year} {2001})\ pp.\
  \bibinfo {pages} {2707--2709 vol.4}\BibitemShut {NoStop}%
\bibitem [{\citenamefont {Mehrling}\ \emph {et~al.}(2017)\citenamefont
  {Mehrling}, \citenamefont {Fonseca}, \citenamefont {Martinez de~la Ossa},\
  and\ \citenamefont {Vieira}}]{PhysRevLett.118.174801}%
  \BibitemOpen
  \bibfield  {author} {\bibinfo {author} {\bibfnamefont {T.~J.}\ \bibnamefont
  {Mehrling}}, \bibinfo {author} {\bibfnamefont {R.~A.}\ \bibnamefont
  {Fonseca}}, \bibinfo {author} {\bibfnamefont {A.}~\bibnamefont {Martinez
  de~la Ossa}},\ and\ \bibinfo {author} {\bibfnamefont {J.}~\bibnamefont
  {Vieira}},\ }\bibfield  {title} {\bibinfo {title} {Mitigation of the hose
  instability in plasma-wakefield accelerators},\ }\href
  {https://doi.org/10.1103/PhysRevLett.118.174801} {\bibfield  {journal}
  {\bibinfo  {journal} {Phys. Rev. Lett.}\ }\textbf {\bibinfo {volume} {118}},\
  \bibinfo {pages} {174801} (\bibinfo {year} {2017})}\BibitemShut {NoStop}%
\bibitem [{\citenamefont {de~la Ossa}\ \emph {et~al.}(2018)\citenamefont {de~la
  Ossa}, \citenamefont {Mehrling},\ and\ \citenamefont
  {Osterhoff}}]{alberto2018stability}%
  \BibitemOpen
  \bibfield  {author} {\bibinfo {author} {\bibfnamefont {A.~M.}\ \bibnamefont
  {de~la Ossa}}, \bibinfo {author} {\bibfnamefont {T.}~\bibnamefont
  {Mehrling}},\ and\ \bibinfo {author} {\bibfnamefont {J.}~\bibnamefont
  {Osterhoff}},\ }\bibfield  {title} {\bibinfo {title} {Intrinsic stabilization
  of the drive beam in plasma wakefield accelerators},\ }\href
  {https://journals.aps.org/prl/accepted/7607eY34Dab12359b9927c86802a99d4a37a9d069}
  {\bibfield  {journal} {\bibinfo  {journal} {Phys. Rev. Lett.}\ } (\bibinfo
  {year} {2018})}\BibitemShut {NoStop}%
\bibitem [{\citenamefont {Pousa}\ \emph {et~al.}(2019)\citenamefont {Pousa},
  \citenamefont {de~la Ossa},\ and\ \citenamefont
  {Assmann}}]{pousa2019intrinsic}%
  \BibitemOpen
  \bibfield  {author} {\bibinfo {author} {\bibfnamefont {A.~F.}\ \bibnamefont
  {Pousa}}, \bibinfo {author} {\bibfnamefont {A.~M.}\ \bibnamefont {de~la
  Ossa}},\ and\ \bibinfo {author} {\bibfnamefont {R.~W.}\ \bibnamefont
  {Assmann}},\ }\bibfield  {title} {\bibinfo {title} {Intrinsic energy spread
  and bunch length growth in plasma-based accelerators due to betatron
  motion},\ }\href@noop {} {\bibfield  {journal} {\bibinfo  {journal}
  {Scientific Reports}\ }\textbf {\bibinfo {volume} {9}},\ \bibinfo {pages} {1}
  (\bibinfo {year} {2019})}\BibitemShut {NoStop}%
\bibitem [{\citenamefont {Cinquegrana}\ \emph {et~al.}(2014)\citenamefont
  {Cinquegrana}, \citenamefont {Cleva}, \citenamefont {Demidovich},
  \citenamefont {Gaio}, \citenamefont {Ivanov}, \citenamefont {Kurdi},
  \citenamefont {Nikolov}, \citenamefont {Sigalotti},\ and\ \citenamefont
  {Danailov}}]{PhysRevSTABCinquegranaPumpProbeJitter2014}%
  \BibitemOpen
  \bibfield  {author} {\bibinfo {author} {\bibfnamefont {P.}~\bibnamefont
  {Cinquegrana}}, \bibinfo {author} {\bibfnamefont {S.}~\bibnamefont {Cleva}},
  \bibinfo {author} {\bibfnamefont {A.}~\bibnamefont {Demidovich}}, \bibinfo
  {author} {\bibfnamefont {G.}~\bibnamefont {Gaio}}, \bibinfo {author}
  {\bibfnamefont {R.}~\bibnamefont {Ivanov}}, \bibinfo {author} {\bibfnamefont
  {G.}~\bibnamefont {Kurdi}}, \bibinfo {author} {\bibfnamefont
  {I.}~\bibnamefont {Nikolov}}, \bibinfo {author} {\bibfnamefont
  {P.}~\bibnamefont {Sigalotti}},\ and\ \bibinfo {author} {\bibfnamefont
  {M.~B.}\ \bibnamefont {Danailov}},\ }\bibfield  {title} {\bibinfo {title}
  {Optical beam transport to a remote location for low jitter pump-probe
  experiments with a free electron laser},\ }\href
  {https://doi.org/10.1103/PhysRevSTAB.17.040702} {\bibfield  {journal}
  {\bibinfo  {journal} {Phys. Rev. ST Accel. Beams}\ }\textbf {\bibinfo
  {volume} {17}},\ \bibinfo {pages} {040702} (\bibinfo {year}
  {2014})}\BibitemShut {NoStop}%
\bibitem [{\citenamefont {Alotaibi}\ \emph {et~al.}(2020)\citenamefont
  {Alotaibi}, \citenamefont {Altuijri}, \citenamefont {Habib}, \citenamefont
  {Hala}, \citenamefont {Hidding}, \citenamefont {Khalil}, \citenamefont
  {McNeil},\ and\ \citenamefont {Traczykowski}}]{alotaibi2020plasma}%
  \BibitemOpen
  \bibfield  {author} {\bibinfo {author} {\bibfnamefont {B.~M.}\ \bibnamefont
  {Alotaibi}}, \bibinfo {author} {\bibfnamefont {R.}~\bibnamefont {Altuijri}},
  \bibinfo {author} {\bibfnamefont {A.}~\bibnamefont {Habib}}, \bibinfo
  {author} {\bibfnamefont {A.}~\bibnamefont {Hala}}, \bibinfo {author}
  {\bibfnamefont {B.}~\bibnamefont {Hidding}}, \bibinfo {author} {\bibfnamefont
  {S.~M.}\ \bibnamefont {Khalil}}, \bibinfo {author} {\bibfnamefont
  {B.}~\bibnamefont {McNeil}},\ and\ \bibinfo {author} {\bibfnamefont
  {P.}~\bibnamefont {Traczykowski}},\ }\bibfield  {title} {\bibinfo {title}
  {Plasma wakefield accelerator driven coherent spontaneous emission from an
  energy chirped electron pulse},\ }\href@noop {} {\bibfield  {journal}
  {\bibinfo  {journal} {New Journal of Physics}\ }\textbf {\bibinfo {volume}
  {22}},\ \bibinfo {pages} {013037} (\bibinfo {year} {2020})}\BibitemShut
  {NoStop}%
\bibitem [{\citenamefont {Emma}\ \emph {et~al.}(2021)\citenamefont {Emma},
  \citenamefont {Xu}, \citenamefont {Fisher}, \citenamefont {Robles},
  \citenamefont {MacArthur}, \citenamefont {Cryan}, \citenamefont {Hogan},
  \citenamefont {Musumeci}, \citenamefont {White},\ and\ \citenamefont
  {Marinelli}}]{emma2021terawatt}%
  \BibitemOpen
  \bibfield  {author} {\bibinfo {author} {\bibfnamefont {C.}~\bibnamefont
  {Emma}}, \bibinfo {author} {\bibfnamefont {X.}~\bibnamefont {Xu}}, \bibinfo
  {author} {\bibfnamefont {A.}~\bibnamefont {Fisher}}, \bibinfo {author}
  {\bibfnamefont {R.}~\bibnamefont {Robles}}, \bibinfo {author} {\bibfnamefont
  {J.}~\bibnamefont {MacArthur}}, \bibinfo {author} {\bibfnamefont
  {J.}~\bibnamefont {Cryan}}, \bibinfo {author} {\bibfnamefont
  {M.}~\bibnamefont {Hogan}}, \bibinfo {author} {\bibfnamefont
  {P.}~\bibnamefont {Musumeci}}, \bibinfo {author} {\bibfnamefont
  {G.}~\bibnamefont {White}},\ and\ \bibinfo {author} {\bibfnamefont
  {A.}~\bibnamefont {Marinelli}},\ }\bibfield  {title} {\bibinfo {title}
  {Terawatt attosecond x-ray source driven by a plasma accelerator},\
  }\href@noop {} {\bibfield  {journal} {\bibinfo  {journal} {APL Photonics}\
  }\textbf {\bibinfo {volume} {6}},\ \bibinfo {pages} {076107} (\bibinfo {year}
  {2021})}\BibitemShut {NoStop}%
\bibitem [{\citenamefont {Kim}\ \emph {et~al.}(2004)\citenamefont {Kim} \emph
  {et~al.}}]{ANLbrightnessWhitePaper2003}%
  \BibitemOpen
  \bibfield  {author} {\bibinfo {author} {\bibfnamefont {K.-J.}\ \bibnamefont
  {Kim}} \emph {et~al.},\ }\href
  {https://www.aps.anl.gov/files/APS-sync/lsnotes/files/APS_1418245.pdf} {\emph
  {\bibinfo {title} {Towards Advanced Electron Beam Brightness Enhancement and
  Conditioning}}},\ \bibinfo {type} {Tech. Rep.}\ \bibinfo {number}
  {ANL/APS/LS-305}\ (\bibinfo {address} {Argonne National Laboratory},\
  \bibinfo {year} {2004})\BibitemShut {NoStop}%
\bibitem [{\citenamefont {Group}(2020)}]{European:2720131}%
  \BibitemOpen
  \bibfield  {author} {\bibinfo {author} {\bibfnamefont {T.~E.~S.}\
  \bibnamefont {Group}},\ }\href {https://doi.org/10.17181/ESU2020Deliberation}
  {\emph {\bibinfo {title} {{Deliberation document on the 2020 Update of the
  European Strategy for Particle Physics}}}},\ \bibinfo {type} {Tech. Rep.}\
  \bibinfo {number} {CERN-ESU-014}\ (\bibinfo {address} {Geneva},\ \bibinfo
  {year} {2020})\BibitemShut {NoStop}%
\bibitem [{\citenamefont {Hidding}\ \emph
  {et~al.}(2019{\natexlab{b}})\citenamefont {Hidding} \emph
  {et~al.}}]{PWFA-FEL}%
  \BibitemOpen
  \bibfield  {author} {\bibinfo {author} {\bibnamefont {Hidding}} \emph
  {et~al.},\ }\href@noop {} {\bibinfo {title} {{STFC PWFA-FEL}: Exploratory
  study of {PWFA}-driven {FEL} at {CLARA}}} (\bibinfo {year}
  {2019}{\natexlab{b}}),\ \bibinfo {note}
  {https://pwfa-fel.phys.strath.ac.uk/}\BibitemShut {NoStop}%
\bibitem [{\citenamefont {Marangos}\ \emph {et~al.}(2020)\citenamefont
  {Marangos} \emph {et~al.}}]{UKXFEL-ScienceCase2020}%
  \BibitemOpen
  \bibfield  {author} {\bibinfo {author} {\bibfnamefont {J.}~\bibnamefont
  {Marangos}} \emph {et~al.},\ }\href@noop {} {\bibinfo {title} {Uk xfel
  science case}} (\bibinfo {year} {2020}),\ \bibinfo {note}
  {https://stfc.ukri.org/files/uk-xfel-science-case/}\BibitemShut {NoStop}%
\bibitem [{\citenamefont {Fano}(1969)}]{Fano1969PhysRev.178.131}%
  \BibitemOpen
  \bibfield  {author} {\bibinfo {author} {\bibfnamefont {U.}~\bibnamefont
  {Fano}},\ }\bibfield  {title} {\bibinfo {title} {Spin orientation of
  photoelectrons ejected by circularly polarized light},\ }\href
  {https://doi.org/10.1103/PhysRev.178.131} {\bibfield  {journal} {\bibinfo
  {journal} {Phys. Rev.}\ }\textbf {\bibinfo {volume} {178}},\ \bibinfo {pages}
  {131} (\bibinfo {year} {1969})}\BibitemShut {NoStop}%
\bibitem [{\citenamefont {Nie}\ \emph {et~al.}(2021)\citenamefont {Nie},
  \citenamefont {Li}, \citenamefont {Morales}, \citenamefont {Patchkovskii},
  \citenamefont {Smirnova}, \citenamefont {An}, \citenamefont {Nambu},
  \citenamefont {Matteo}, \citenamefont {Marsh}, \citenamefont {Tsung},
  \citenamefont {Mori},\ and\ \citenamefont {Joshi}}]{JoshiSpin2021}%
  \BibitemOpen
  \bibfield  {author} {\bibinfo {author} {\bibfnamefont {Z.}~\bibnamefont
  {Nie}}, \bibinfo {author} {\bibfnamefont {F.}~\bibnamefont {Li}}, \bibinfo
  {author} {\bibfnamefont {F.}~\bibnamefont {Morales}}, \bibinfo {author}
  {\bibfnamefont {S.}~\bibnamefont {Patchkovskii}}, \bibinfo {author}
  {\bibfnamefont {O.}~\bibnamefont {Smirnova}}, \bibinfo {author}
  {\bibfnamefont {W.}~\bibnamefont {An}}, \bibinfo {author} {\bibfnamefont
  {N.}~\bibnamefont {Nambu}}, \bibinfo {author} {\bibfnamefont
  {D.}~\bibnamefont {Matteo}}, \bibinfo {author} {\bibfnamefont {K.~A.}\
  \bibnamefont {Marsh}}, \bibinfo {author} {\bibfnamefont {F.}~\bibnamefont
  {Tsung}}, \bibinfo {author} {\bibfnamefont {W.~B.}\ \bibnamefont {Mori}},\
  and\ \bibinfo {author} {\bibfnamefont {C.}~\bibnamefont {Joshi}},\ }\bibfield
   {title} {\bibinfo {title} {In situ generation of high-energy spin-polarized
  electrons in a beam-driven plasma wakefield accelerator},\ }\href
  {https://doi.org/10.1103/PhysRevLett.126.054801} {\bibfield  {journal}
  {\bibinfo  {journal} {Phys. Rev. Lett.}\ }\textbf {\bibinfo {volume} {126}},\
  \bibinfo {pages} {054801} (\bibinfo {year} {2021})}\BibitemShut {NoStop}%
\bibitem [{\citenamefont {Yakimenko}\ \emph {et~al.}(2019)\citenamefont
  {Yakimenko}, \citenamefont {Meuren}, \citenamefont {Del~Gaudio},
  \citenamefont {Baumann}, \citenamefont {Fedotov}, \citenamefont {Fiuza},
  \citenamefont {Grismayer}, \citenamefont {Hogan}, \citenamefont {Pukhov},
  \citenamefont {Silva},\ and\ \citenamefont
  {White}}]{Yakimenko2019PhysRevLett.122.190404}%
  \BibitemOpen
  \bibfield  {author} {\bibinfo {author} {\bibfnamefont {V.}~\bibnamefont
  {Yakimenko}}, \bibinfo {author} {\bibfnamefont {S.}~\bibnamefont {Meuren}},
  \bibinfo {author} {\bibfnamefont {F.}~\bibnamefont {Del~Gaudio}}, \bibinfo
  {author} {\bibfnamefont {C.}~\bibnamefont {Baumann}}, \bibinfo {author}
  {\bibfnamefont {A.}~\bibnamefont {Fedotov}}, \bibinfo {author} {\bibfnamefont
  {F.}~\bibnamefont {Fiuza}}, \bibinfo {author} {\bibfnamefont
  {T.}~\bibnamefont {Grismayer}}, \bibinfo {author} {\bibfnamefont {M.~J.}\
  \bibnamefont {Hogan}}, \bibinfo {author} {\bibfnamefont {A.}~\bibnamefont
  {Pukhov}}, \bibinfo {author} {\bibfnamefont {L.~O.}\ \bibnamefont {Silva}},\
  and\ \bibinfo {author} {\bibfnamefont {G.}~\bibnamefont {White}},\ }\bibfield
   {title} {\bibinfo {title} {Prospect of studying nonperturbative qed with
  beam-beam collisions},\ }\href
  {https://doi.org/10.1103/PhysRevLett.122.190404} {\bibfield  {journal}
  {\bibinfo  {journal} {Phys. Rev. Lett.}\ }\textbf {\bibinfo {volume} {122}},\
  \bibinfo {pages} {190404} (\bibinfo {year} {2019})}\BibitemShut {NoStop}%
\bibitem [{\citenamefont {Chen}\ \emph {et~al.}(1989)\citenamefont {Chen},
  \citenamefont {Rajagopalan},\ and\ \citenamefont
  {Rosenzweig}}]{chen:1989prd}%
  \BibitemOpen
  \bibfield  {author} {\bibinfo {author} {\bibfnamefont {P.}~\bibnamefont
  {Chen}}, \bibinfo {author} {\bibfnamefont {S.}~\bibnamefont {Rajagopalan}},\
  and\ \bibinfo {author} {\bibfnamefont {J.}~\bibnamefont {Rosenzweig}},\
  }\bibfield  {title} {\bibinfo {title} {Final focusing and enhanced disruption
  from an underdense plasma lens in a linear collider},\ }\href
  {https://doi.org/10.1103/PhysRevD.40.923} {\bibfield  {journal} {\bibinfo
  {journal} {Physical Review D}\ }\textbf {\bibinfo {volume} {40}},\ \bibinfo
  {pages} {923} (\bibinfo {year} {1989})}\BibitemShut {NoStop}%
\bibitem [{\citenamefont {Chen}\ \emph {et~al.}(1990)\citenamefont {Chen},
  \citenamefont {Oide}, \citenamefont {Sessler},\ and\ \citenamefont
  {Yu}}]{chen:1990prl}%
  \BibitemOpen
  \bibfield  {author} {\bibinfo {author} {\bibfnamefont {P.}~\bibnamefont
  {Chen}}, \bibinfo {author} {\bibfnamefont {K.}~\bibnamefont {Oide}}, \bibinfo
  {author} {\bibfnamefont {A.~M.}\ \bibnamefont {Sessler}},\ and\ \bibinfo
  {author} {\bibfnamefont {S.~S.}\ \bibnamefont {Yu}},\ }\bibfield  {title}
  {\bibinfo {title} {Plasma-based adiabatic focuser},\ }\href
  {https://doi.org/10.1103/PhysRevLett.64.1231} {\bibfield  {journal} {\bibinfo
   {journal} {Physical Review Letters}\ }\textbf {\bibinfo {volume} {64}},\
  \bibinfo {pages} {1231} (\bibinfo {year} {1990})}\BibitemShut {NoStop}%
\bibitem [{\citenamefont {Doss}\ \emph {et~al.}(2019)\citenamefont {Doss},
  \citenamefont {Adli}, \citenamefont {Ariniello}, \citenamefont {Cary},
  \citenamefont {Corde}, \citenamefont {Hidding}, \citenamefont {Hogan},
  \citenamefont {Hunt-Stone}, \citenamefont {Joshi}, \citenamefont {Marsh},
  \citenamefont {Rosenzweig}, \citenamefont {Vafaei-Najafabadi}, \citenamefont
  {Yakimenko},\ and\ \citenamefont {Litos}}]{Doss2019}%
  \BibitemOpen
  \bibfield  {author} {\bibinfo {author} {\bibfnamefont {C.}~\bibnamefont
  {Doss}}, \bibinfo {author} {\bibfnamefont {E.}~\bibnamefont {Adli}}, \bibinfo
  {author} {\bibfnamefont {R.}~\bibnamefont {Ariniello}}, \bibinfo {author}
  {\bibfnamefont {J.}~\bibnamefont {Cary}}, \bibinfo {author} {\bibfnamefont
  {S.}~\bibnamefont {Corde}}, \bibinfo {author} {\bibfnamefont
  {B.}~\bibnamefont {Hidding}}, \bibinfo {author} {\bibfnamefont
  {M.}~\bibnamefont {Hogan}}, \bibinfo {author} {\bibfnamefont
  {K.}~\bibnamefont {Hunt-Stone}}, \bibinfo {author} {\bibfnamefont
  {C.}~\bibnamefont {Joshi}}, \bibinfo {author} {\bibfnamefont
  {K.}~\bibnamefont {Marsh}}, \bibinfo {author} {\bibfnamefont
  {J.}~\bibnamefont {Rosenzweig}}, \bibinfo {author} {\bibfnamefont
  {N.}~\bibnamefont {Vafaei-Najafabadi}}, \bibinfo {author} {\bibfnamefont
  {V.}~\bibnamefont {Yakimenko}},\ and\ \bibinfo {author} {\bibfnamefont
  {M.}~\bibnamefont {Litos}},\ }\bibfield  {title} {\bibinfo {title}
  {Laser-ionized, beam-driven, underdense, passive thin plasma lens},\
  }\bibfield  {journal} {\bibinfo  {journal} {Physical Review Accelerators and
  Beams}\ }\textbf {\bibinfo {volume} {22}},\ \href
  {https://doi.org/10.1103/physrevaccelbeams.22.111001}
  {10.1103/physrevaccelbeams.22.111001} (\bibinfo {year} {2019})\BibitemShut
  {NoStop}%
\bibitem [{\citenamefont {Wittig}\ \emph {et~al.}(2016)\citenamefont {Wittig},
  \citenamefont {Karger}, \citenamefont {Knetsch}, \citenamefont {Xi},
  \citenamefont {Deng}, \citenamefont {Rosenzweig}, \citenamefont {Bruhwiler},
  \citenamefont {Smith}, \citenamefont {Sheng}, \citenamefont {Jaroszynski},
  \citenamefont {Manahan},\ and\ \citenamefont {Hidding}}]{WITTIG201683}%
  \BibitemOpen
  \bibfield  {author} {\bibinfo {author} {\bibfnamefont {G.}~\bibnamefont
  {Wittig}}, \bibinfo {author} {\bibfnamefont {O.~S.}\ \bibnamefont {Karger}},
  \bibinfo {author} {\bibfnamefont {A.}~\bibnamefont {Knetsch}}, \bibinfo
  {author} {\bibfnamefont {Y.}~\bibnamefont {Xi}}, \bibinfo {author}
  {\bibfnamefont {A.}~\bibnamefont {Deng}}, \bibinfo {author} {\bibfnamefont
  {J.~B.}\ \bibnamefont {Rosenzweig}}, \bibinfo {author} {\bibfnamefont
  {D.~L.}\ \bibnamefont {Bruhwiler}}, \bibinfo {author} {\bibfnamefont
  {J.}~\bibnamefont {Smith}}, \bibinfo {author} {\bibfnamefont {Z.-M.}\
  \bibnamefont {Sheng}}, \bibinfo {author} {\bibfnamefont {D.~A.}\ \bibnamefont
  {Jaroszynski}}, \bibinfo {author} {\bibfnamefont {G.~G.}\ \bibnamefont
  {Manahan}},\ and\ \bibinfo {author} {\bibfnamefont {B.}~\bibnamefont
  {Hidding}},\ }\bibfield  {title} {\bibinfo {title} {Electron beam
  manipulation, injection and acceleration in plasma wakefield accelerators by
  optically generated plasma density spikes},\ }\href
  {https://doi.org/https://doi.org/10.1016/j.nima.2016.02.027} {\bibfield
  {journal} {\bibinfo  {journal} {Nuclear Instruments and Methods in Physics
  Research Section A: Accelerators, Spectrometers, Detectors and Associated
  Equipment}\ }\textbf {\bibinfo {volume} {829}},\ \bibinfo {pages} {83 }
  (\bibinfo {year} {2016})},\ \bibinfo {note} {2nd European Advanced
  Accelerator Concepts Workshop - EAAC 2015}\BibitemShut {NoStop}%
\end{thebibliography}%


%

\end{document}